\documentclass[10pt]{iopart}
\usepackage{graphics}
\begin{document}

\title{New Physically Realistic Solutions for Charged Fluid Spheres}

\author{J Burke and D Hobill}

\address{Department of Physics and Astronomy,
University of Calgary, Calgary, Alberta, Canada, T2N 1N4}
\ead{hobill@crag.ucalgary.ca}

\begin{abstract}
{A new class of solutions to the coupled, spherically symmetric Einstein-Maxwell 
equations for a compact
material source is constructed.  Some of these solutions can be made to satisfy a number of
requirements for being physically relevant, including having a causal
speed of sound. In the case of vanishing charge these solutions reduce to those
found by Bayin and Tolman. Only the latter can be considered as having 
physically realistic properties.}
\end{abstract}

\vspace{2pc}
   
\section {Introduction} 

One of the most fundamental problems in general relativity is the construction
of
exact solutions to the Einstein equations (or the Einstein equations coupled to 
other equations associated with fundamental physical theories). An even more
challenging problem is is to find solutions that might
represent physically realizable systems that could possibly occur in
nature.  It is well known that of the many solutions to the systems of equations 
derived from the Einstein equations lead to unphysical behaviour.  In many cases,
the spacetimes constructed may be truly singular even in the presence of matter or 
other fields. Alternatively negative energy 
densities and/or
negative pressures are often required in order to meet regularity conditions or boundary conditions
that are imposed upon the metric or other geometric objects. 
In other cases where energy densities and
pressures are positive, often one finds that the speed of sound violates causality by
being greater than the speed of light.  For example constant density 
configurations, where the pressure has a radial dependence to ensure that the
interior solution remains static will lead to infinite sound speeds.  Charged dust solutions on
the other hand have zero sound speed. They require such a fine balance between
the mass and charge distributions that they are almost always unstable to any
perturbations. They  either undergo gravitational collapse or fly apart due to
electrostatic repulsion.

This paper will present a new solution to the spherically symmetric time independent
 Einstein-Maxwell system of equations that govern the behaviour of the space-time
in the interior of a charged fluid sphere. At the outer boundary the solutions will 
be matched to the external vacuum solution for the field equations  i.e. the
Reissner-Nordstr\"{o}m solution. It will be demonstrated that some of these
solutions have parameter values that lead to physically realistic properties. 

While the total number of known exact interior solutions to the Einstein-Maxwell equations is
much smaller than their uncharged counterparts, the number of solutions that
can be said to be physically relevant is small.  A review of such solutions
has been provided by Ivanov [1] and this makes a good starting place for understanding the different 
approaches that have been taken and the results obtained from those approaches. In some instances
corrections to some of the original papers are provided.
The most commonly used methods employ 
either isotropic coordinates (where the problem was first discussed by Papapetrou [2]
 and Majumdar [3]) or curvature (i.e. Schwarzschild-like) coordinates.  

As mentioned in the introductory paragraph above, this paper will be primarily
concerned with the construction of physically ``interesting'' solutions that might 
arise from regular initial conditions for charge and mass densities along with realistic
pressure distributions.  The motivations for this are to find compact charged fluid
configurations that might represent realistic sources for the exterior vacuum
Reissner-Nordstr\"{o}m solutions. In particular one can ask whether it is
possible to find realistic solutions for the extreme case where
the total charge is equal to the total mass.  A second motivation is to determine
the possible end states for systems that have collapsed gravitationally 
but due to the repulsive nature of the electrostatic interaction are unable to form
black holes and result in stable configurations.   

The approach to be taken here is to write the Einstein-Maxwell system in
curvature coordinates.  The Einstein equations will yield three equations
for two metric functions in terms of the mass density, the charge aspect and
the fluid pressure.  The Maxwell equations simply state the the electrostatic
Maxwell field is just the Coulomb field in a spherically symmetric space time.
The solution will be given in terms of explicit closed-form functions of the
radial coordinate for the two metric coefficients, a mass density, a
charge aspect (the integrated charge density) and the pressure of the fluid.

The review by Ivanov outlines that various ans\"{a}tse that have been utilized 
to find exact solutions. Essentially one gives two out of the five functions and uses the
Einstein-Maxwell equations to construct the remainder.  Of course some methods
of solution are easier than others, particularly if one is not concerned with 
ensuring that the solution have a physical interpretation.  The simplest method
is to give the functional dependence of the metric functions and construct the
remaining variables from the derivatives on the ``left-hand-side'' of the differential
equations.  On the other hand giving the mass and/or charge distributions along
with an equation of state of the form $P = P(\rho)$ requires
solving a set of nonlinear coupled differential equations. But the method has the
advantage that there is some control over the physics. Other methods use a
combination of the two approaches.  

The requirements that make a solution physically relevant has had much discussion
in the literature. (See e.g.~ the reviews by Delgaty and Lake [4] and Finch and Skea [5] that
analyze over one hundred known solutions for uncharged, spherically symmetric, fluid solutions
in general relativity.)  Those that we expect to meet are the following:
\begin{enumerate}
\item The metric coefficients are regular everywhere including at the origin $r=0$.
\item The metric functions match to the Reissner-Nordstr\"{o}m functions at the fluid-vacuum interface.
\item The mass density is positive and decreasing outward toward the boundary.
\item The integrated charge and mass increase outward to give the RN parameters at the boundary.
\item The pressure $P$ is positive and finite everywhere inside the fluid 
\item The pressure vanishes at the fluid boundary with the vacuum.
\item The speed of sound $v_{\rm s} = (dP/d\rho)^{1/2}$ is causal ($0\leq v \leq 1$).
\item The speed of sound $v_{\rm s} = (dP/d\rho)^{1/2}$ monotonically decreases
with increasing $r$.
\item Both the pressure and mass density are decreasing functions of $r$:
$dP/dr < 0$ and $d\rho/dr <0$.
\end{enumerate}

This list is not necessarily exhaustive but does include all of the
requirements presented by Delgaty and Lake in their extensive review of exact
fluid sphere solutions. Supplementary 
conditions that define what Finch and Skea call ``interesting
solutions'' have been reviewed in that manuscript. These extra conditions include: an explicit 
equation of state in the form $P = P(\rho)$, and non-numerical solutions to the differential equations that
governing the behaviour of some of the dependent functions. While the latter
condition will be met in what follows, the we have not succeeded in providing
an equation of state outside of giving both the pressure and mass density in
terms of functions of a radial coordinate.   

In comparing the behaviour of 127 known solutions against the requirements listed
above, Delgaty and Lake found that only 9 of those satisfied all of the
conditions.  Clearly such solutions are rare.  Many solutions have pathologies
associated with negative pressures, and/or non causal sound speeds. As will be 
shown in what follows many of the new solutions also demonstrate such
pathological behaviour.

The outline for this manuscript is as follows: the next section introduces the
coordinates, the line element, and the form of the Einstein-Maxwell equations to be solved.
A strategy for solving for the unknown functions is then presented.
Section 3 then applies the method to obtain a general form of
the solutions in terms of analytic functions. Special cases are analyzed to determine which solutions and what 
parameters lead to physically relevant behaviour by meeting all of the
conditions listed above simultaneously. Finally, the last section
concludes with some discussions.   

\section{The Einstein-Maxwell Equations}   

In curvature coordinates, the line element of a spherically symmetric, static
spacetime can be written in the form:
\begin{equation} ds^2  = e^\nu dt^2 - e^\lambda dr^2 - r^2 d\Omega^2
\label{metric1}
\end{equation}
where $d\Omega^2 = d\theta ^2 + \sin^2\theta d\varphi^2$ is the metric on 
the unit two-sphere in terms of spherical polar angles $\theta$ and $\varphi$,
and $\nu$ and $\lambda$ are $r$-dependent functions.  

The coupled Einstein Maxwell equations can be written in the following form:
\begin{eqnarray}
 \frac{\lambda _r}{r}e^{-\lambda} + \frac{1}{r^2}(1-e^{-\lambda}) & = &
\kappa \rho + \frac{q^2}{r^4}  \label{fe1} \\
 \frac{\nu _r}{r}e^{-\lambda} - \frac{1}{r^2}(1-e^{-\lambda}) & = &
\kappa P - \frac{q^2}{r^4}  \label{fe2}\\
 e^{-\lambda} \left[ \frac{\nu _{r r}}{2} - \frac{\lambda _r
\nu _r}{4} + \frac{(\nu_r)^2}{4} + \frac{\nu_r -
\lambda_r}{2r} \right] & = & \kappa P + \frac{q^2}{r^4}  \label{fe3}
\end{eqnarray}
where the subscript ($_r$) represents a derivative with respect to $r$ and $\kappa = 8
\pi G/c^4$. In what follows, geometric units are used where $G=c=1$.

The charge function $q(r)$ is obtained by integrating the charge density
$\sigma(r)$ over a spherical proper volume:
$$ q(r) = \frac{\kappa}{2}\int^r_0 \sigma(r) r^2 e^{\lambda/2} dr. $$
In what follows, it will be assumed  that the charge function is known and
this will be used to obtain the charge density once a solution for the metric
function $\lambda(r)$ is determined.

The spherical symmetry of the problem allows only a radial component to the
electric field which can be computed from an electrostatic potential, $\phi(r)$.
This leads directly to the one non-zero component of the Maxwell field tensor:
$$ F_{01} = \frac{\partial \phi}{\partial r} = -\frac{q}{r^2} e^{(\nu + \lambda)/2}. $$

Equation (2) provides a single equation for $\lambda$ and since the the left-hand-side
is a total derivative of the function $r(1-e^{-\lambda})$ multiplied by $r^{-2}$
one introduces the mass aspect: 
$$ M(r) = \frac{\kappa}{2} \int^r_0 \rho r^2 dr + \frac{1}{2} \int^r_0
\frac{q^2}{r^2} dr $$
which leads directly to an expression for $\lambda(r)$:
\begin{equation}
 z \equiv e^{-\lambda} = 1- \frac{2M}{r} \label{zdef} 
\end{equation}
Therefore if the mass density and the charge aspect are given as functions of
$r$ one can find an expression for the mass aspect and therefore for the
function $z(r)$.  

Assuming that the metric function $\nu$ has been determined one can immediately obtain an
expression for the pressure by taking a sum of the equations (2) and (3). Then
one obtains:
\begin{equation}
 \kappa (P + \rho) = \frac{e^{-\lambda}}{r} (\nu_r + \lambda _r) =
\frac{z}{r}\nu_r  - \frac{z_r}{r} \label{pressurep} 
\end{equation}

Finally using equations (2) - (4) one can obtain the following equations by
introducing the function $y = e^{\nu/2}$:
\begin{eqnarray}
 2r^2 z y_{rr} + (r^2 z_r - 2rz)y_r + \left( r z_r - 2z + 2 - \frac{4q^2}{r^2} \right)y & = & 0 \label{y1} \\ 
 2r^2 z y_{rr} + (r^2 z_r - 2rz)y_r + ( 5r z_r + 2z -2 +4 \kappa \rho r^2)y & = & 0 \label{y2} \\
 2r^2 z y_{rr} + (r^2 z_r + 6rz)y_r + ( r z_r + 2z -2 -4 \kappa P r^2) y & = &
0 \label{y3} 
\end{eqnarray}

While the third equation (9) couples to equation (6) the first two equations are
equivalent and lead to second order linear differential equations for the
function $y$. 

The procedure for constructing a solution is as follows:  ($i$) give the mass 
density and the charge aspect functions as explicit functions in the radial
coordinate $r$; ($ii$) compute the mass aspect which leads to  an explicit
solution for the function $z(r)$; ($iii$) obtain a solution for the function
$y(r)$ using the differential equation (7) or (8); ($iv$) finally solve for
the pressure, $p(r)$ using equation (6).

The advantages to this procedure are that one can begin with a radial
dependence that guarantees that some of the requirements that the solution
be physically significant are met from the very beginning.  Here it assumed
that solutions for the conditions on the metric functions will follow if
the input is physically motivated.  Given that there will be two constants
of integration associated with the solution for $y$ and some free parameters
associated with the choices for $\sigma$ and $\rho$, there should be enough
freedom to ensure that the remaining equations are satisfied.

Of course the difficulty of solving the equations for $y(r)$ depends on the 
choices that are made for $\rho(r)$ and $q(r)$.  In the remainder of this paper
we present a choice that leads to some easily solvable equations after making 
the appropriate set of coordinate transformations.  We then explore these
solutions to determine whether or not they can meet the conditions for 
physical acceptability.

\section{A Set of Solutions}

One must begin with defining the mass density as a function of the radial
coordinate and we choose:
\begin{equation}
\rho =\rho _{0}-\rho _{1}r^{2}-\rho _{2}r^{4}  \label{rho-eqn-r}
\end{equation}
The charge function must also be given and in order to ensure that it is
concentrated at the outer edge of the sphere it is given as a simple power
function:
\begin{equation}
q=q_{2}r^{4}.  \label{q-eqn-r}
\end{equation}
With these given, the first metric function we can solve for is $\lambda $.
Using the substitution $z=e^{-\lambda }$, we have
\[
z=1-\frac{2M(r)}{r},
\]
where the mass function $M(r)$ is given by
\begin{equation}
M(r)=\frac{1}{2}\int\limits_{0}^{r}\left( \kappa \rho +\frac{q^{2}}{r^{4}}%
\right) r^{2}dr.  \label{big-M}
\end{equation}
We integrate and obtain
\[
z=1-ar^{2}-br^{4}-cr^{6},
\]
where $a$, $b$, and $c$ are constants:

\begin{equation}
a=\frac{1}{3}\kappa \rho _{0};  \label{a-cons}
\end{equation}

\begin{equation}
b=\frac{1}{5}\kappa \rho _{1};  \label{b-cons}
\end{equation}
and
\[
c=\frac{1}{7}\left( q_{2}^{2}-\kappa \rho _{2}\right) .
\]

To solve for the second metric function we use the Einstein equations
written as linear second-order differential equations for $y=e^{\nu /2}$. In
particular, where subscripts denote derivatives with respect to the given
variable,
\[
2r^{2}zy_{rr}+\left( r^{2}z_{r}-2rz\right) y_{r}+\left( rz_{r}-2z+2-4\frac{%
q^{2}}{r^{2}}\right) y=0.
\]

A change of variable to $x=r^{2}$ gives a mass density from (\ref{rho-eqn-r}%
) and a charge function from (\ref{q-eqn-r}) of, respectively,
\[
\rho =\rho _{0}-\rho _{1}x-\rho _{2}x^{2}
\]
and
\[
q=q_{2}x^{2}.
\]
\bigskip The differential equation is now

\[
zy_{xx}+\frac{1}{2}z_{x}y_{x}-\left[ \frac{\kappa \rho _{1}}{10}+\frac{1}{7}%
\left( 8q_{2}^{2}-\kappa \rho _{2}\right) x\right] y=0,
\]
where
\begin{equation}
z=1-ax-bx^{2}-cx^{3}.  \label{z-eqn-x}
\end{equation}
We choose $\kappa \rho _{2}=8q_{2}^{2}$ so that $\rho _{2}>0$ and the
constant $c$ becomes
\begin{equation}
c=-q_{2}^{2}=-\frac{1}{8}\kappa \rho _{2}<0  \label{c-cons}
\end{equation}
and the differential equation, with another change of variable
\[
\xi =\int\limits_{0}^{x_{0}}\frac{dx}{\sqrt{z}}=\int\limits_{0}^{x_{0}}\frac{%
dx}{\sqrt{1-ax-bx^{2}-cx^{3}}},
\]
can now be written as the linear second-order constant coefficient
differential equation of a simple harmonic oscillator:
\begin{equation}
y_{\xi \xi }+dy=0.  \label{DE-sho}
\end{equation}
The constant $d$ depends on $\rho _{1}$:
\begin{equation}
d=\frac{1}{20}\kappa \rho _{1}=-\frac{b}{4}.  \label{d-cons}
\end{equation}
Equation (\ref{DE-sho}) has the following three types of solutions:
\begin{equation}
y=C_{1}+C_{2}\xi  \label{y-lin}
\end{equation}

\begin{equation}
y=C_{1}e^{\sqrt{-d}\xi }+C_{2}e^{-\sqrt{-d}\xi }  \label{y-exp}
\end{equation}
and
\begin{equation}
y=C_{1}\sin \left( \sqrt{d}\xi \right) +C_{2}\cos \left( \sqrt{d}\xi \right)
\label{y-trig}
\end{equation}
for $d=0$, $d<0$, and $d>0$, respectively.

The final function to be found is the pressure $P$. We have the equation
\begin{equation}
\kappa \left( P+\rho \right) =\frac{e^{-\lambda }}{r}\left( \nu _{r}+\lambda
_{r}\right) =\frac{z}{r}\nu _{r}-\frac{z_{r}}{r}  \label{P+rho-eqn}
\end{equation}
which can also be written in terms of the new coordinates $x$ and $\xi $:
\begin{equation}
\kappa P=4\sqrt{z}\frac{y_{\xi }}{y}-2z_{x}-\kappa \rho .  \label{P-eqn-xi}
\end{equation}
The constants $C_{1}$ and $C_{2}$ are determined by conditions on $y$ and
the pressure. At the boundary of the fluid sphere, which is determined by $%
P\left( x_{0}\right) =0$, the metric functions must match the external
Reissner-Nordstr\"{o}m metric, given by
\begin{equation}
e^{\nu }=e^{-\lambda }=1-\frac{2m(x_{0})}{r}-\frac{q^{2}(x_{0})}{r^{2}}.
\label{RN}
\end{equation}
That is, at the boundary we have
\begin{equation}
y(\xi _{0})=\sqrt{z(x_{0})}.  \label{Junction-RN}
\end{equation}

A possible second condition comes from setting the boundary where $\rho
\left( x_{0}\right) =0$. In this case, equation (\ref{P+rho-eqn}) implies
that the derivatives of $\nu $ and $\lambda $ with respect to $r$ must be
equal. This can be written as
\begin{equation}
2y_{\xi }(x_{0})=z_{x}(x_{0})  \label{metric primes equal}
\end{equation}

This second condition, however, is not necessarily required. The mass
density does not have to vanish at the boundary. In this case the boundary
conditions must be modified. We still impose the junction condition with the
external Reissner-Nordstr\"{o}m metric, equation (\ref{Junction-RN}),
however, since the mass density does not go to zero at the boundary,
equation (\ref{metric primes equal}) no longer applies. Instead we solve for
the pressure, given by equation (\ref{P-eqn-xi}), going to zero. We obtain
\[
\frac{y_{\xi _{0}}}{y(\xi _{0})}=\left( \frac{2z_{x}(x_{0})+\kappa \rho
(x_{0})}{4\sqrt{z(x_{0})}}\right)
\]
and since $y(\xi _{0})=\sqrt{z(x_{0})}$ from the junction condition, we have
\begin{equation}
y_{\xi _{0}}=\frac{1}{4}\left( 2z_{x}(x_{0})+\kappa \rho (x_{0})\right) .
\label{yprime-condition}
\end{equation}

The charge density $\sigma $ can be found with $q$ and $z$ determined. The
charge density is
\begin{eqnarray}
\kappa \sigma &=&\frac{2q_{r}}{r^{2}}e^{-\lambda /2}  \label{sigma-eqn} \\
&=&8\sqrt{-c}\sqrt{x}\sqrt{z}.  \nonumber
\end{eqnarray}
In addition, the total mass of the fluid sphere can be found using the
relation
\[
m(r)=M(r)+\frac{q^{2}}{2r^{2}}.
\]
where $M(r)$ is given in equation (\ref{big-M}). We solve and find the total
mass out to the boundary $x_{0}$ to be
\begin{equation}
m(x_{0})=\frac{a}{2}x_{0}^{3/2}+\frac{b}{2}x_{0}^{5/2}  \label{m-eqn-x0}
\end{equation}
while the total charge to the boundary is
\begin{equation}
q(x_{0})=\sqrt{-c}x_{0}^{2}.  \label{q-eqn-x0}
\end{equation}
Yet another condition that could be used is to match the metric functions to
the extreme Reissner-Nordstr\"{o}m solution, where we equate equations (\ref
{m-eqn-x0}) and (\ref{q-eqn-x0}):
\begin{equation}
\frac{a}{2}x_{0}^{3/2}+\frac{b}{2}x_{0}^{5/2}=\sqrt{-c}x_{0}^{2}
\label{extremeRN-cond}
\end{equation}

We now consider the following cases for the solutions of equation (\ref
{DE-sho}).

\bigskip

\subsection{Case $d=0$ ($b=\protect\rho _{1}=0$)}

\bigskip When $\rho _{1}=0$ the equation for $y$ is linear in $\xi $,
equation (\ref{y-lin}). We note that if the mass density does not go to zero
at the boundary, it is possible to have a positive pressure solution. That
is, the contribution from the mass must be greater than from the charge. In
this case, we use the junction condition (\ref{Junction-RN}) and equation (%
\ref{yprime-condition}) to solve for the constants $C_{1}$ and $C_{2}$. We
find
\[
C_{1}=\sqrt{z(x_{0})}-C_{2}\xi _{0},
\]
as before, and
\[
C_{2}=\frac{a}{4}+\frac{c}{2}x_{0}^{2}.
\]

\begin{figure}[h]
\begin{center}
{\scalebox{.30}{\rotatebox{-90}{\includegraphics{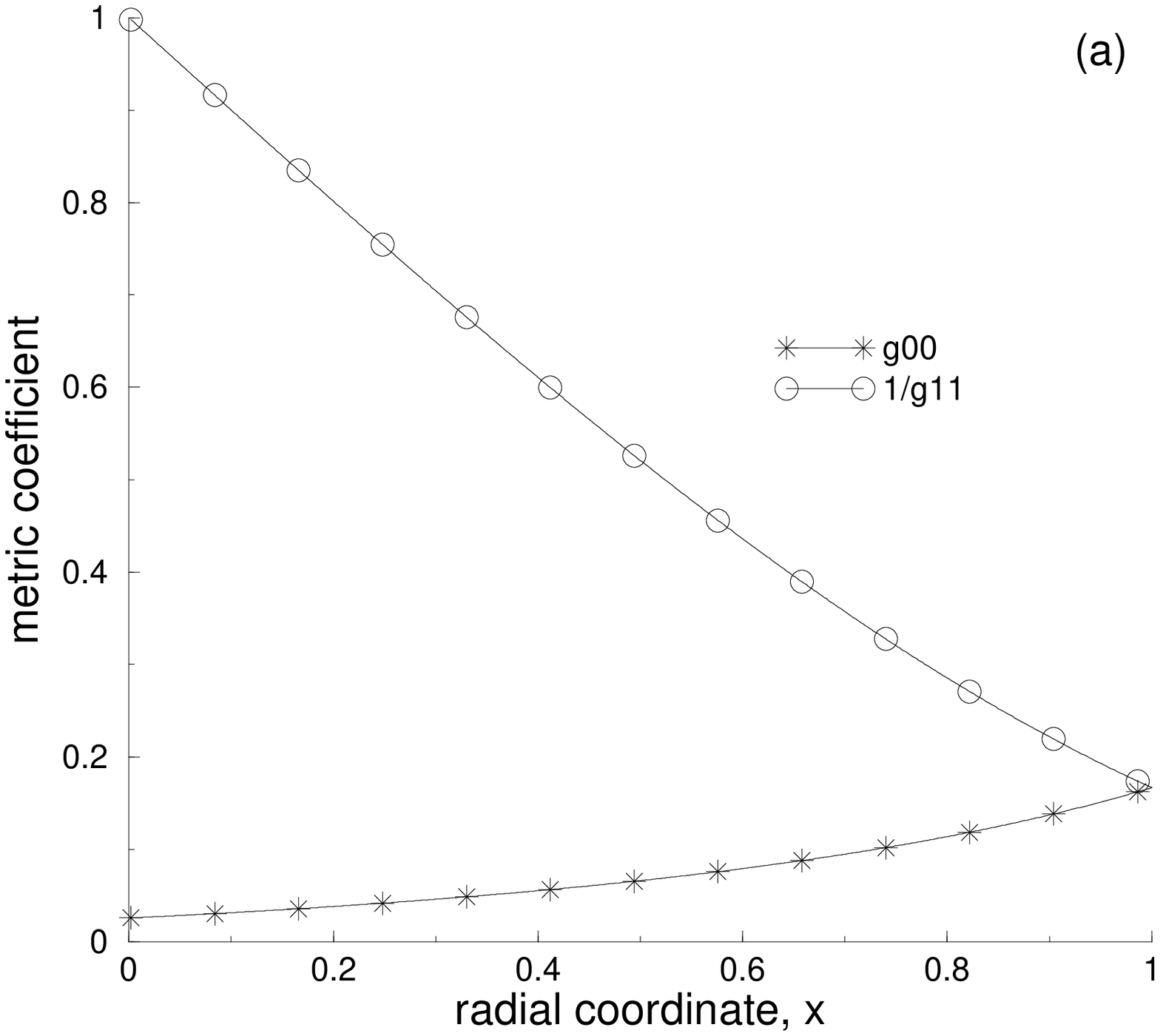}}
\rotatebox{-90}{\includegraphics{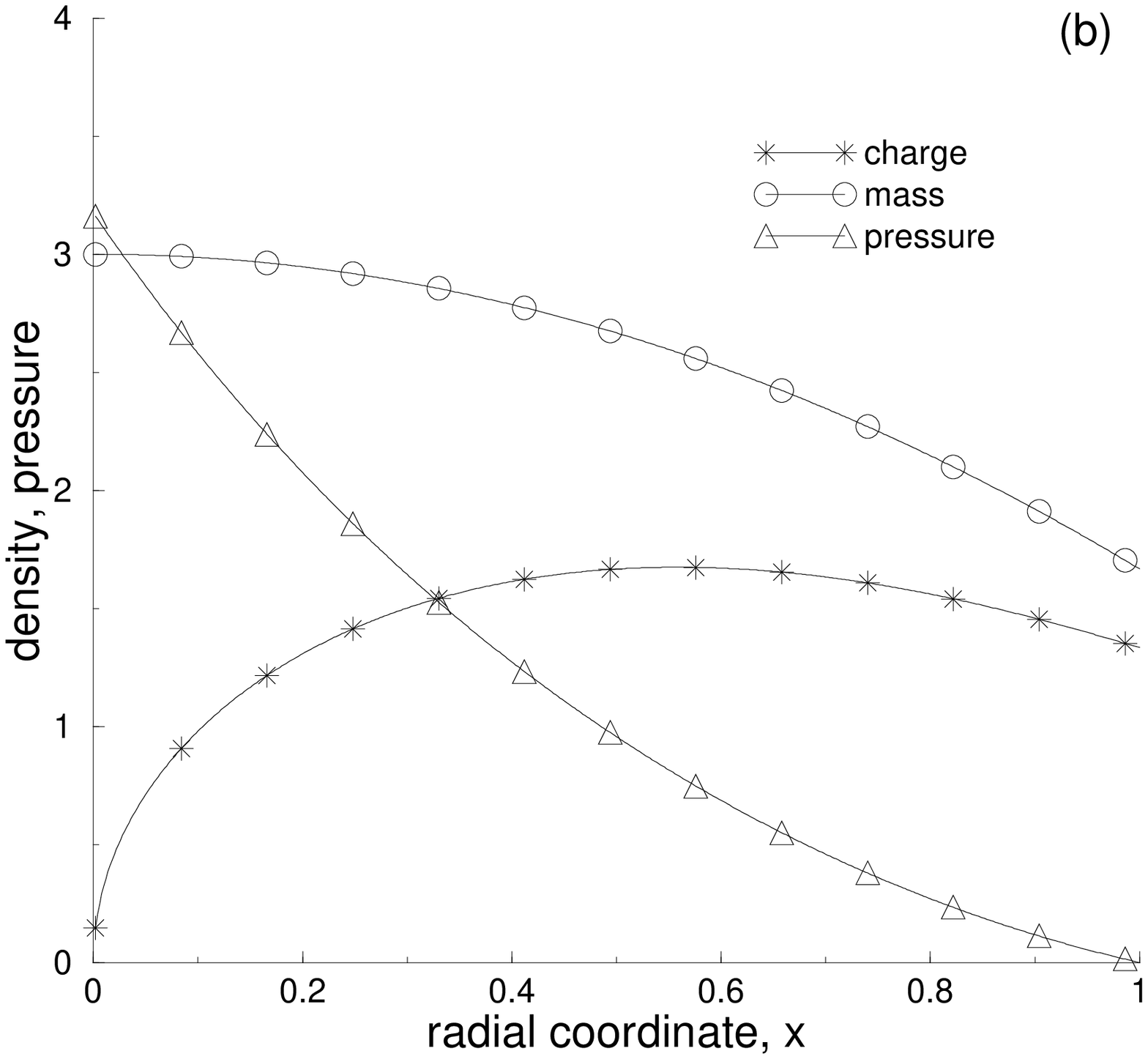}}}
\scalebox{.30}{\rotatebox{-90}{\includegraphics{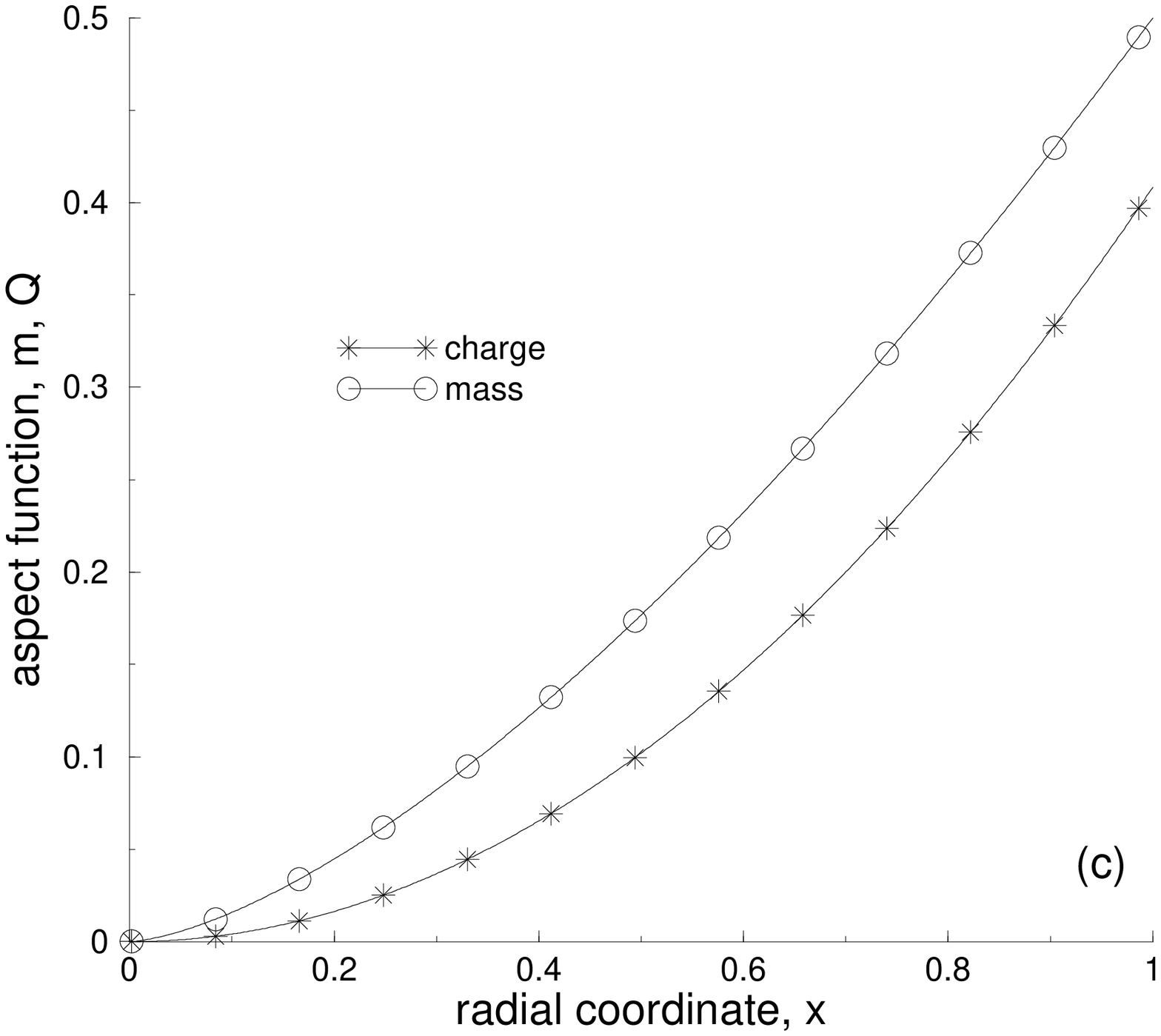}}
\rotatebox{-90}{\includegraphics{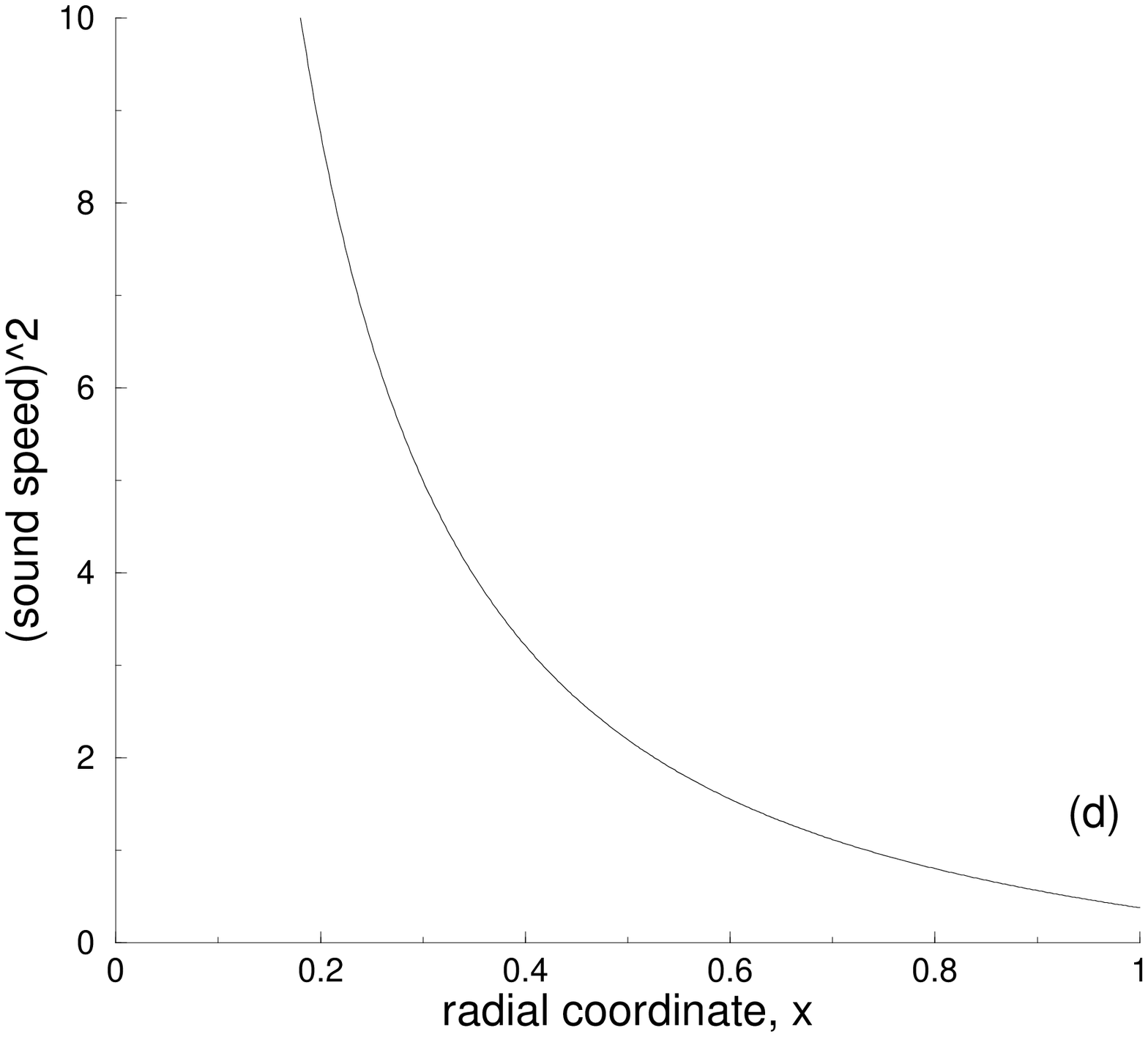}}}} 
\end{center}
\caption{ Plots of: (a) the metric functions $y^2$ and $z$; (b) the pressure
$P$, the mass and charge densities $\kappa\rho$ and 
$\kappa\sigma$; (c) the mass and charge functions $m(r)$
and $q(r)$  and (d) $dP/d\rho$ for $b=0$.}
\label{cubic-li}
\end{figure}

Figure (\ref{cubic-li}) shows a solution where the mass density is always
greater than the charge density and the pressure starts positive and
monotonically goes to zero at the boundary. The case shown has the following
choices for $a$, $c$, and $x_{0}$:
\begin{eqnarray*}
a &=&1 \\
c &=&-\frac{1}{6} \\
x_{0} &=&1
\end{eqnarray*}

\bigskip The speed of sound in the fluid sphere is also of interest. This
speed should be positive and causal, that is
\begin{equation}
0\leq \frac{dP}{d\rho }\leq 1.  \label{ssp-condition}
\end{equation}
Figure (\ref{cubic-li}) is a plot of the speed of sound in this medium
which indicates that the speed is certainly not causal in this case.

\begin{figure}[h]
\begin{center}
{\scalebox{.30}{\rotatebox{-90}{\includegraphics{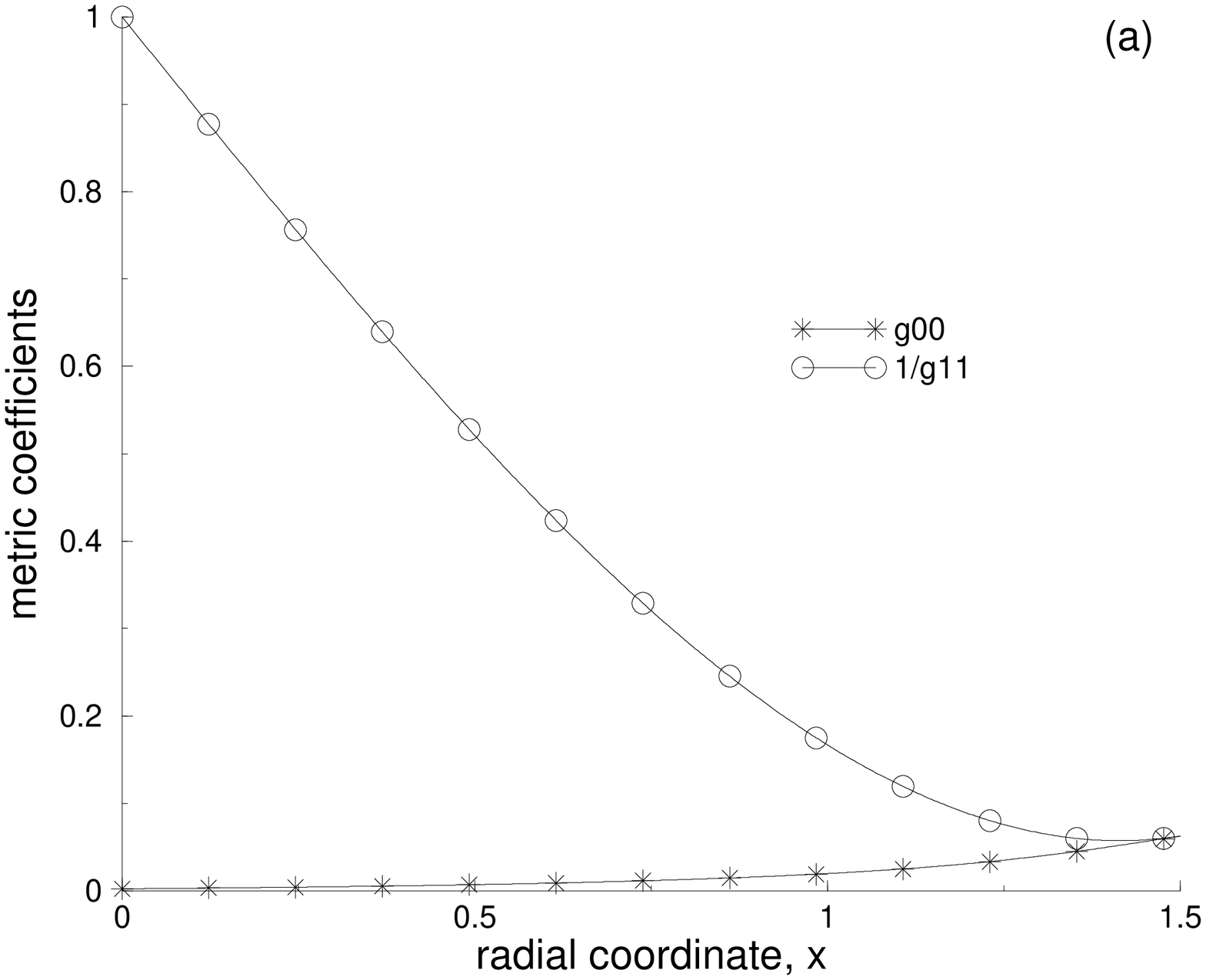}}
\rotatebox{-90}{\includegraphics{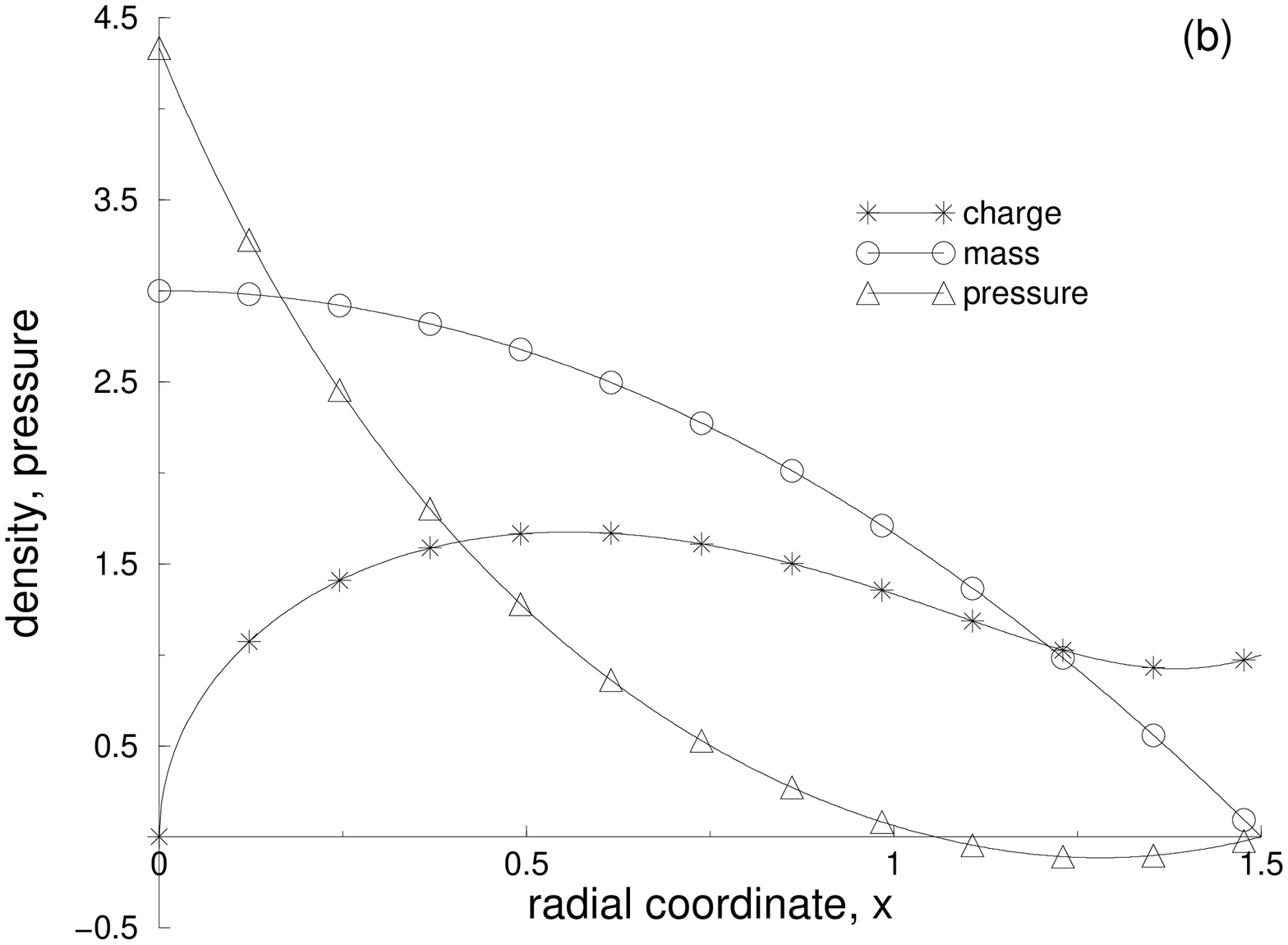}}}
\scalebox{.30}{\rotatebox{-90}{\includegraphics{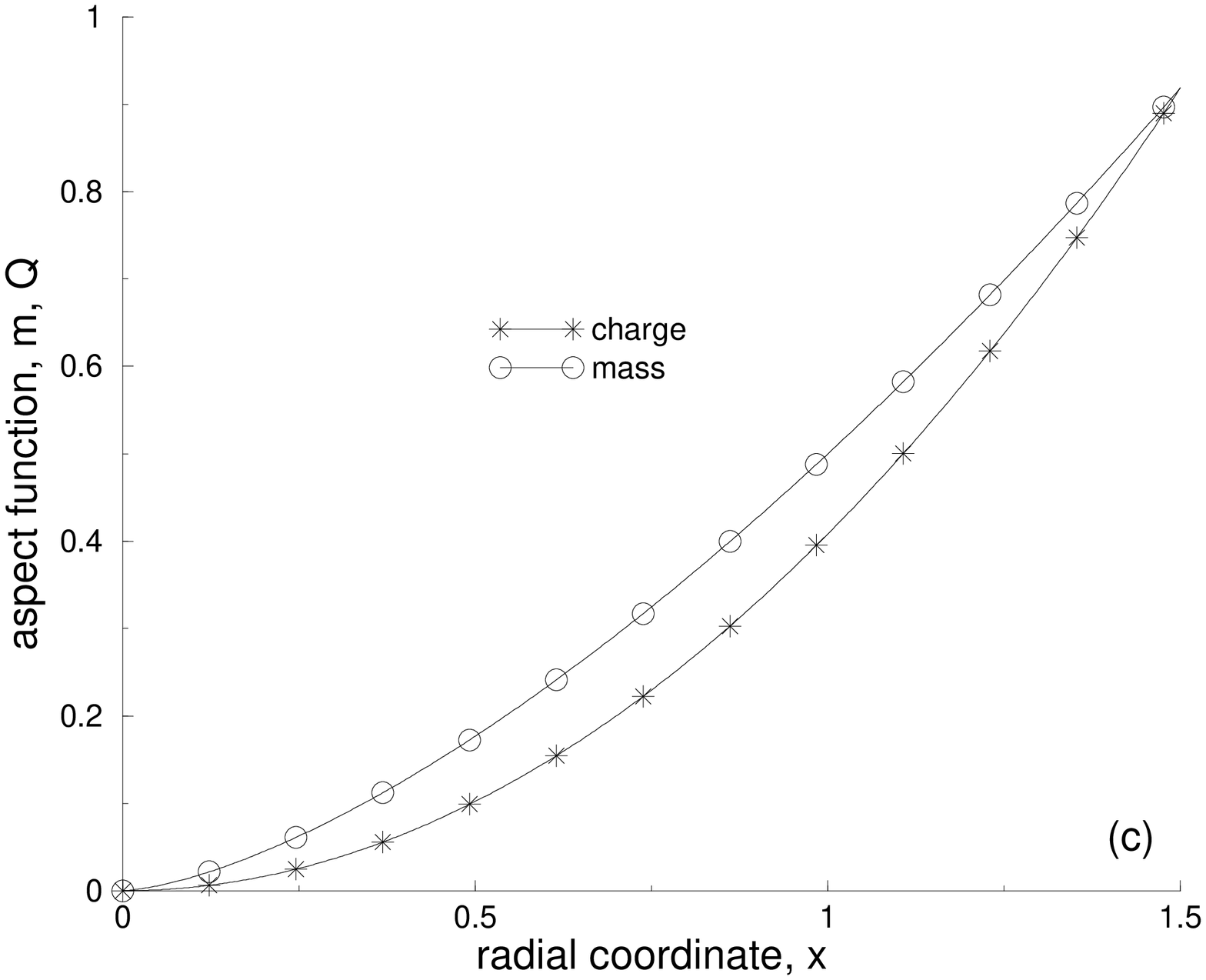}}
\rotatebox{-90}{\includegraphics{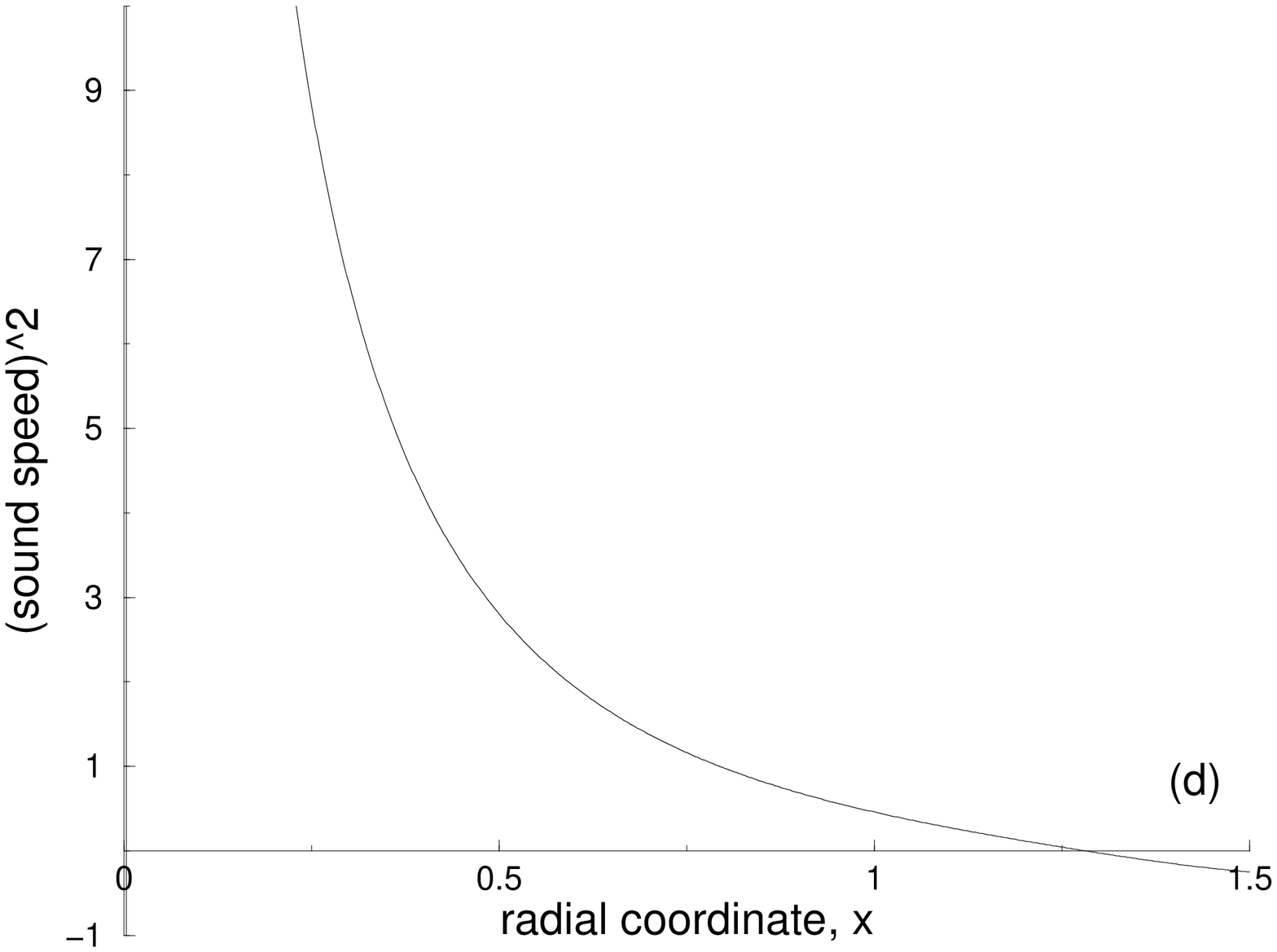}}}} 
\end{center}
\caption{ Plots of: (a) the metric functions $y^2$ and $z$; (b) the pressure
$P$, the mass and charge densities $\kappa\rho$ and 
$\kappa\sigma$; (c) the mass and charge functions $m(r)$
and $q(r)$  and (d) $dP/d\rho$ for 
the extreme Reissner-Nordstr\"{o}m solution and $b=0$.}
\label{c-eRN-li}
\end{figure}

If this solution is matched to the extreme Reissner-Nordstr\"{o}m solution,
we impose the condition given in equation (\ref{extremeRN-cond}). Where $b=0$%
, the boundary $x_{0}$ is solved for:
\[
x_{0}=\frac{1}{2}\frac{a^{2}}{(-c)}.
\]
Unfortunately, this radial boundary also produces negative pressures over
part of the fluid sphere. Figure (\ref{c-eRN-li}) shows the example with
\begin{eqnarray*}
a &=&1 \\
c &=&-\frac{1}{6} \\
x_{0} &=&1
\end{eqnarray*}

The speed of sound is singular at the center of the sphere and decreases
when both the pressure and mass density are decreasing The speed is negative
when the pressure increases. The speed is plotted in Figure (\ref{c-eRN-li}).

If we impose the condition $\rho (x_{0})=0$, the boundary of the fluid
sphere is
\[
x_{0}=\sqrt{\frac{\rho _{0}}{\rho _{2}}}=\sqrt{\frac{3a}{8(-c)}}
\]
and equation (\ref{metric primes equal}) implies
\[
C_{2}=\frac{a}{16}.
\]
The junction condition (\ref{Junction-RN}) produces
\[
C_{1}=\sqrt{z(x_{0})}-\frac{a}{16}\xi _{0}.
\]

\begin{figure}[h]
\begin{center}
{\scalebox{.30}{\rotatebox{-90}{\includegraphics{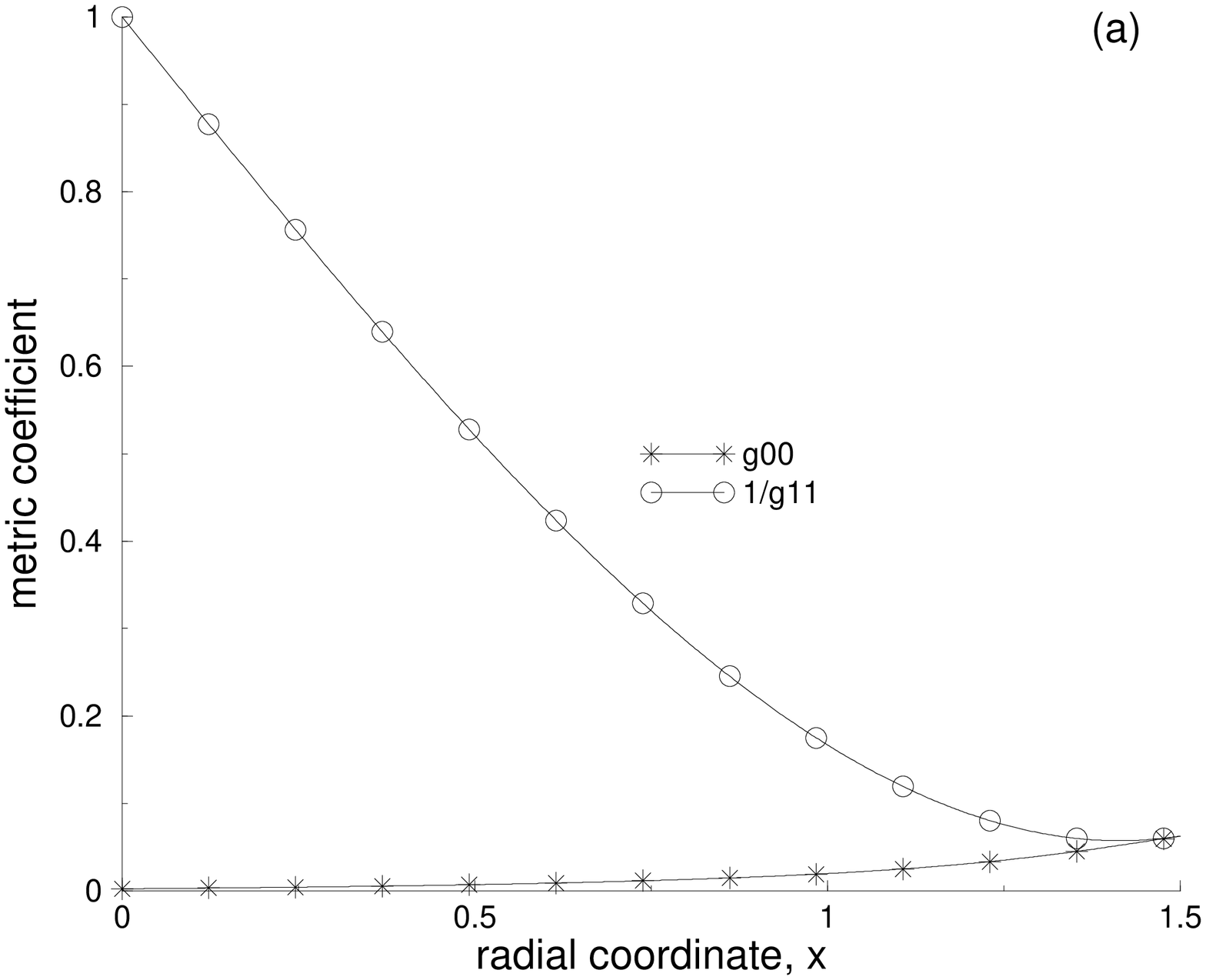}}
\rotatebox{-90}{\includegraphics{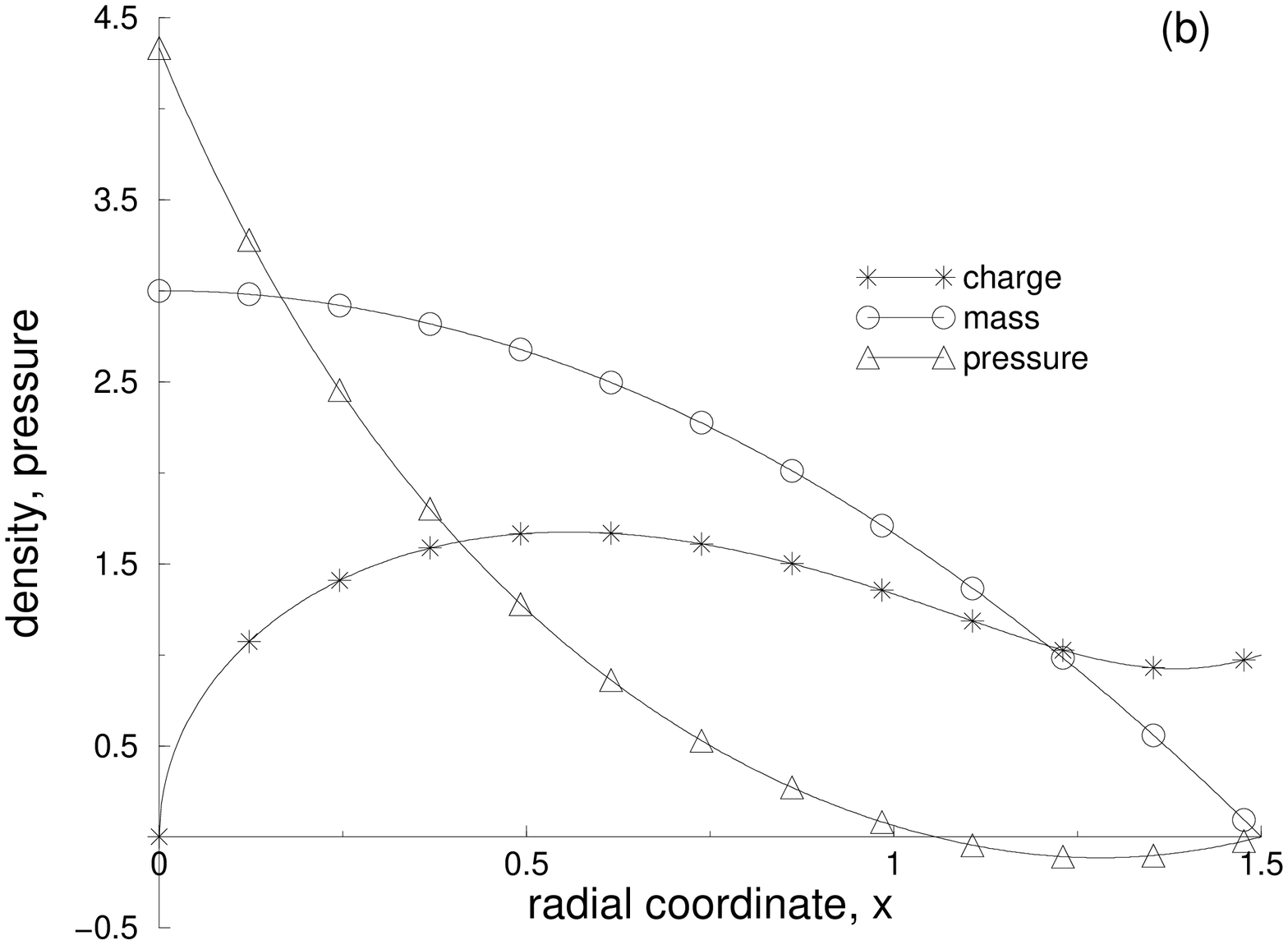}}}
\scalebox{.30}{\rotatebox{-90}{\includegraphics{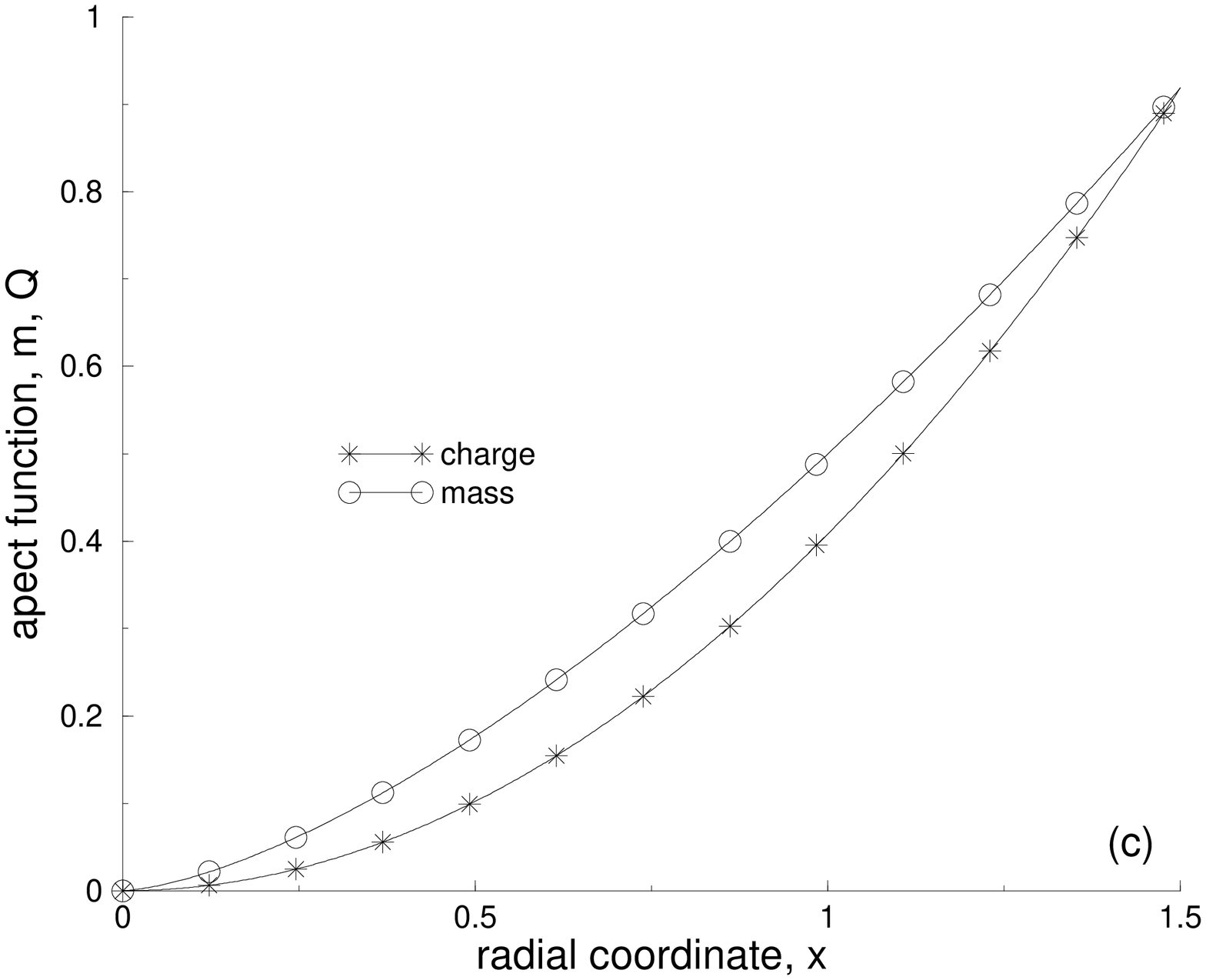}}
\rotatebox{-90}{\includegraphics{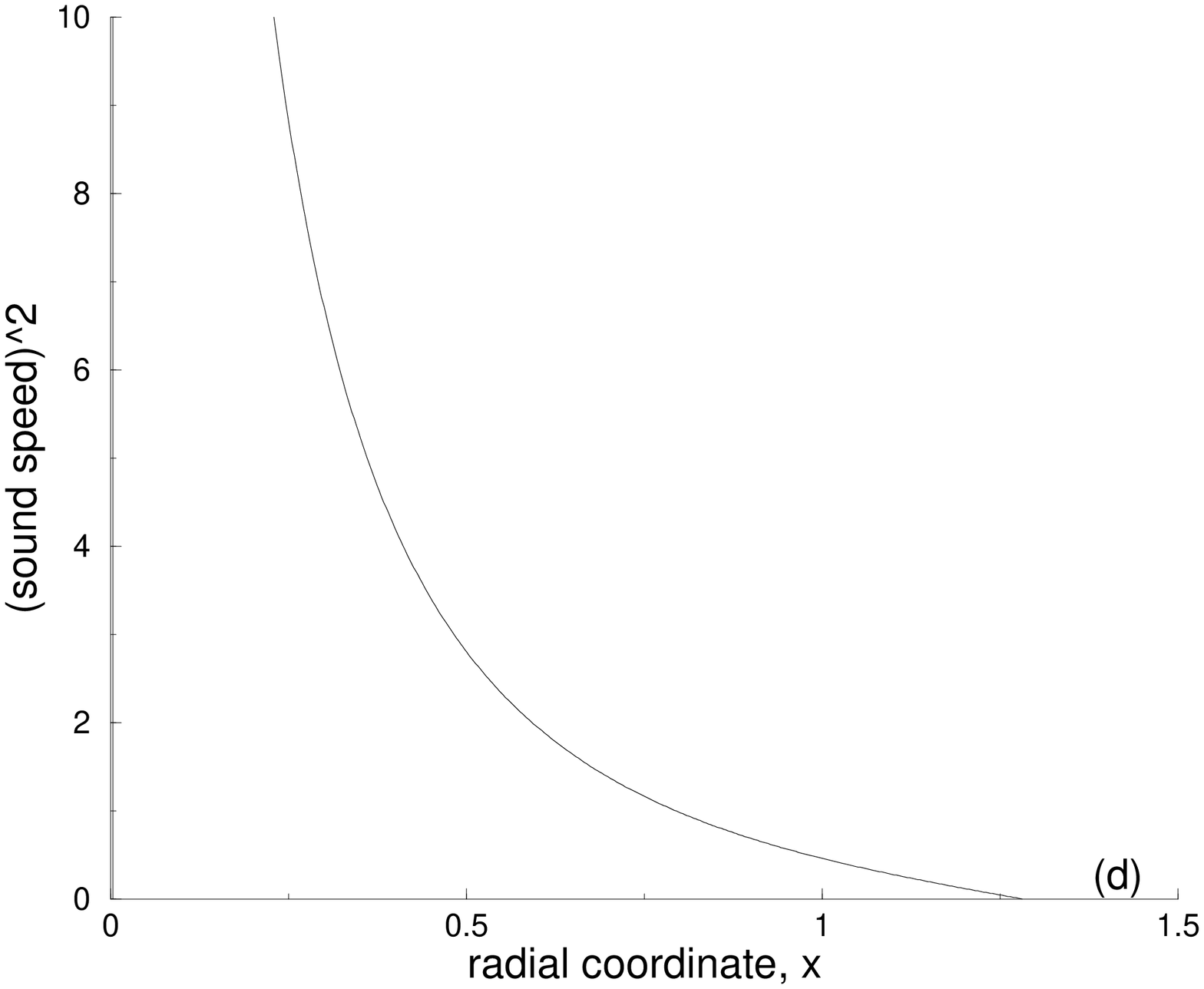}}}} 
\end{center}
\caption{ Plots of: (a) the metric functions $y^2$ and $z$; (b) the pressure
$P$, the mass and charge densities $\kappa\rho$ and 
$\kappa\sigma$; (c) the mass and charge functions $m(r)$
and $q(r)$  and (d) $dP/d\rho$ for 
$b=0$ and $\protect\rho(x_{0})=0$.}
\label{c-rho0-li}
\end{figure}

In this case, the pressure, from equation (\ref{P-eqn-xi}), reads
\begin{equation}
\kappa P=4\sqrt{z}\left( \frac{C_{2}}{C_{1}+C_{2}\xi }\right) -a-2cx^{2}.
\label{P-eqn-linear-xi}
\end{equation}
The pressure will be positive at the center ($x=\xi =0$) so long as $C_{1}<%
\frac{1}{4}$. Unfortunately, having the two conditions on the pressure, that
the pressure is positive at the center and vanishes at the boundary, is not
sufficient to prove that this is a positive pressure case. Numerical
evaluation of the pressure shows that it has an additional point between the
center and the boundary at which it goes to zero. That is, the pressure
starts positive, then becomes negative before vanishing at the boundary. We
illustrate a particular example of this behavior in Figure (\ref{c-rho0-li})
with the following values for the constants $a$ and $c$:
\begin{eqnarray*}
a &=&1 \\
c &=&-\frac{1}{6}
\end{eqnarray*}

\subsection{Case $d<0$ ($b>0$, $\protect\rho _{1}<0$)}

\begin{figure}[h]
\begin{center}
{\scalebox{.30}{\rotatebox{-90}{\includegraphics{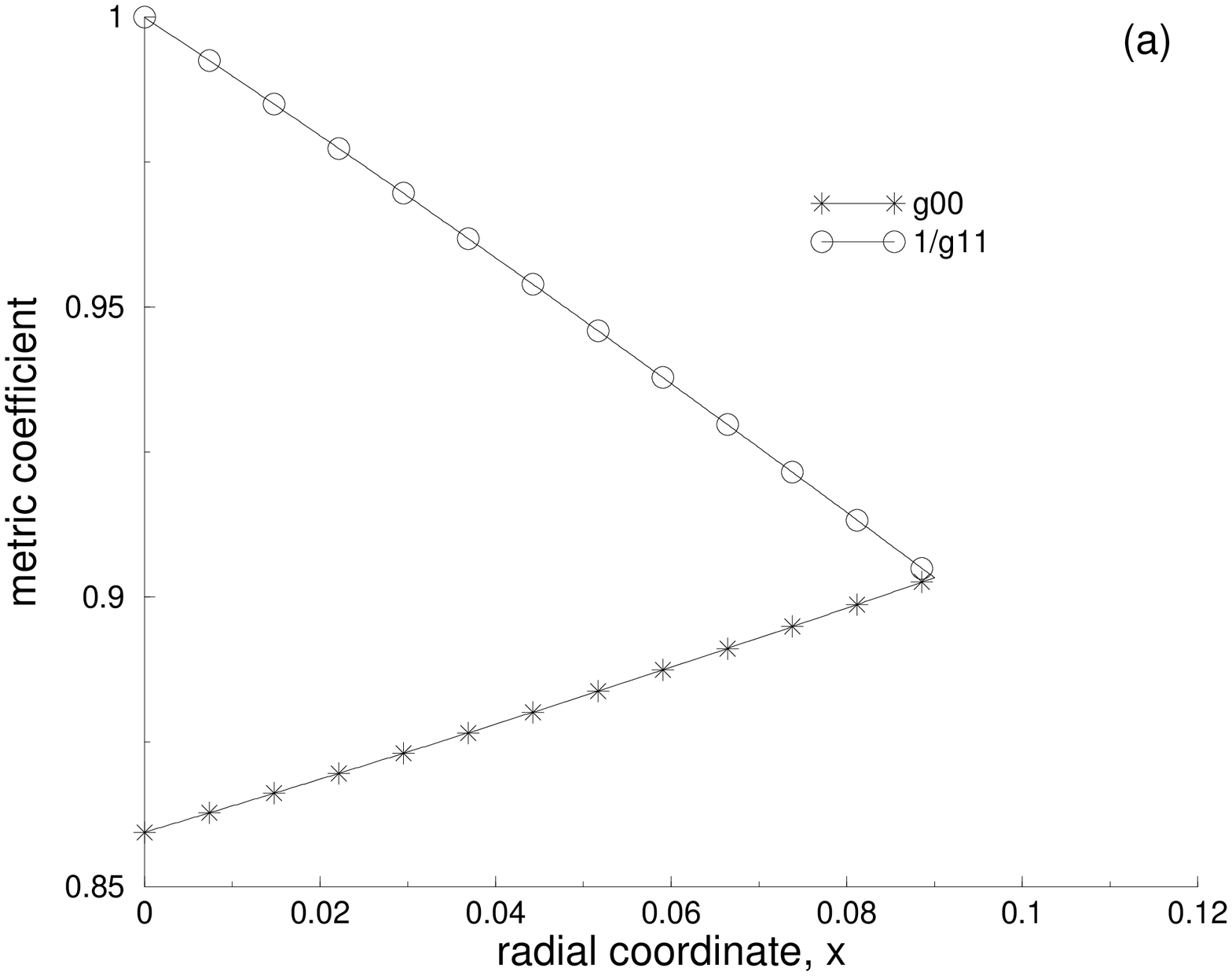}}
\rotatebox{-90}{\includegraphics{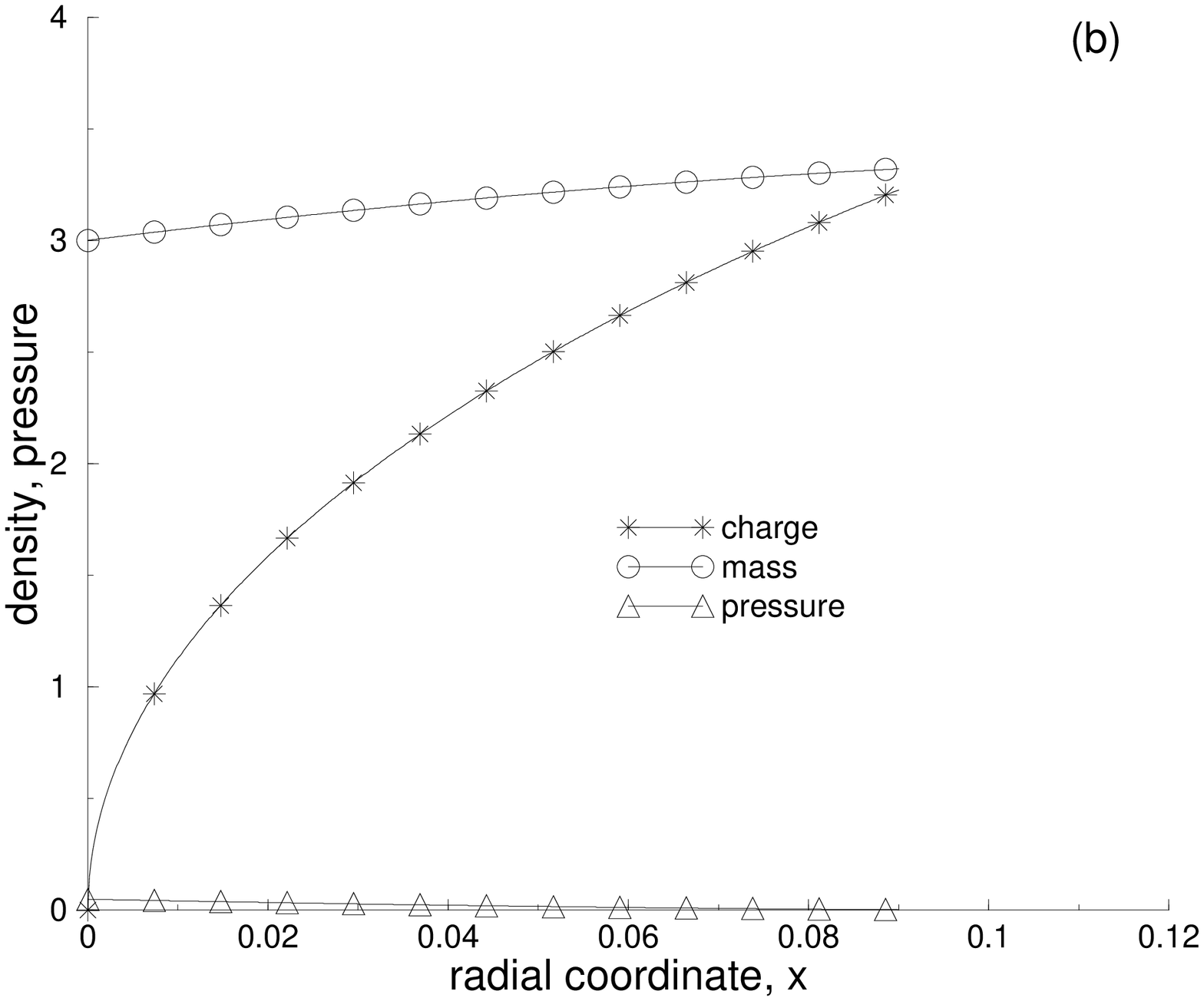}}}
\scalebox{.30}{\rotatebox{-90}{\includegraphics{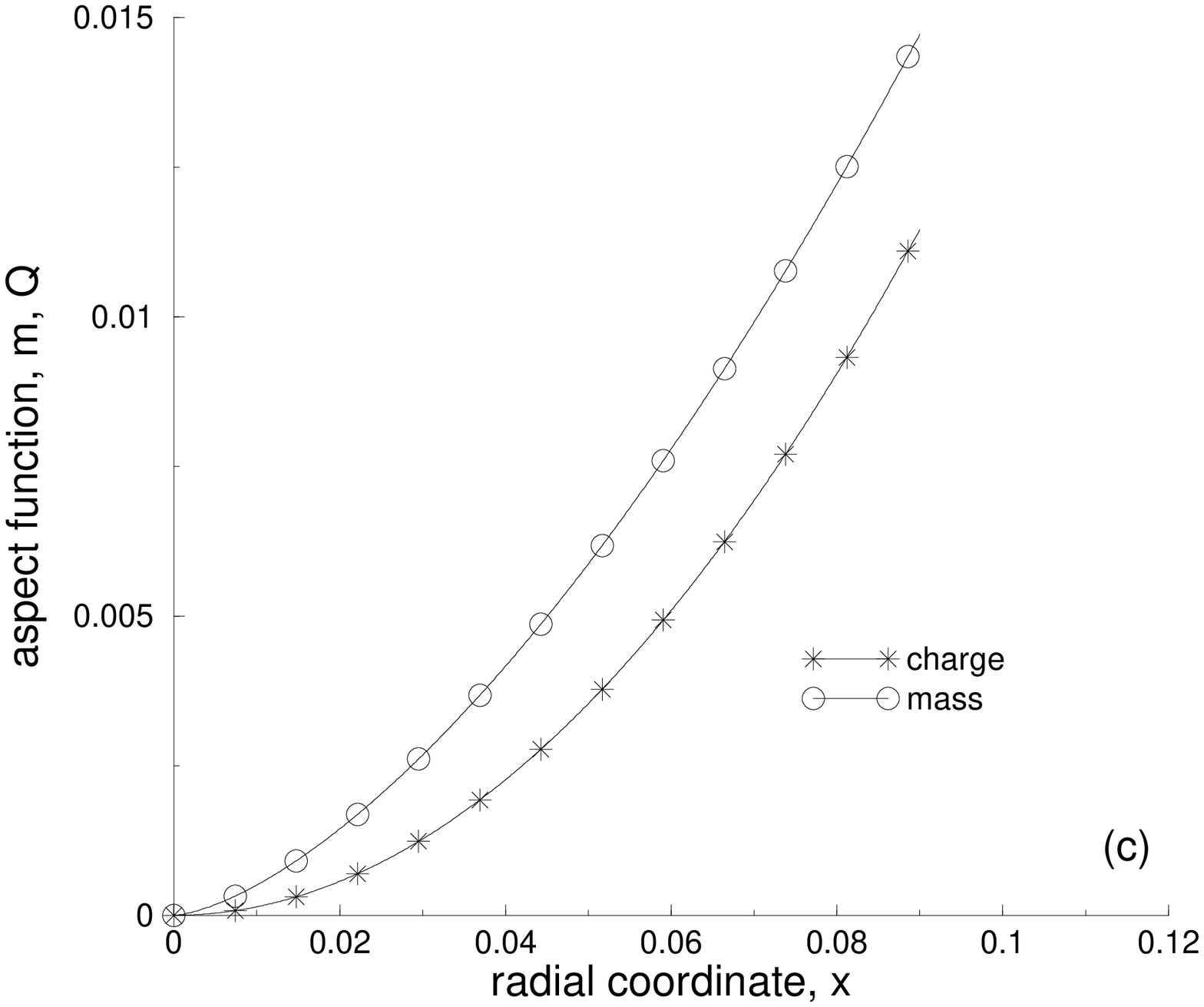}}
\rotatebox{-90}{\includegraphics{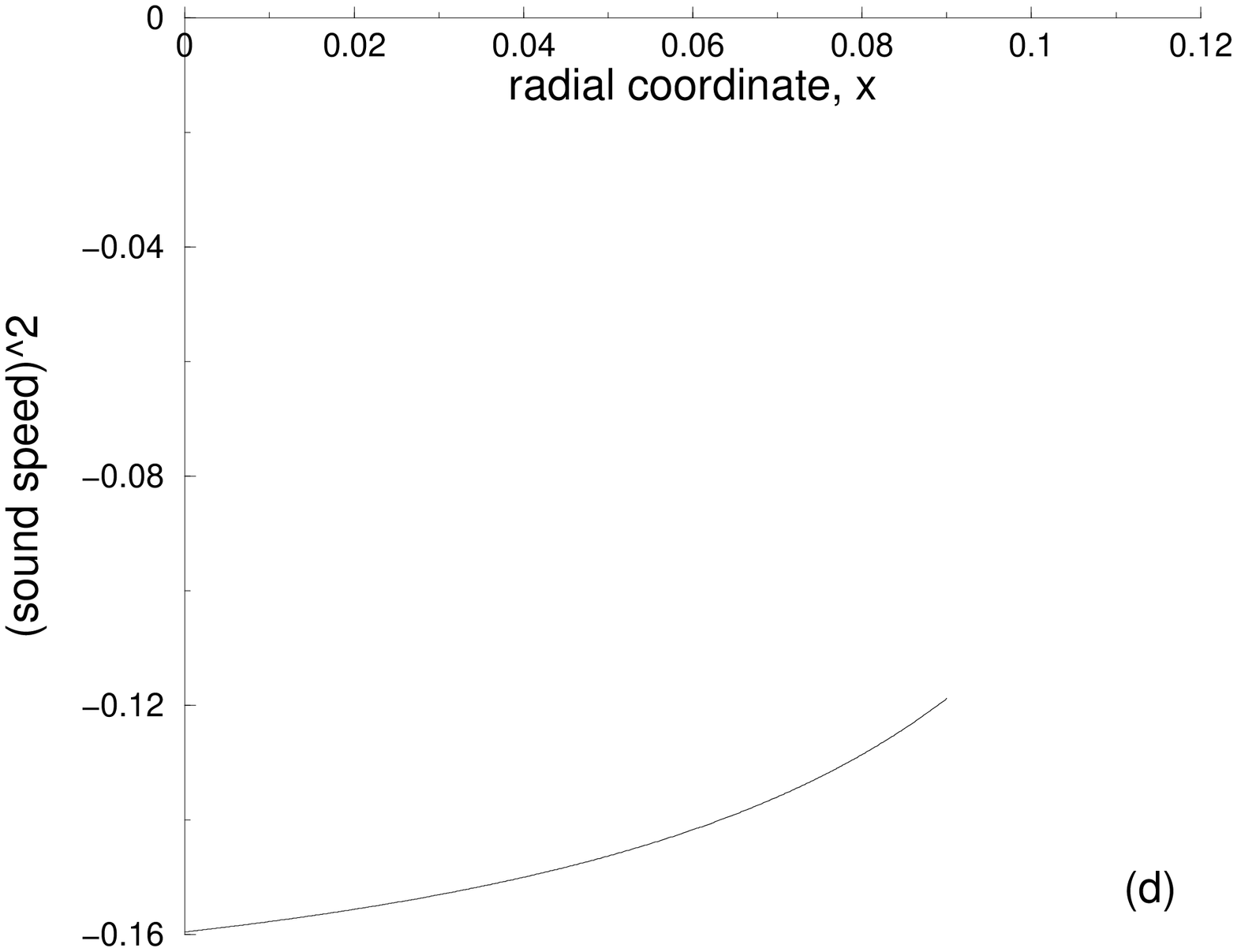}}}} 
\end{center}
\caption{ Plots of: (a) the metric functions $y^2$ and $z$; (b) the pressure
$P$, the mass and charge densities $\kappa\rho$ and 
$\kappa\sigma$; (c) the mass and charge functions $m(r)$
and $q(r)$  and (d) $dP/d\rho$ for 
 $b>0$.}
\label{cubic-exp}
\end{figure}

This case involves the real exponential solution for $y$ given by equation (%
\ref{y-exp}). If we do not impose the condition that the mass density goes
to zero at the boundary, we can solve for the constants $C_{1}$ and $C_{2}$
using the junction condition (\ref{Junction-RN}) and equation (\ref
{yprime-condition}). These two equations result in
\begin{eqnarray*}
C_{1} &=&\frac{1}{2}e^{-\sqrt{-d}\xi _{0}}\left[ \sqrt{z(x_{0})}+\frac{1}{%
\sqrt{-d}}\left( \frac{a}{4}+\frac{b}{4}x_{0}+\frac{c}{2}x_{0}^{2}\right) %
\right] \\
C_{2} &=&\frac{1}{2}e^{\sqrt{-d}\xi _{0}}\left[ \sqrt{z(x_{0})}-\frac{1}{%
\sqrt{-d}}\left( \frac{a}{4}+\frac{b}{4}x_{0}+\frac{c}{2}x_{0}^{2}\right) %
\right]
\end{eqnarray*}
These constants do give positive pressures as shown in the following
example. Choose the following values for the constants $a$, $b$, $c$, and $%
x_{0}$:
\begin{eqnarray*}
a &=&1 \\
b &=&1 \\
c &=&-2 \\
x_{0} &=&0.09
\end{eqnarray*}
Our results are shown in Figure (\ref{cubic-exp}).

Once again, the speed of sound in this medium $\frac{dP}{d\rho }$ is
evaluated and, as shown in Figure (\ref{cubic-exp}), is negative for this
case. This is due to the fact that the mass density increases over the range
of interest while the pressure decreases.

\begin{figure}[h]
\begin{center}
{\scalebox{.30}{\rotatebox{-90}{\includegraphics{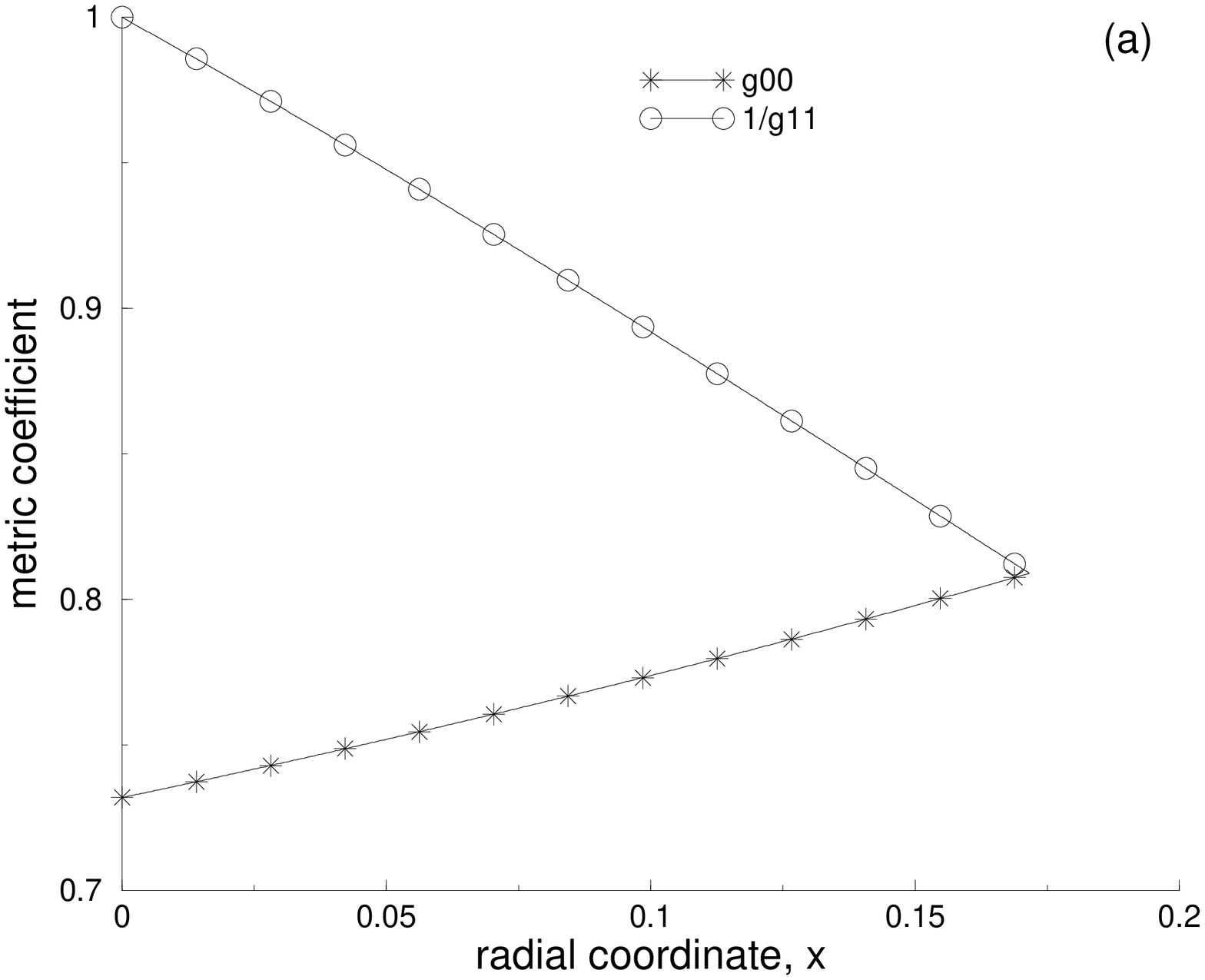}}
\rotatebox{-90}{\includegraphics{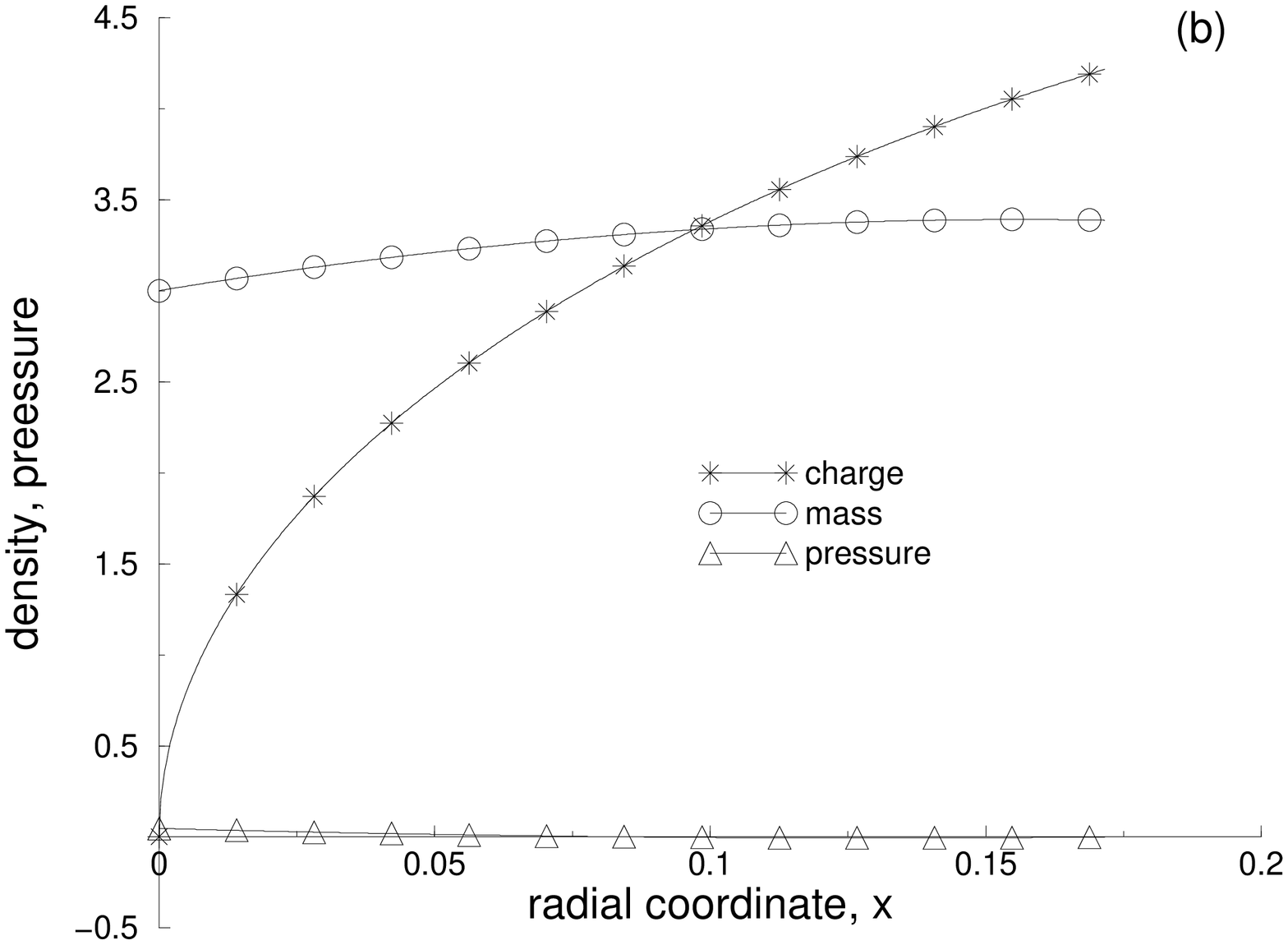}}}
\scalebox{.30}{\rotatebox{-90}{\includegraphics{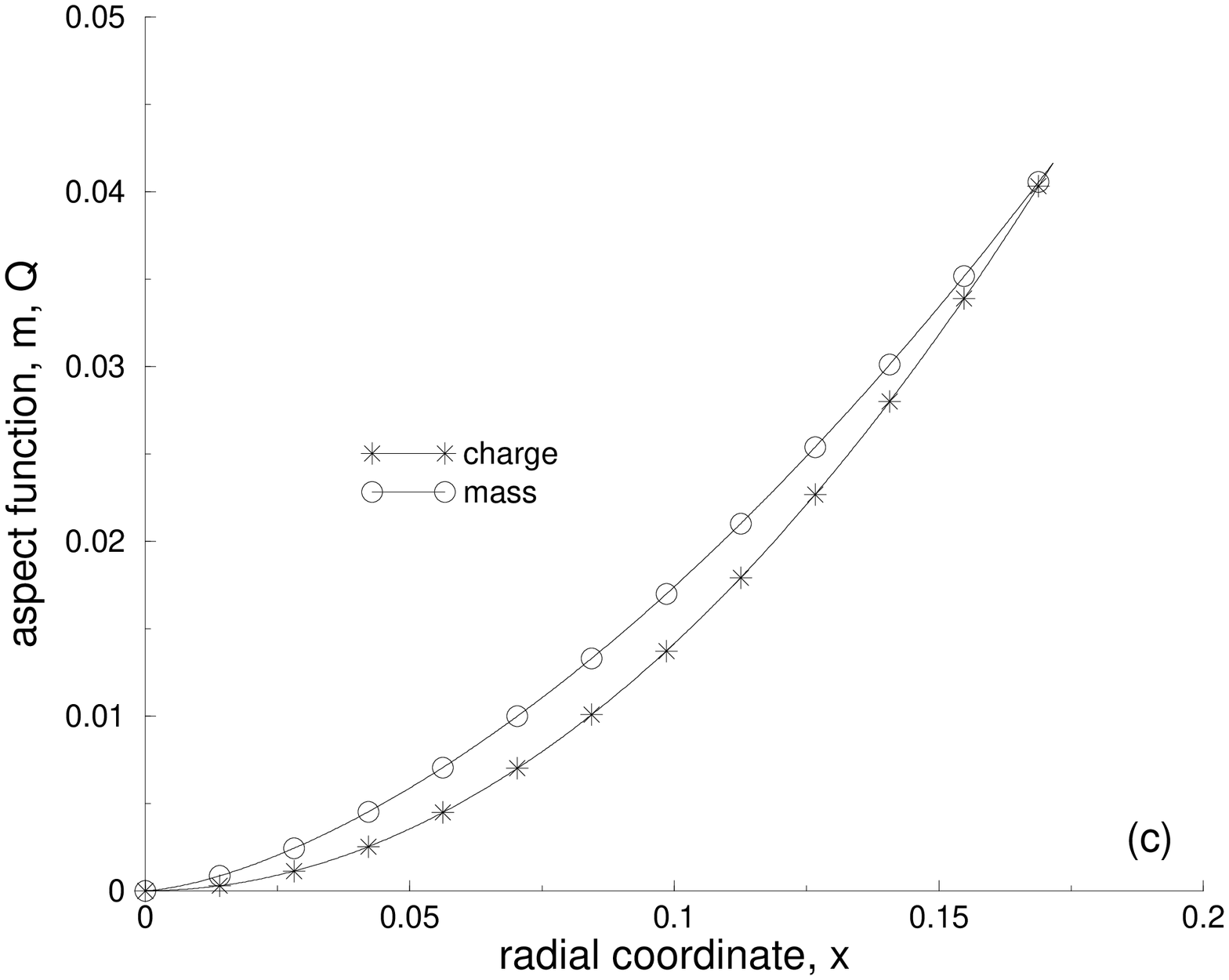}}
\rotatebox{-90}{\includegraphics{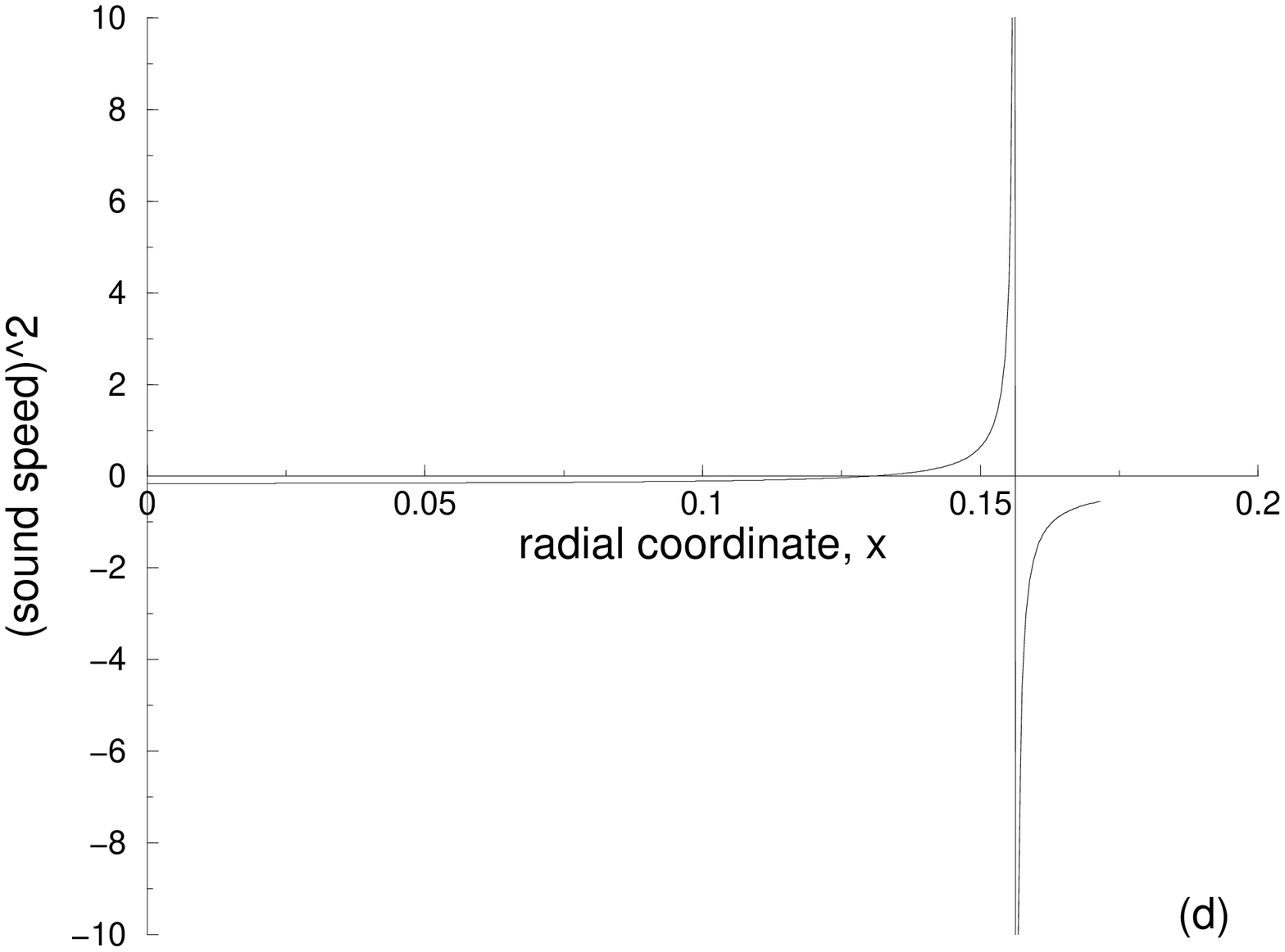}}}} 
\end{center}
\caption{ Plots of: (a) the metric functions $y^2$ and $z$; (b) the pressure
$P$, the mass and charge densities $\kappa\rho$ and 
$\kappa\sigma$; (c) the mass and charge functions $m(r)$
and $q(r)$  and (d) $dP/d\rho$ for 
extreme Reissner-Nordstr\"{o}m solution and $ b>0$.}
\label{c-eRN-exp}
\end{figure}

A match to the extreme Reissner-Nordstr\"{o}m solution leads to a boundary
given by the solution to equation (\ref{extremeRN-cond}):
\[
x_{0}=\left( \frac{\sqrt{-c}}{b}\pm \frac{1}{b}\sqrt{-c-ab}\right) ^{2}
\]
where the minus sign case is chosen to ensure that the mass density remains
positive over the region of the fluid sphere. Using the same values for the
other constants as the previous example, Figure (\ref{c-eRN-exp}) is
obtained.

The speed of sound becomes singular when $d\rho =0$.

\begin{figure}[h]
\begin{center}
{\scalebox{.30}{\rotatebox{-90}{\includegraphics{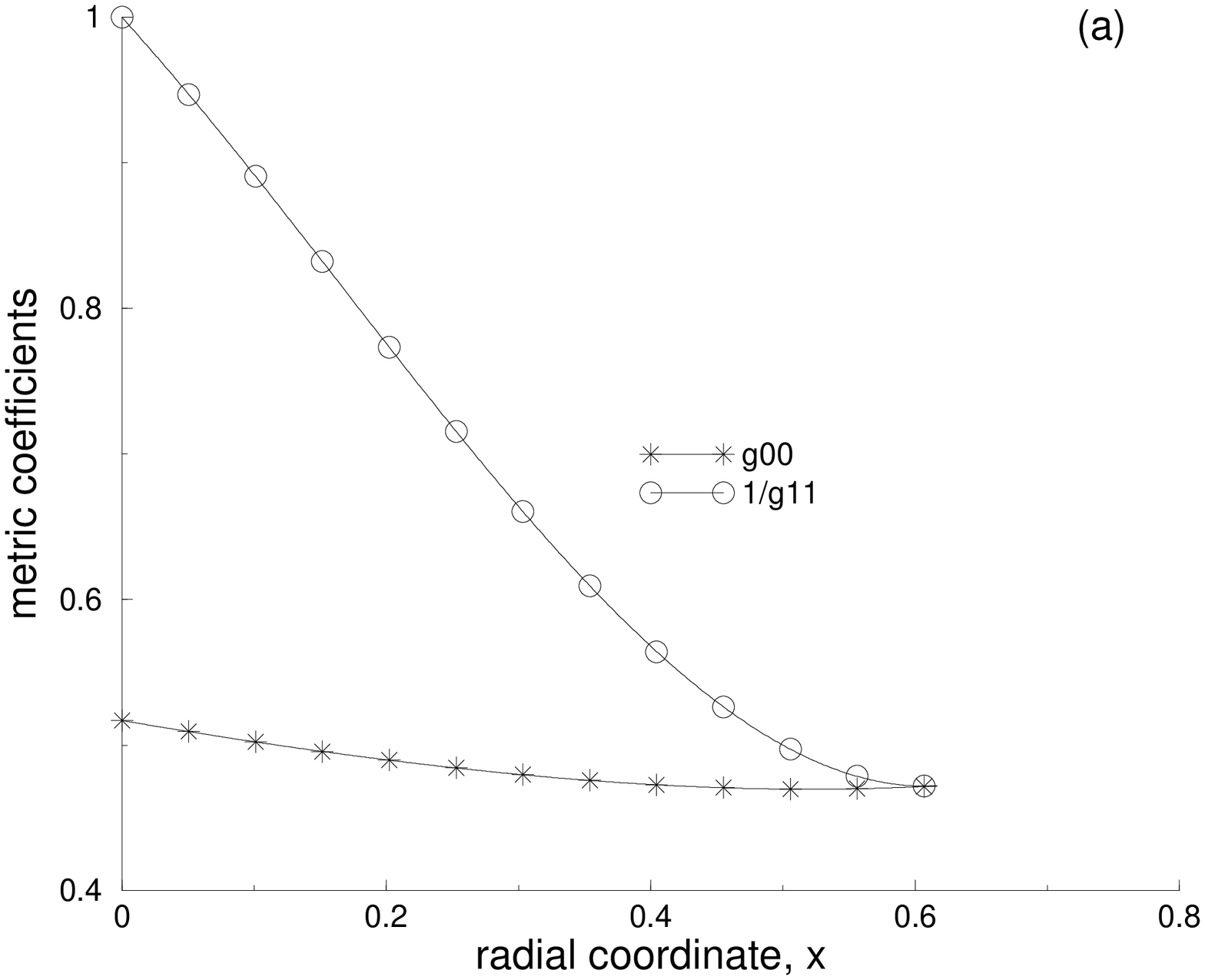}}
\rotatebox{-90}{\includegraphics{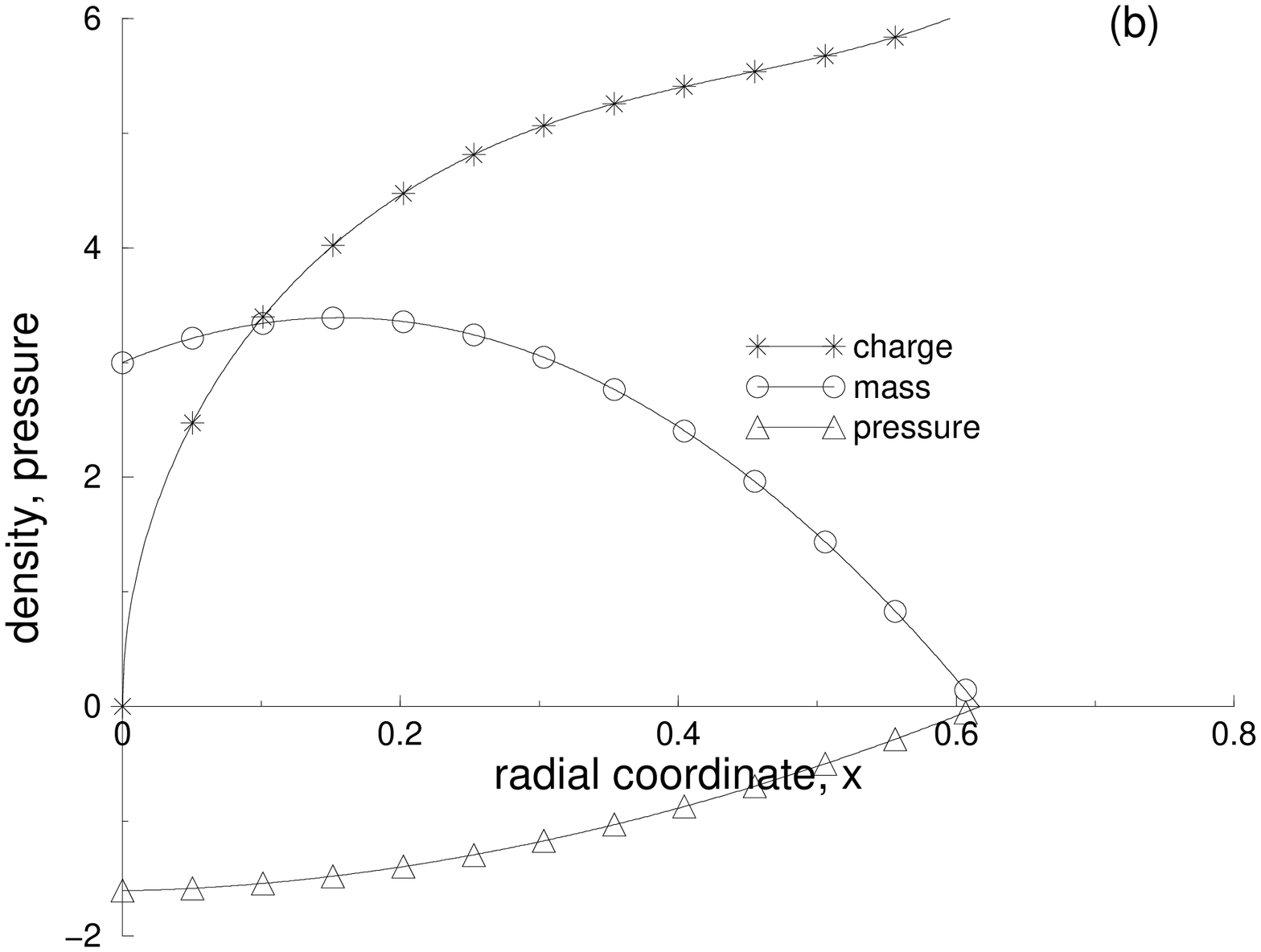}}}
\scalebox{.30}{\rotatebox{-90}{\includegraphics{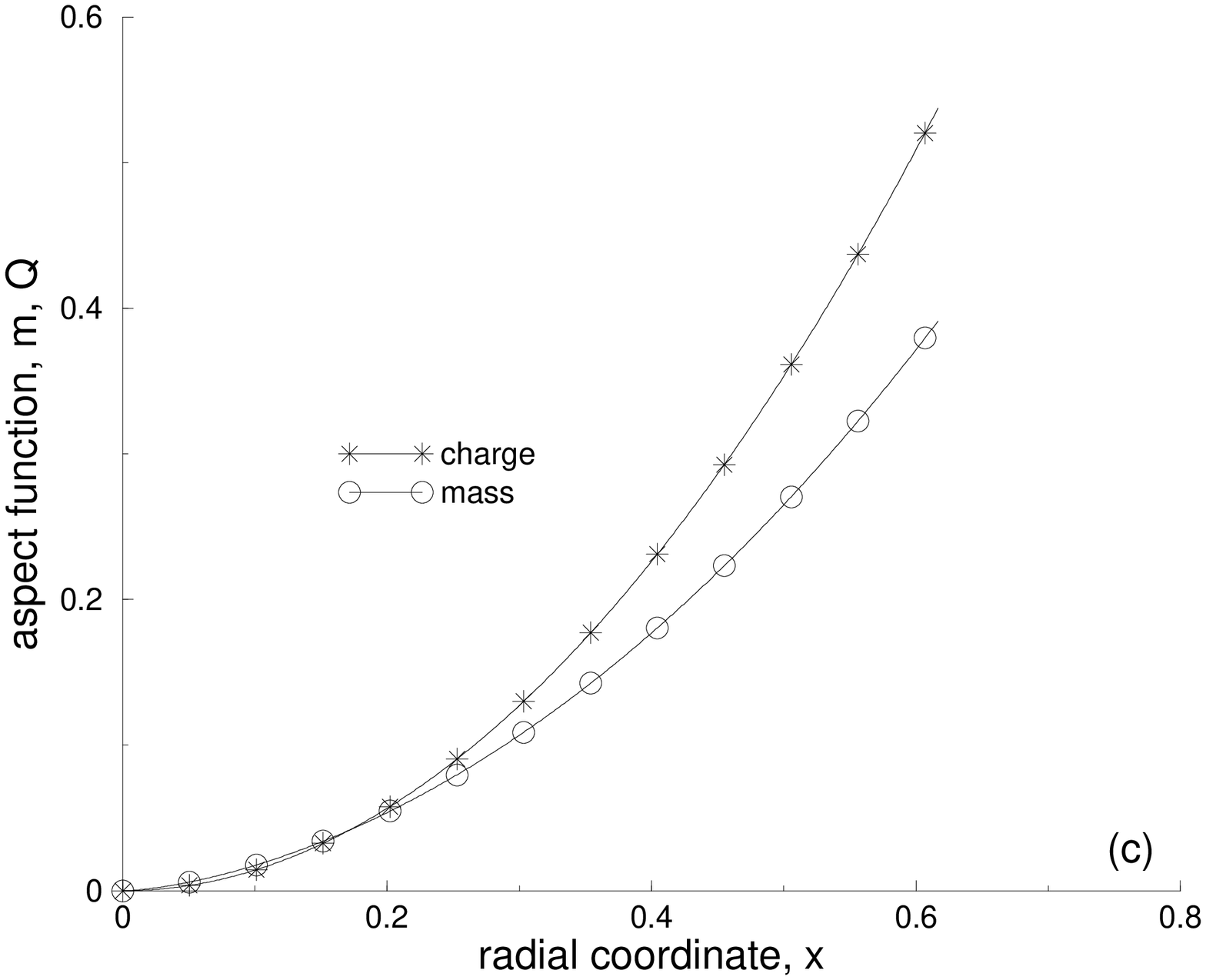}}
\rotatebox{-90}{\includegraphics{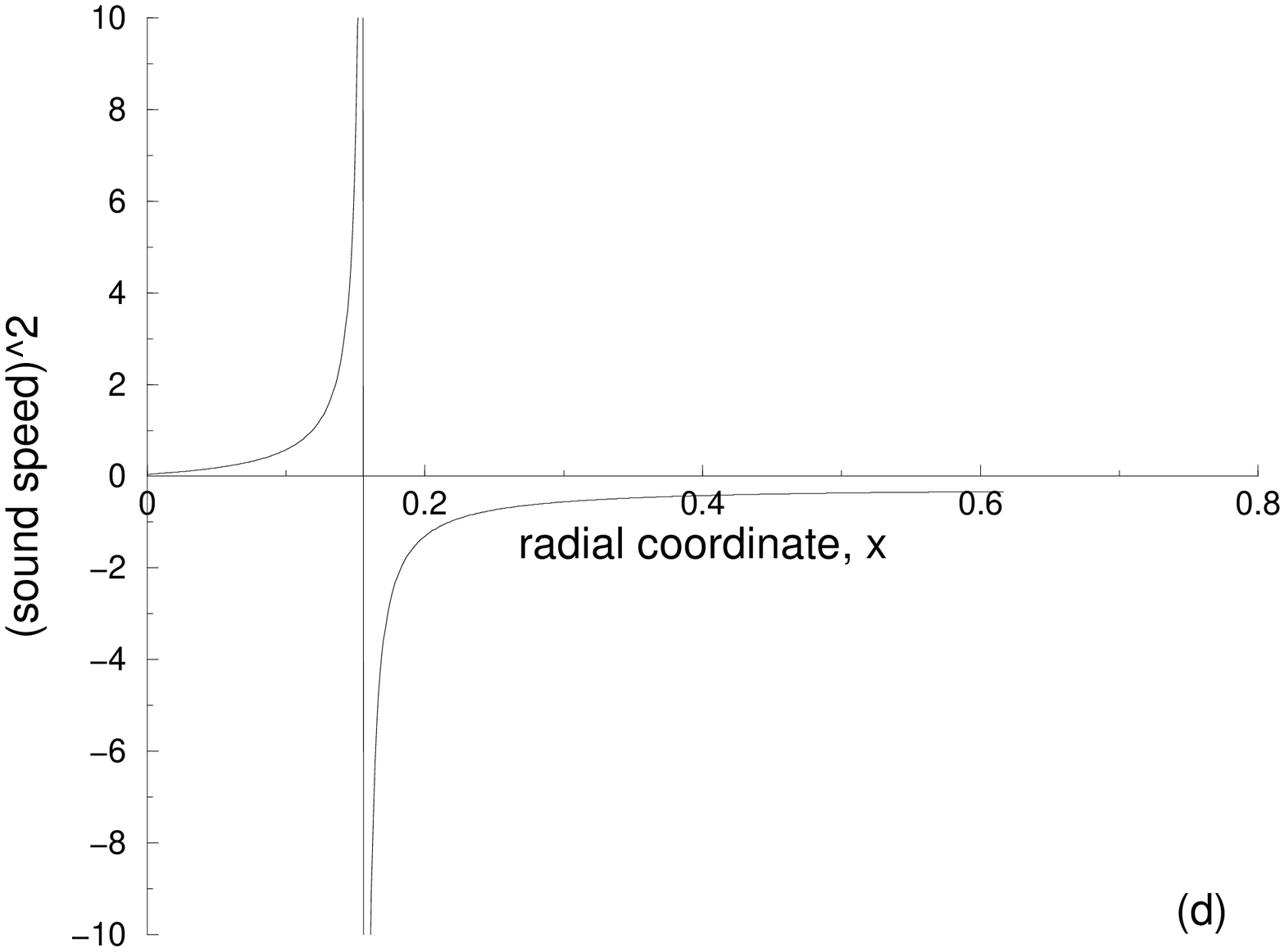}}}} 
\end{center}
\caption{ Plots of: (a) the metric functions $y^2$ and $z$; (b) the pressure
$P$, the mass and charge densities $\kappa\rho$ and 
$\kappa\sigma$; (c) the mass and charge functions $m(r)$
and $q(r)$  and (d) $dP/d\rho$ for 
$b>0$ and $\protect\rho (x_{0})=0$.} \label{c-rho0-exp}
\end{figure}

If the mass density is forced to go to zero at the boundary of the fluid
sphere ($\rho (x_{0})=0$), we solve for the boundary:
\begin{eqnarray}
x_{0} &=&-\frac{\rho _{1}}{2\rho _{2}}+\frac{1}{2\rho _{2}}\sqrt{\rho
_{1}^{2}+4\rho _{0}\rho _{2}}  \label{x0bnd-exp-rh0} \\
&=&\frac{5b}{16(-c)}+\frac{1}{16(-c)}\sqrt{25b^{2}+96a(-c)}  \nonumber
\end{eqnarray}
The two conditions given by equations (\ref{Junction-RN}) and (\ref{metric
primes equal}) allow us to solve for the constants $C_{1}$ and $C_{2}$,
\begin{eqnarray*}
C_{1} &=&\frac{1}{2}e^{-\sqrt{-d}\xi _{0}}\left[ \sqrt{z(x_{0})}+\frac{%
z_{x}(x_{0})}{2\sqrt{-d}}\right] \\
C_{2} &=&\frac{1}{2}e^{\sqrt{-d}\xi _{0}}\left[ \sqrt{z(x_{0})}-\frac{%
z_{x}(x_{0})}{2\sqrt{-d}}\right]
\end{eqnarray*}
The pressure, from equation (\ref{P-eqn-xi}), is now
\begin{equation}
\kappa P=4\sqrt{-dz}\left( \frac{C_{1}e^{\sqrt{-d}\xi }-C_{2}e^{-\sqrt{-d}%
\xi }}{C_{1}e^{\sqrt{-d}\xi }+C_{2}e^{-\sqrt{-d}\xi }}\right) -a-bx-2cx^{2}
\label{P-eqn-exp-xi}
\end{equation}
The condition for the pressure to be positive at the center of the fluid
sphere is
\[
\frac{C_{1}-C_{2}}{C_{1}+C_{2}}>\frac{a}{2\sqrt{b}}.
\]
This clearly requires $C_{1}>C_{2}$ but this sets no clear condition on $%
z_{x}\left( x_{0}\right) $ due to the presence of the real exponentials in $%
C_{1}$ and $C_{2}$. The pressures appear to be negative when the
contribution from the charge overwhelms the contribution from the mass,
especially near the boundary where the mass density is virtually nil. The
constants $a$, $b$, and $c$ are the same as in the previous example while $%
x_{0}$ is now given by equation (\ref{x0bnd-exp-rh0}). Figure (\ref
{c-rho0-exp}) shows this negative pressure case.
The speed of sound in this case becomes singular at the point of maximum
mass density (where $d\rho =0$).

\begin{figure}[h]
\begin{center}
{\scalebox{.30}{\rotatebox{-90}{\includegraphics{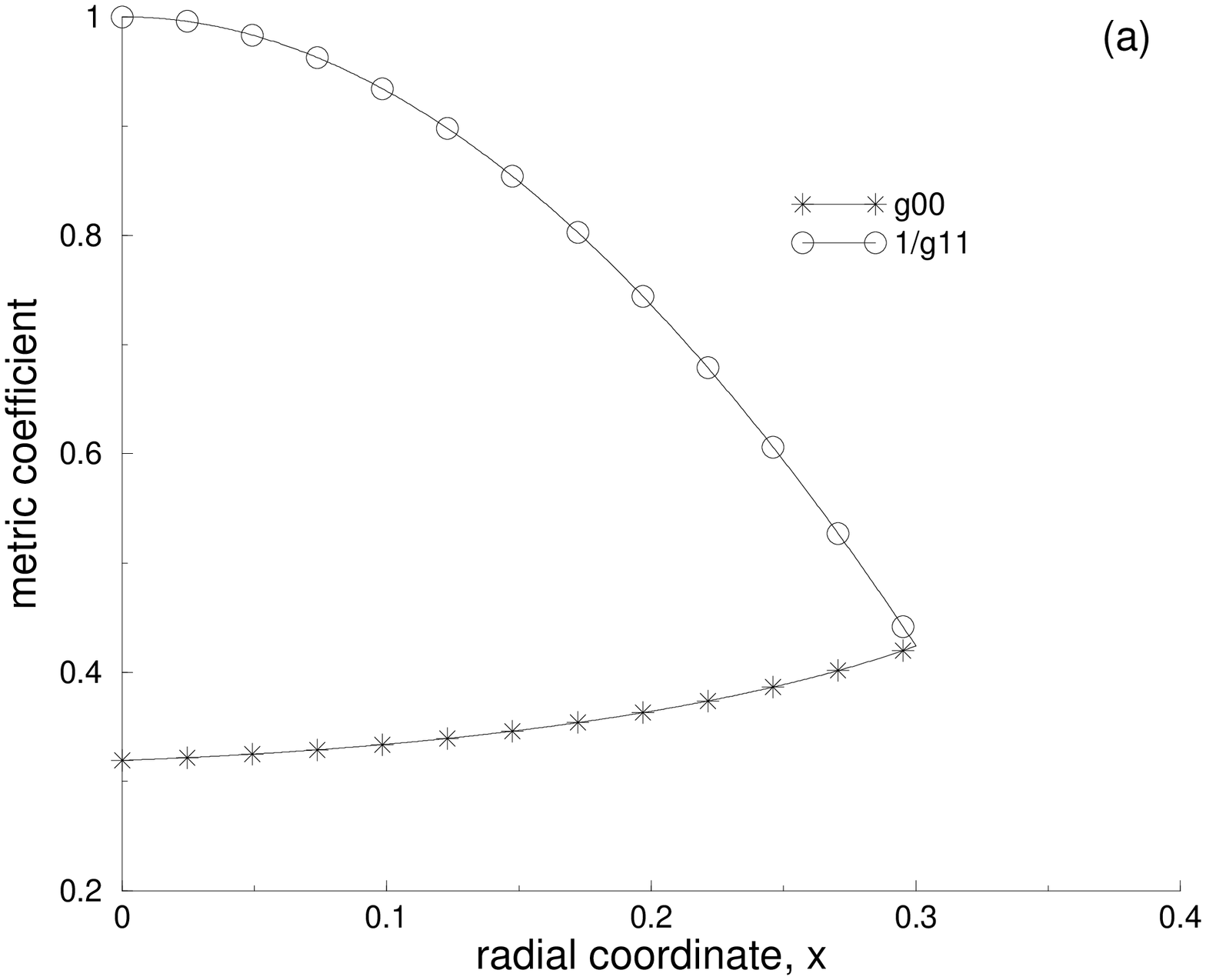}}
\rotatebox{-90}{\includegraphics{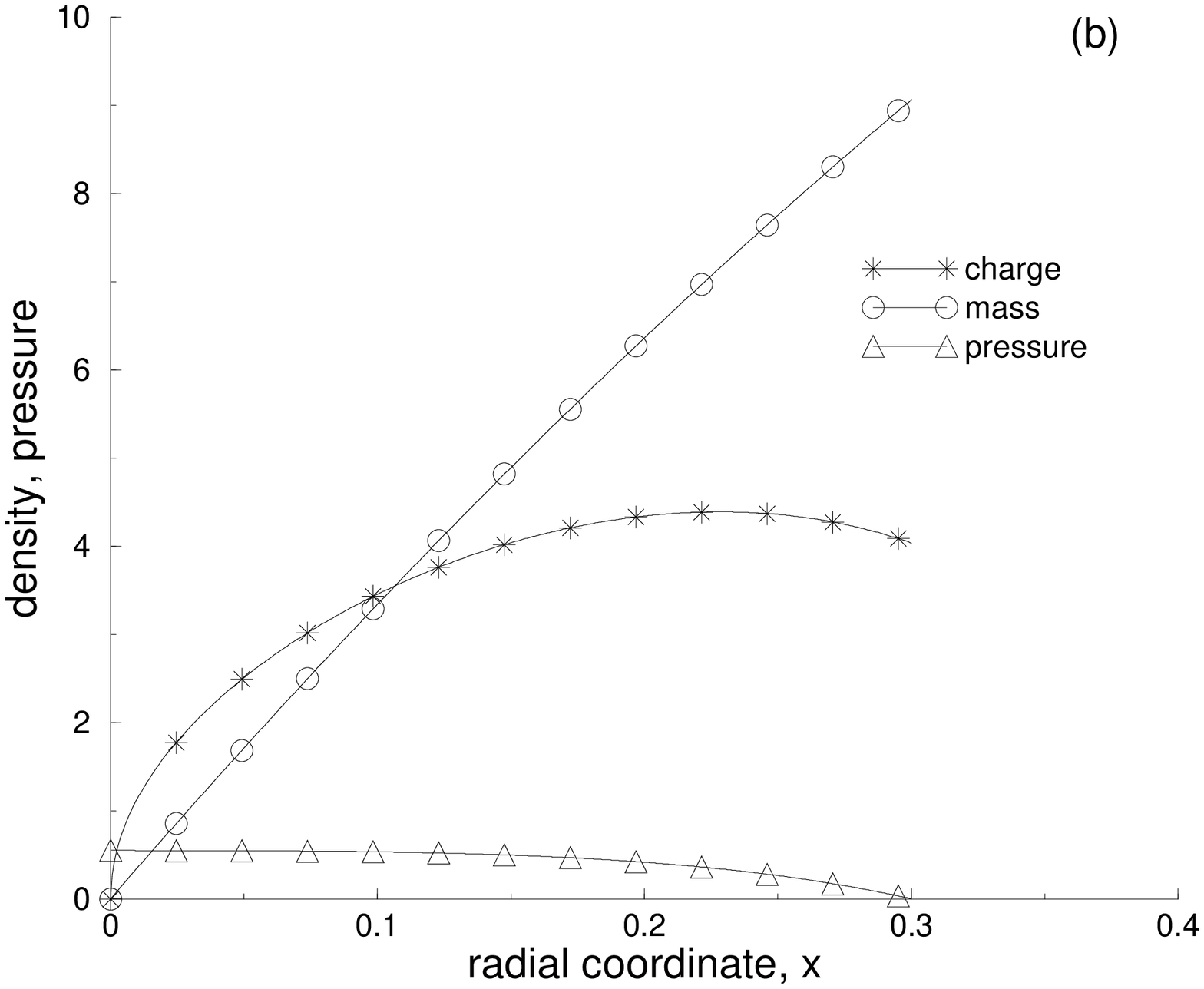}}}
\scalebox{.30}{\rotatebox{-90}{\includegraphics{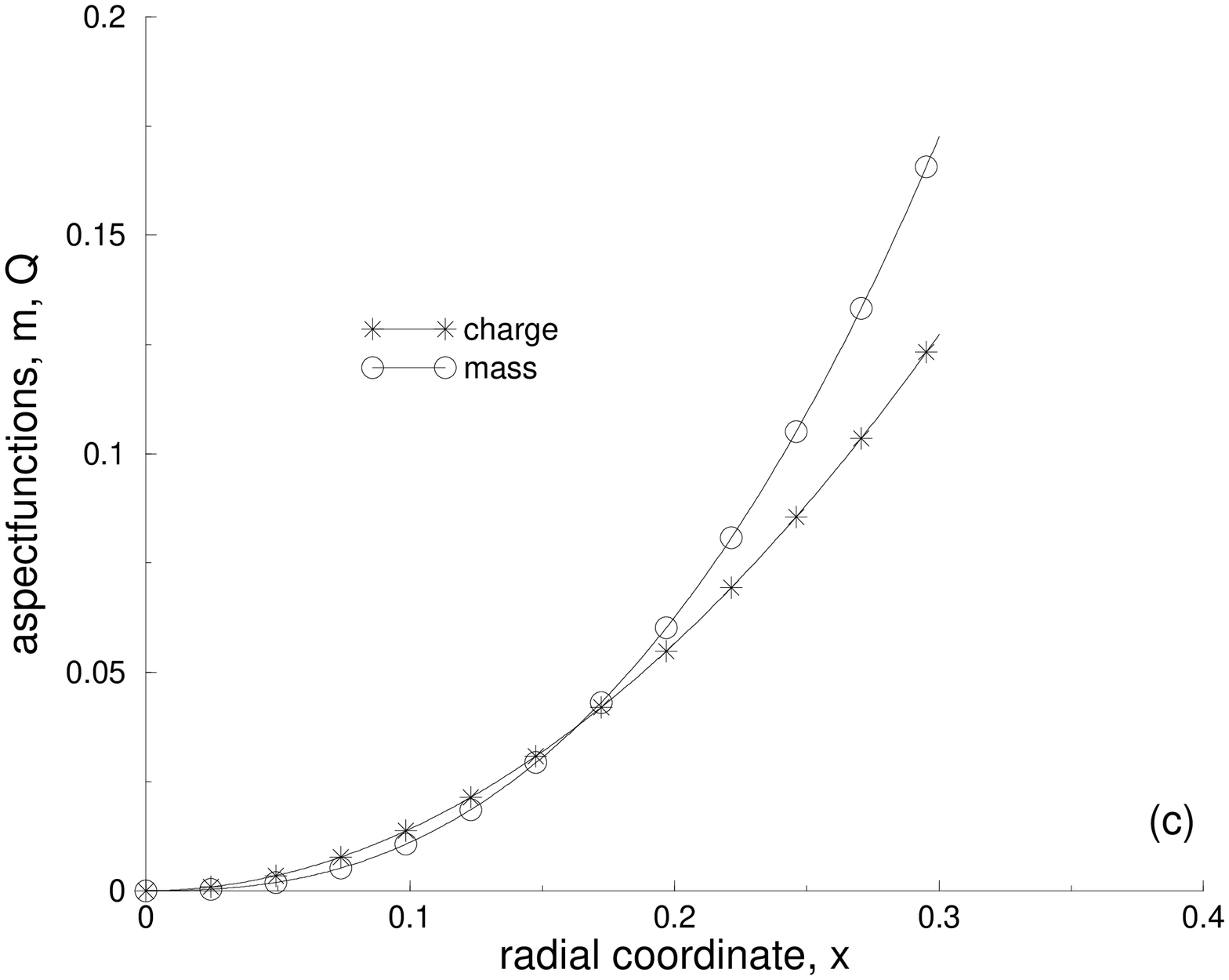}}
\rotatebox{-90}{\includegraphics{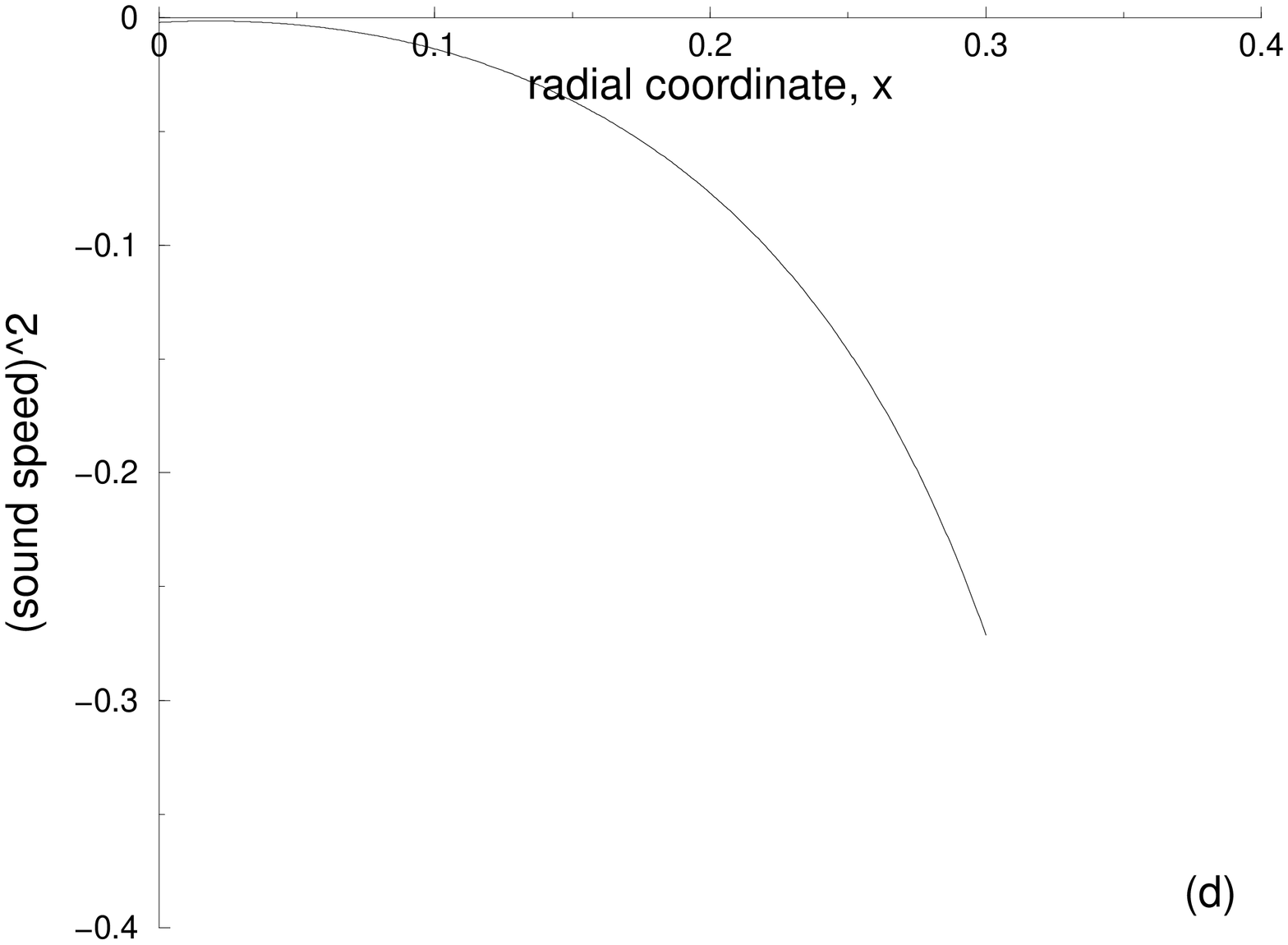}}}}
\end{center}
\caption{ Plots of: (a) the metric functions $y^2$ and $z$; (b) the pressure
$P$, the mass and charge densities $\kappa\rho$ and 
$\kappa\sigma$; (c) the mass and charge functions $m(r)$
and $q(r)$  and (d) $dP/d\rho$ for 
$a=0$ and $b>0$.} 
\label{cubic-exp-a0}
\end{figure}

We can also investigate the sub-case of $\rho _{0}=0$ with both conditions
on $\rho (x_{0})$. Thus, both the mass and charge densities are zero at the
center of the fluid sphere. This is the only relevant case in which to
examine this sub-case as the mass density will remain positive for a large
enough choice of $\rho _{1}$. When $\rho (x_{0})>0$ we desire the mass
density to be large relative to the charge density, so the following values
for the constants $a$, $b$, $c$, and $x_{0}$ are chosen:
\begin{eqnarray*}
a &=&0 \\
b &=&10 \\
c &=&-2 \\
x_{0} &=&0.3
\end{eqnarray*}
We do in fact obtain positive pressures as shown in Figure (\ref
{cubic-exp-a0}).
The speed of sound in this case remains negative due to the increasing mass
density with a decreasing pressure. The speed is shown in Figure (\ref
{cubic-exp-a0}).

\begin{figure}[h]
\begin{center}
{\scalebox{.30}{\rotatebox{-90}{\includegraphics{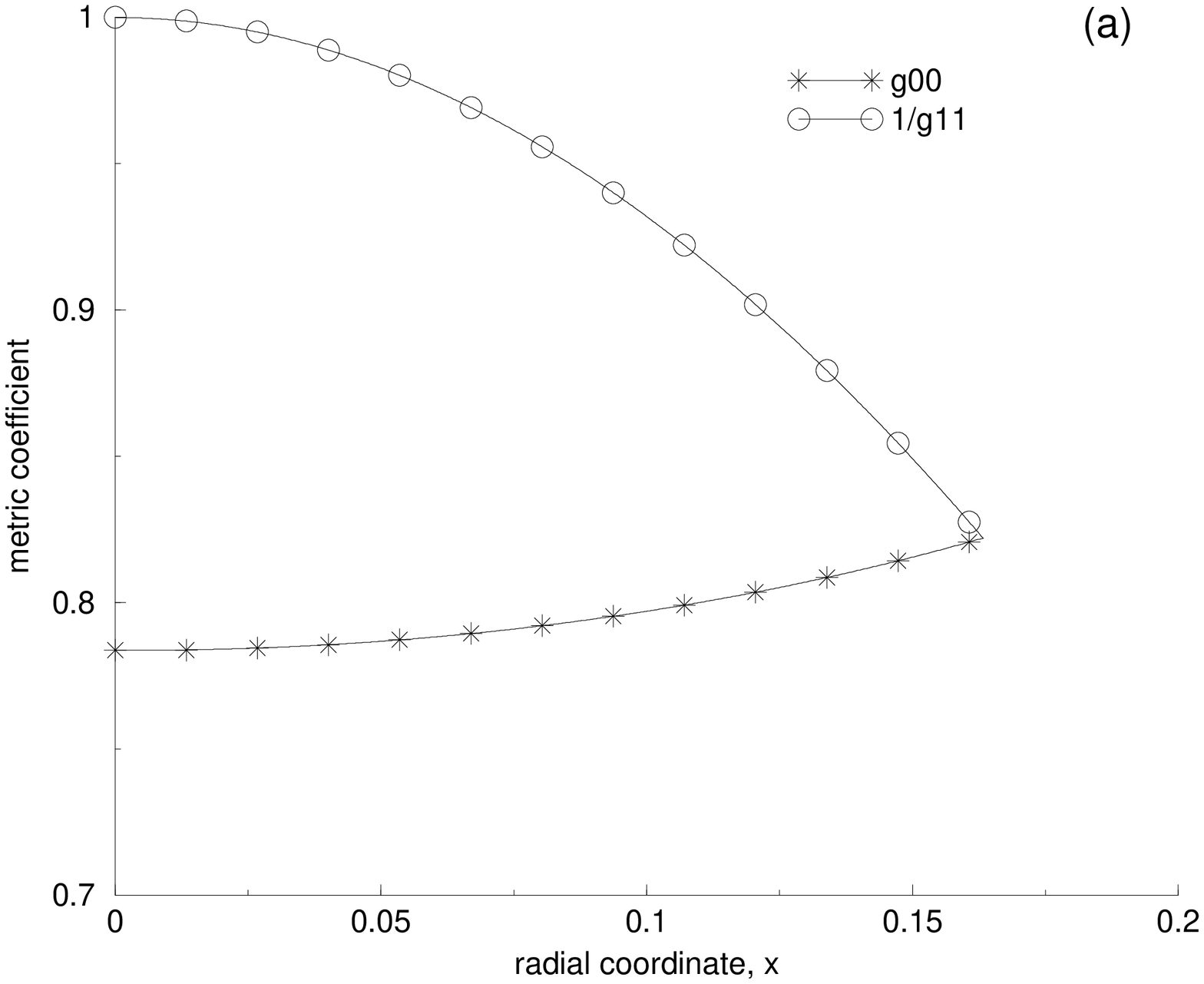}}
\rotatebox{-90}{\includegraphics{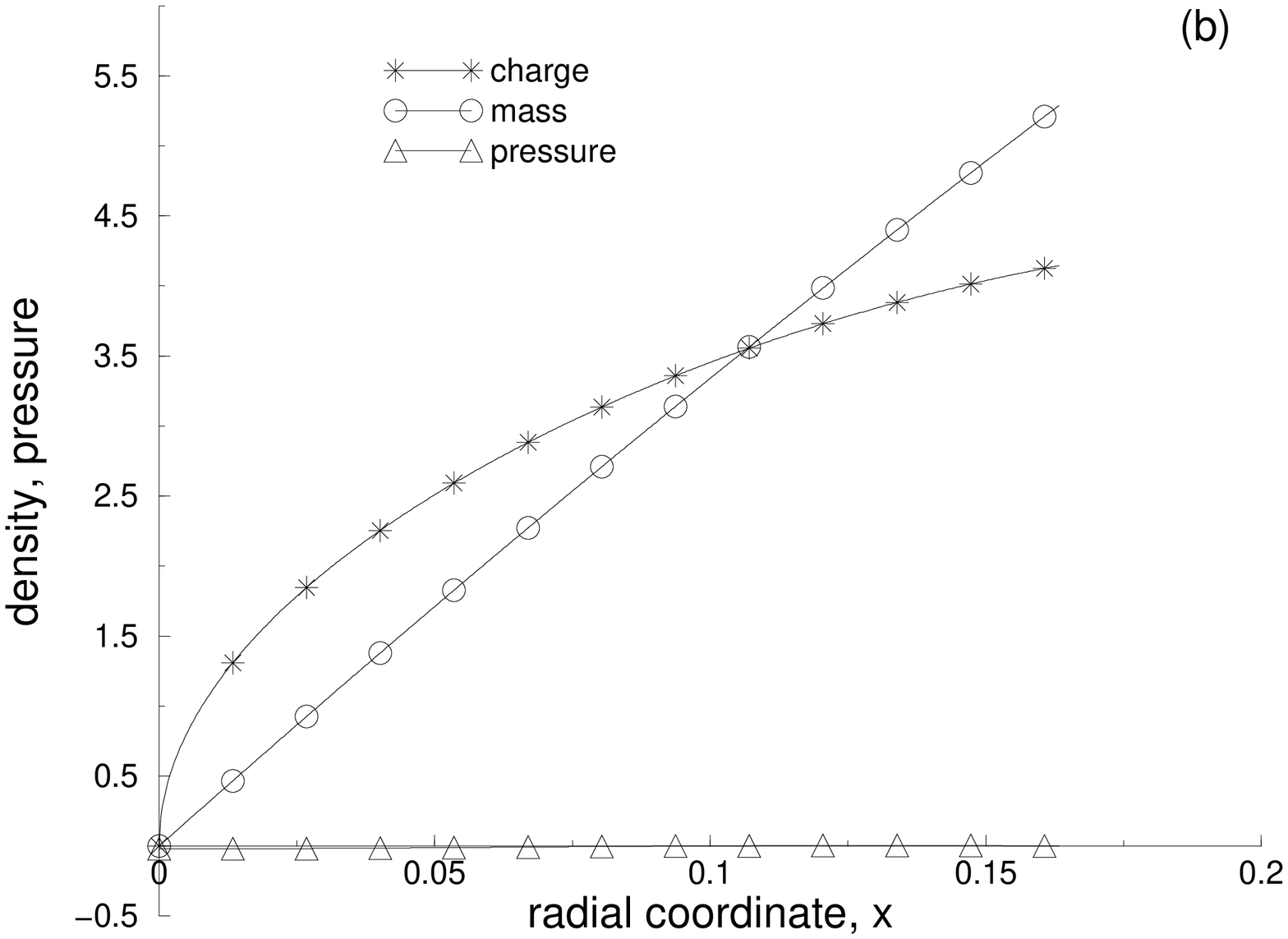}}}
\scalebox{.30}{\rotatebox{-90}{\includegraphics{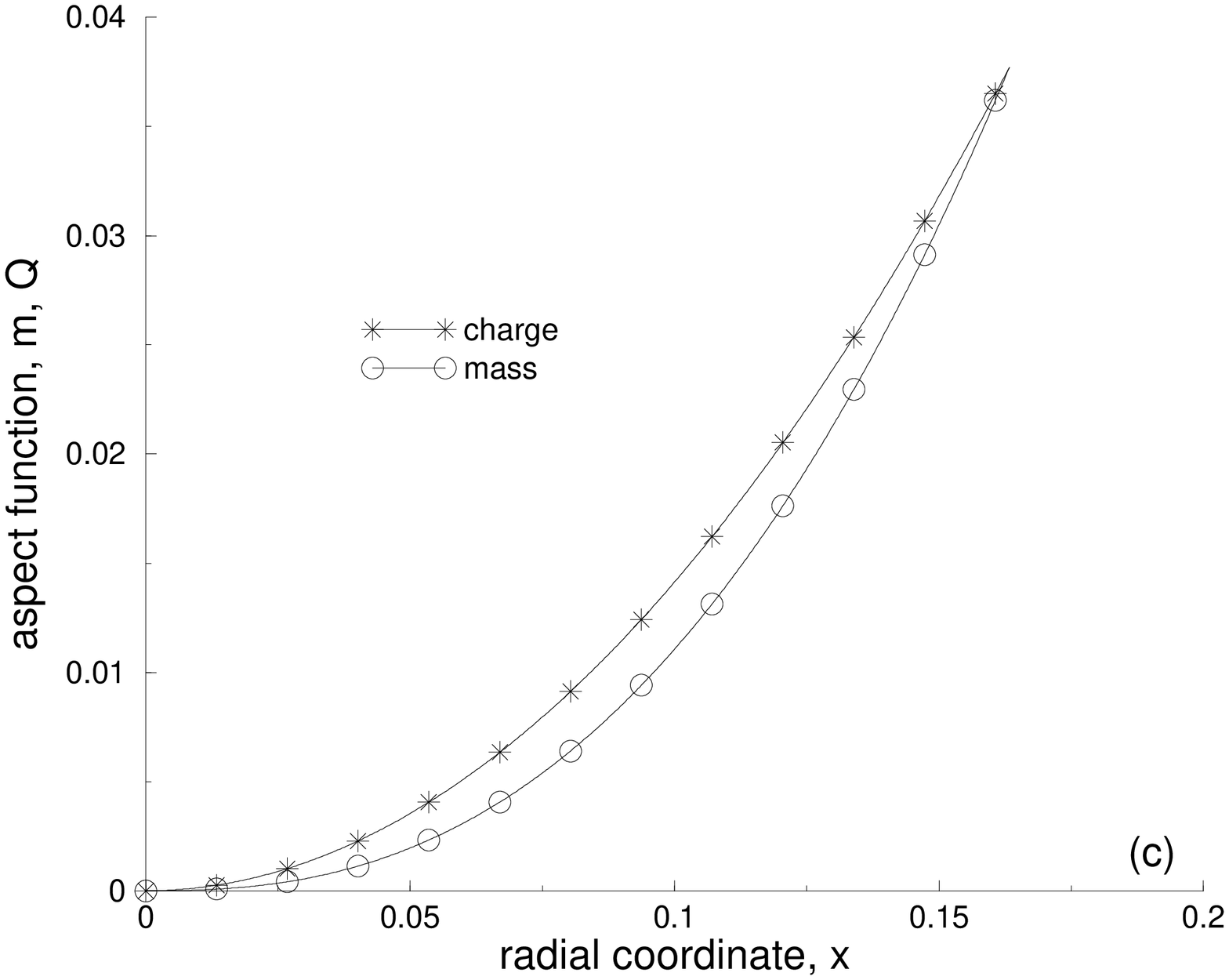}}
\rotatebox{-90}{\includegraphics{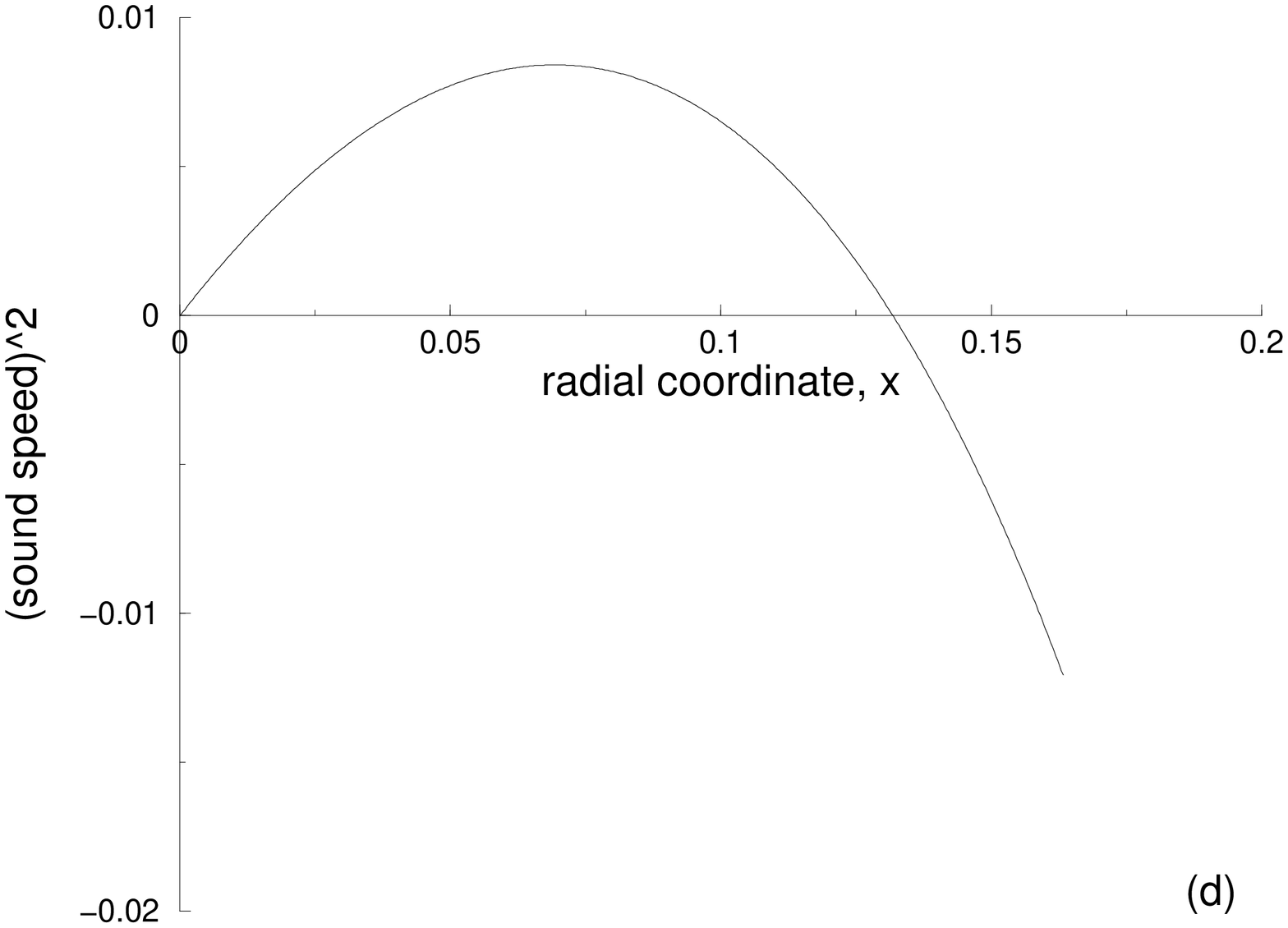}}}}
\end{center}
\caption{ Plots of: (a) the metric functions $y^2$ and $z$; (b) the pressure
$P$, the mass and charge densities $\kappa\rho$ and 
$\kappa\sigma$; (c) the mass and charge functions $m(r)$
and $q(r)$  and (d) $dP/d\rho$ for 
extreme Reissner-Nordstr\"{o}m solution, $a=0$, and $%
b>0$.}
\label{c-eRN-exp-a0}
\end{figure}

The extreme Reissner-Nordstr\"{o}m solution has a boundary given from
equation (\ref{extremeRN-cond}) of:
\[
x_{0}=\frac{4(-c)}{b^{2}}
\]
This leads to negative pressures over part of the sphere as in the example
given in Figure (\ref{c-eRN-exp-a0}) where the following values for the
constants are used
\begin{eqnarray*}
a &=&0 \\
b &=&7 \\
c &=&-2
\end{eqnarray*}
The speed of sound in this case begins positive yet becomes negative as the
pressure increases and then decreases. This speed is given in Figure (\ref
{c-eRN-exp-a0}).

\begin{figure}[h]
\begin{center}
{\scalebox{.30}{\rotatebox{-90}{\includegraphics{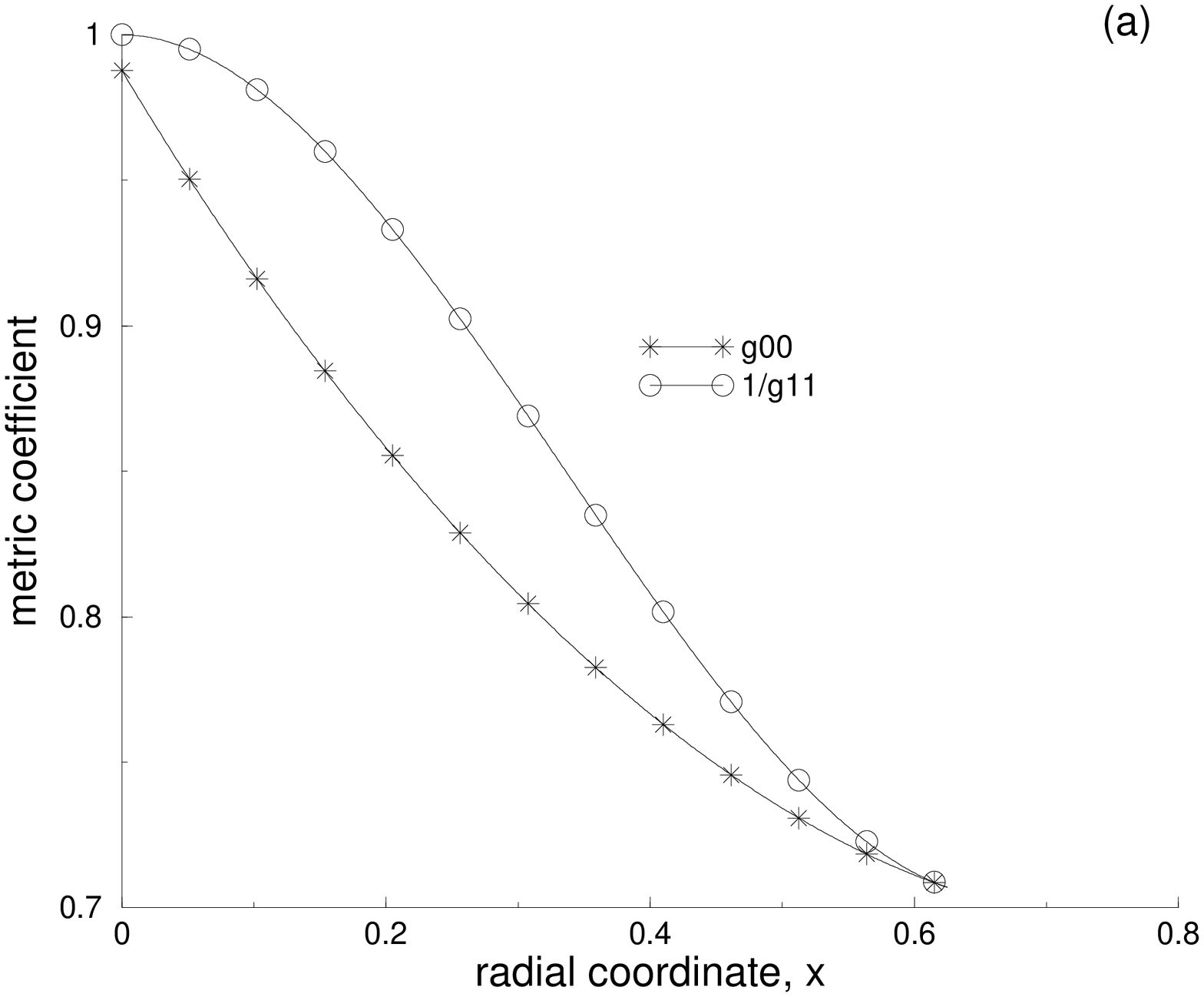}}
\rotatebox{-90}{\includegraphics{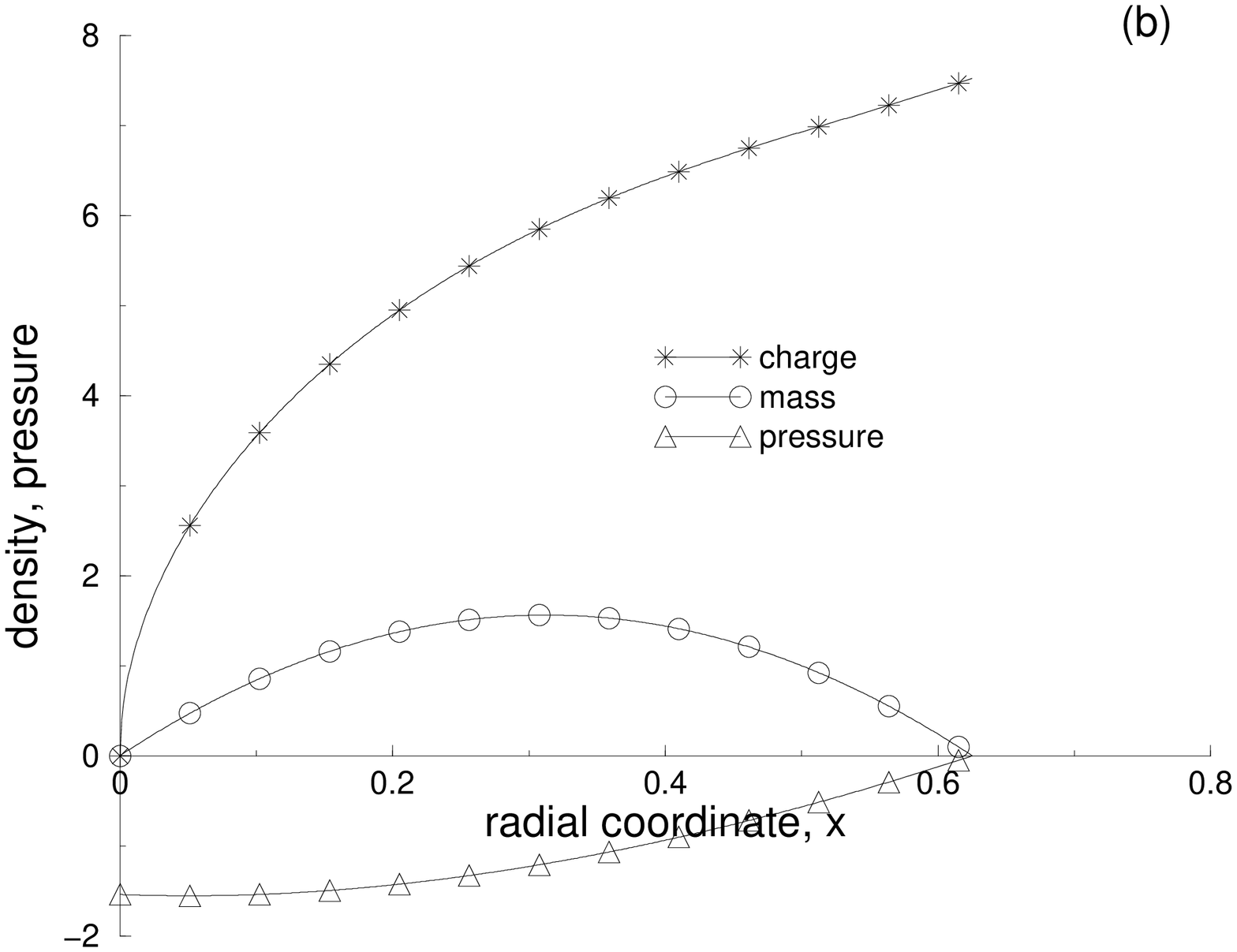}}}
\scalebox{.30}{\rotatebox{-90}{\includegraphics{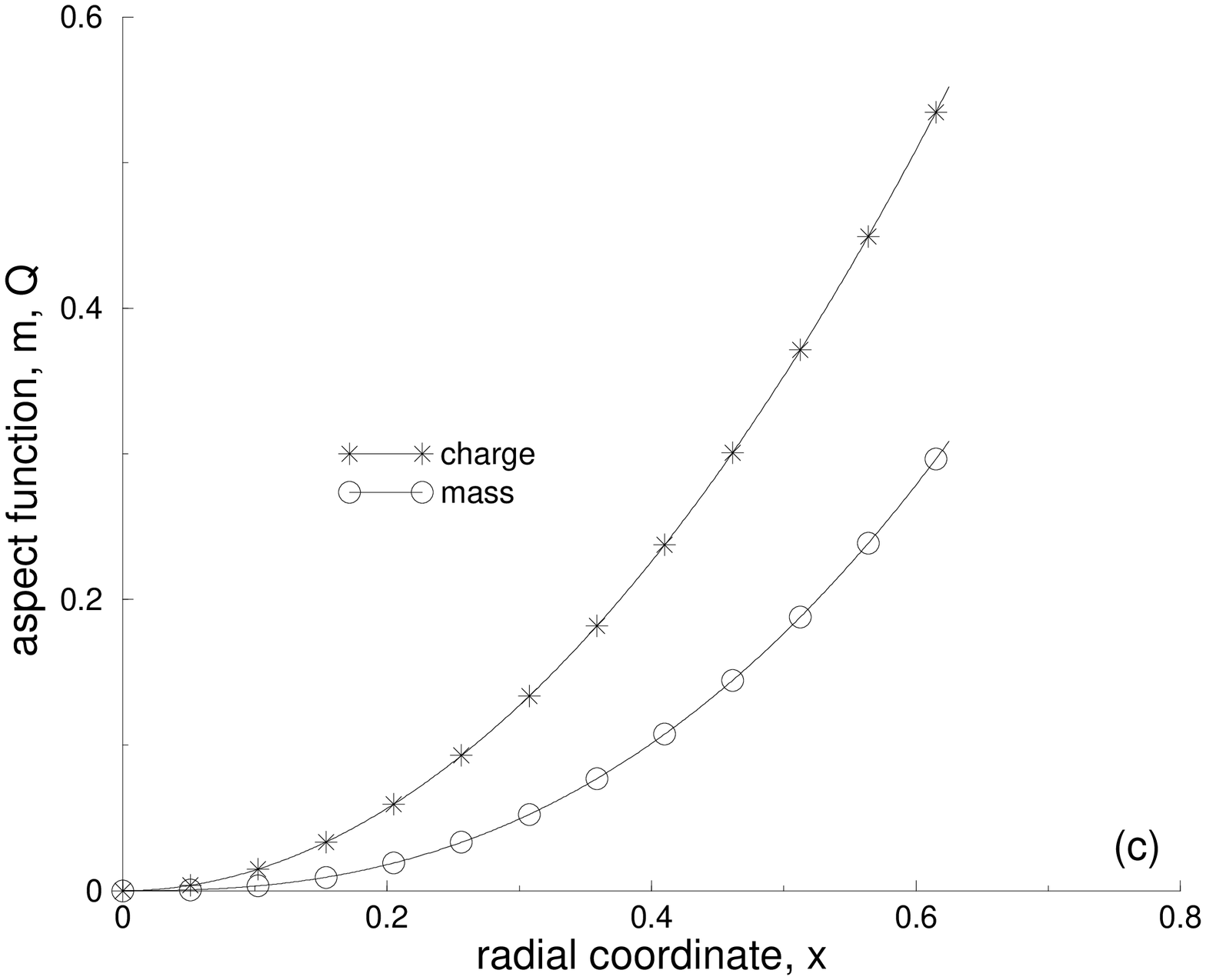}}
\rotatebox{-90}{\includegraphics{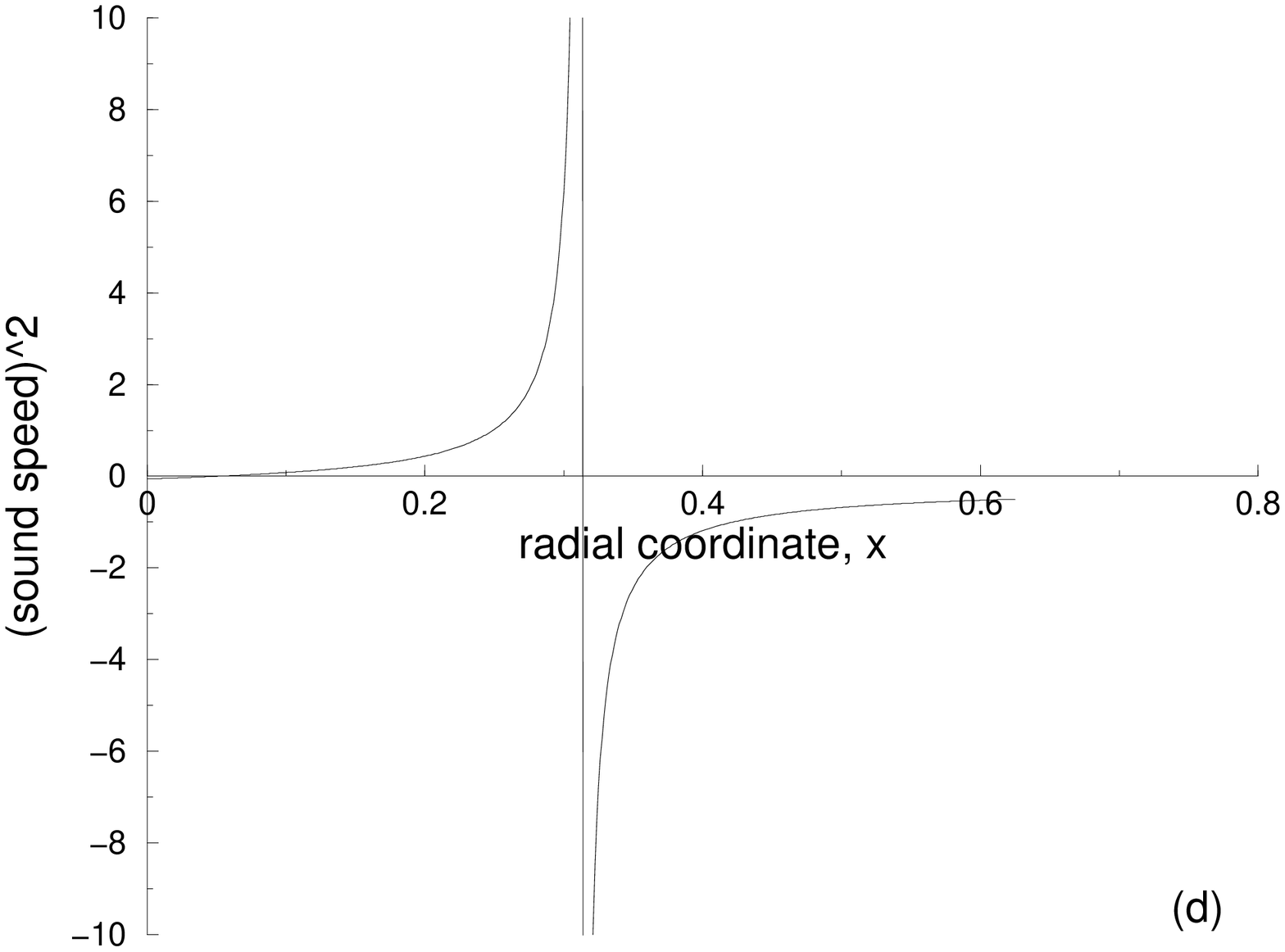}}}}
\end{center}
\caption{ Plots of: (a) the metric functions $y^2$ and $z$; (b) the pressure
$P$, the mass and charge densities $\kappa\rho$ and 
$\kappa\sigma$; (c) the mass and charge functions $m(r)$
and $q(r)$  and (d) $dP/d\rho$ for 
$a=0$, $b>0$, and $\protect\rho (x_{0})=0$.} 
\label{c-rho0-exp-a0}
\end{figure}

However, when $\rho (x_{0})=0$ and $\rho _{0}=0$ the mass density remains
below the charge density and, hence, negative pressures result. For example,
we choose the following values for $b$ and $c$:
\begin{eqnarray*}
b &=&2 \\
c &=&-2.
\end{eqnarray*}
The resulting pressure is shown in Figure (\ref{c-rho0-exp-a0}).
\begin{figure}[h]
\begin{center}
{\scalebox{.30}{\rotatebox{-90}{\includegraphics{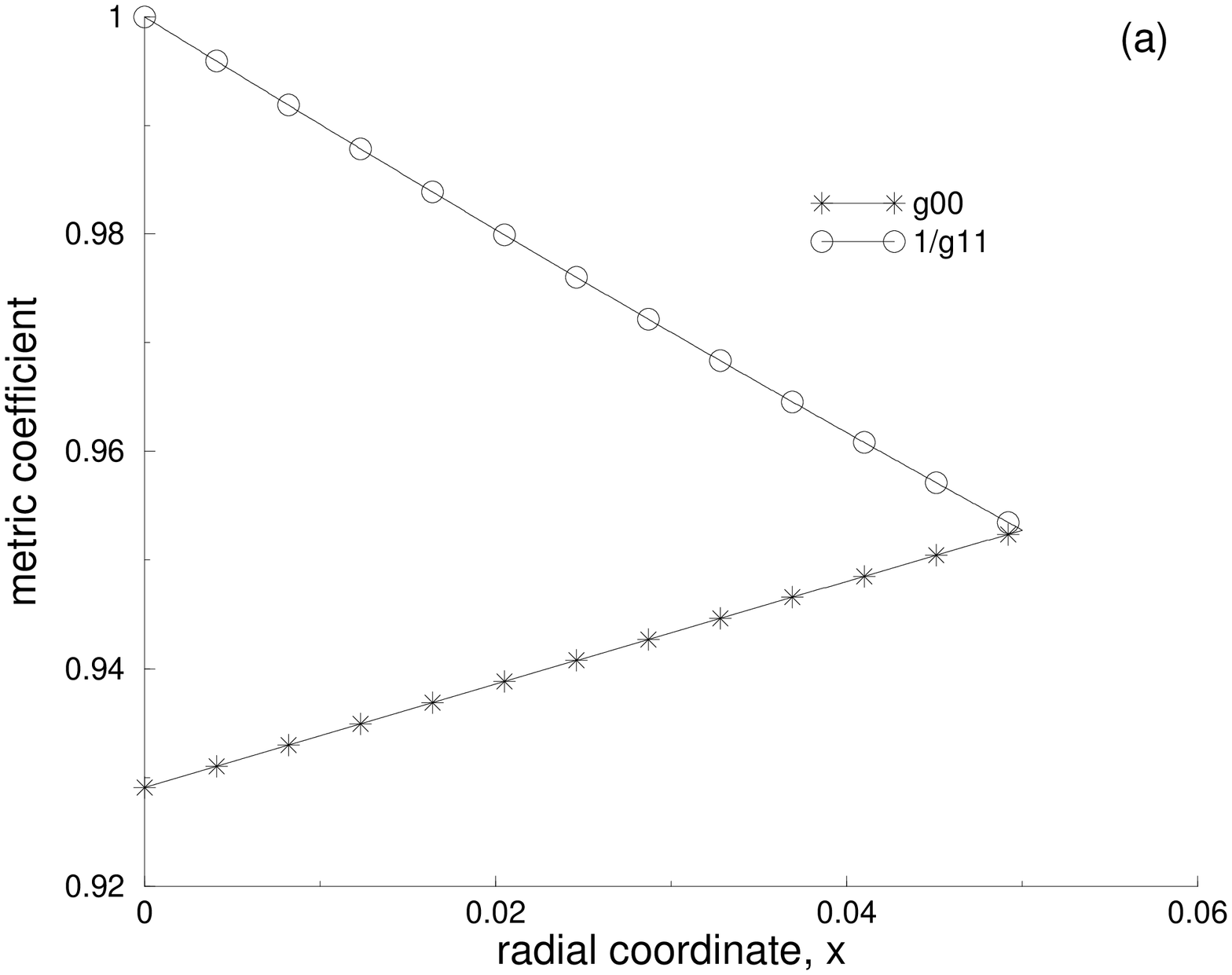}}
\rotatebox{-90}{\includegraphics{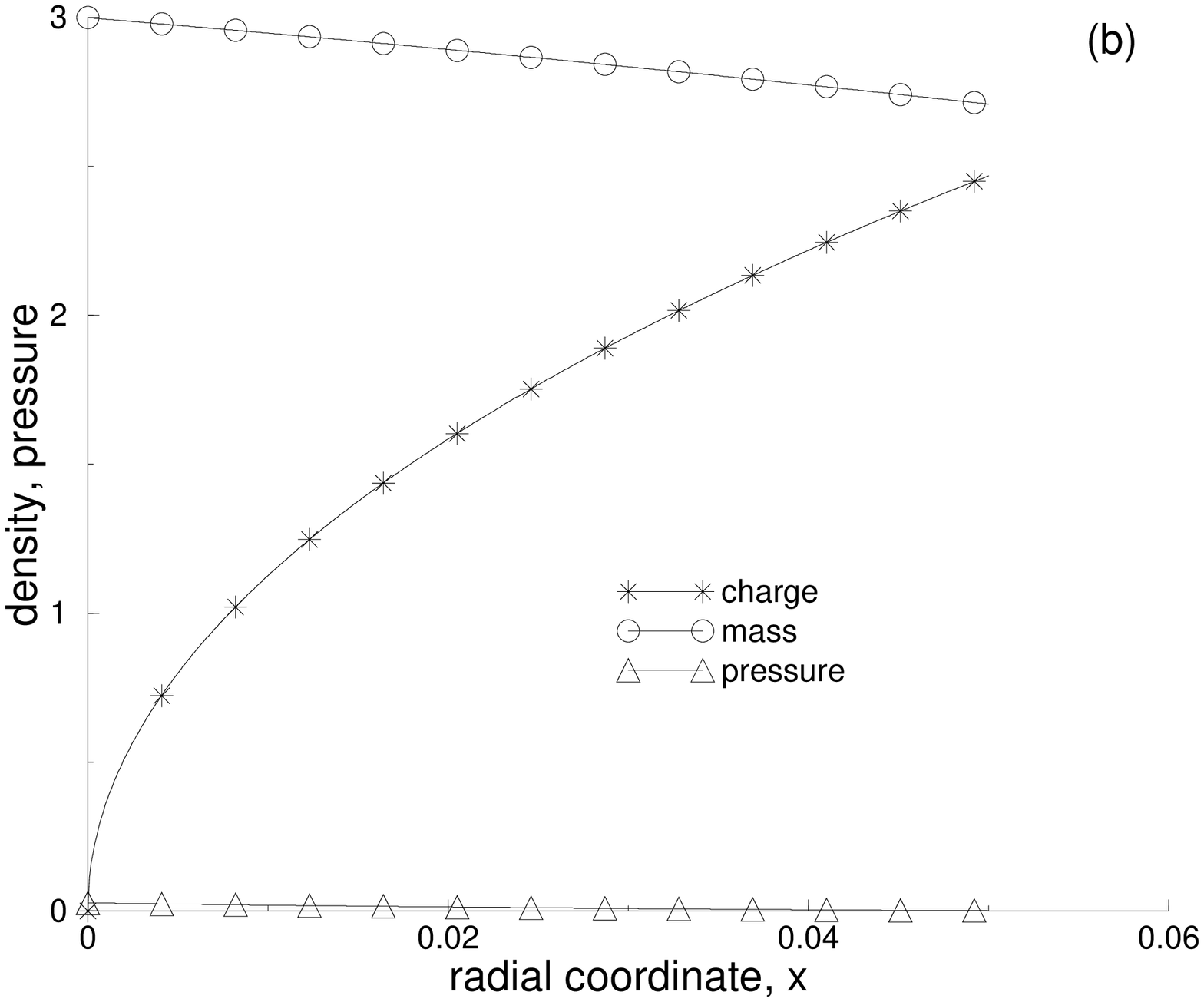}}}
\scalebox{.30}{\rotatebox{-90}{\includegraphics{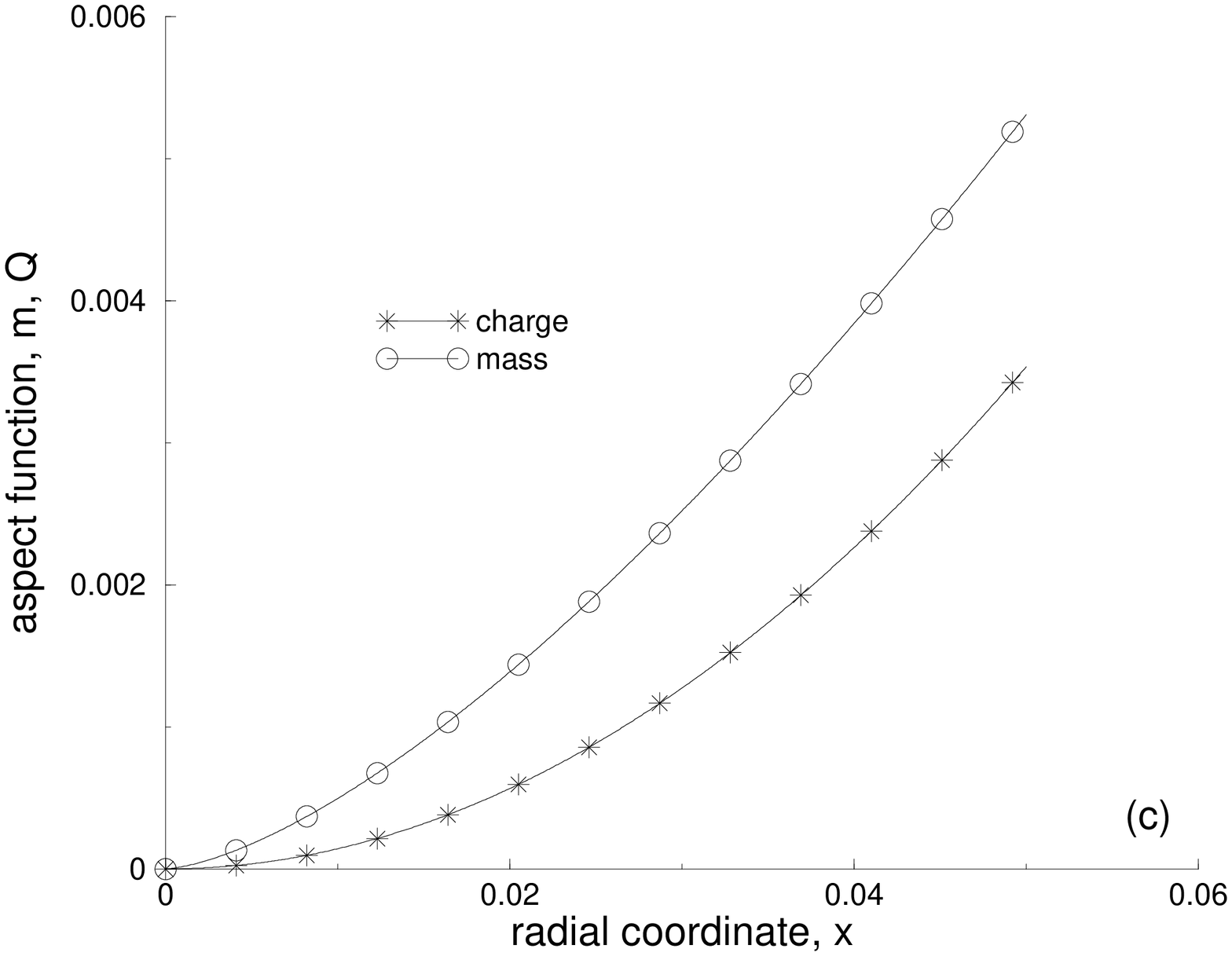}}
\rotatebox{-90}{\includegraphics{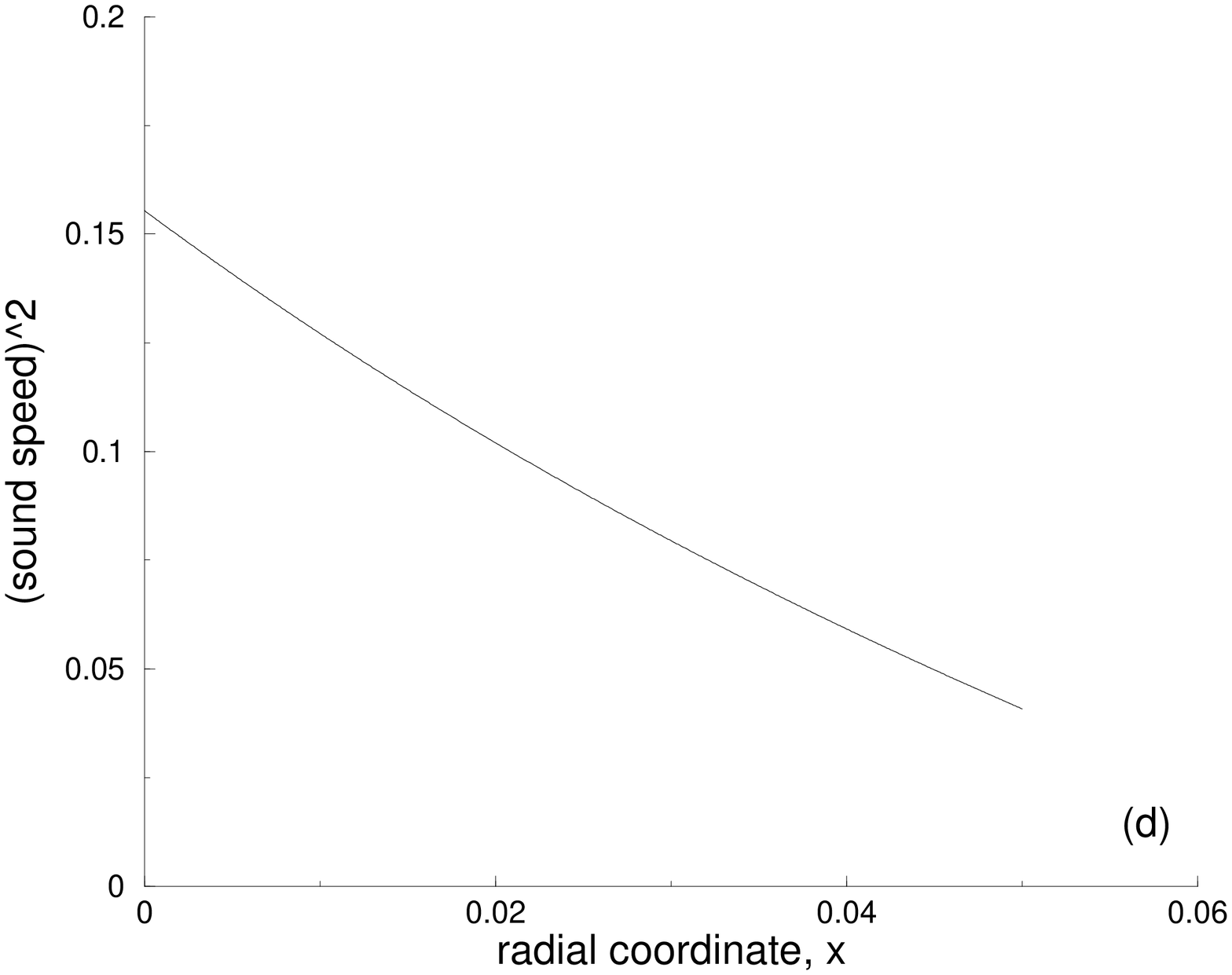}}}}
\end{center}
\caption{ Plots of: (a) the metric functions $y^2$ and $z$; (b) the pressure
$P$, the mass and charge densities $\kappa\rho$ and 
$\kappa\sigma$; (c) the mass and charge functions $m(r)$
and $q(r)$  and (d) $dP/d\rho$ for 
$b<0$.}
\label{cubic-trig}
\end{figure}

The speed of sound in this case becomes singular when the mass density peaks
(i.e. $d\rho =0$).

\subsection{Case $d>0$ ($b<0$, $\protect\rho _{1}>0$)}

\bigskip This case involves the sine and cosine solution for $y$ given by
equation (\ref{y-trig}). Now, if we use the junction condition (\ref
{Junction-RN}) and equation (\ref{yprime-condition}), the constants $C_{1}$
and $C_{2}$ become
\begin{eqnarray*}
C_{1} &=&\sqrt{z(x_{0})}\sin \sqrt{\frac{b}{4}}\xi _{0}+\frac{1}{\sqrt{b}}%
\left( \frac{a}{2}+\frac{b}{2}x_{0}+cx_{0}^{2}\right) \cos \sqrt{\frac{b}{4}}%
\xi _{0} \\
C_{2} &=&\sqrt{z(x_{0})}\cos \sqrt{\frac{b}{4}}\xi _{0}-\frac{1}{\sqrt{b}}%
\left( \frac{a}{2}+\frac{b}{2}x_{0}+cx_{0}^{2}\right) \sin \sqrt{\frac{b}{4}}%
\xi _{0}.
\end{eqnarray*}
We obtain positive pressures in this case as well. The following example,
Figure(\ref{cubic-trig}), uses the following values for the constants $a$, $%
b $, $c$, and $x_{0}$:
\begin{eqnarray*}
a &=&1 \\
b &=&-1 \\
c &=&-2 \\
x_{0} &=&0.05
\end{eqnarray*}


The speed of sound in this case, as shown in Figure (\ref{cubic-trig}), does
follow the conditions given by equation (\ref{ssp-condition}).

Thus, this case satisfies the conditions for a physically realistic
solution, that: at the center of the sphere, the mass $m$ vanishes and $%
e^{\lambda}=1$; the metric functions $y^2$ and $z$ match the external
Reissner-Nordstr\"{o}m metric at the boundary; the pressure is positive and
monotonically goes to zero at the boundary; the mass density is positive;
and the speed of sound in this medium is both positive and causal.

\begin{figure}[h]
\begin{center}
{\scalebox{.30}{\rotatebox{-90}{\includegraphics{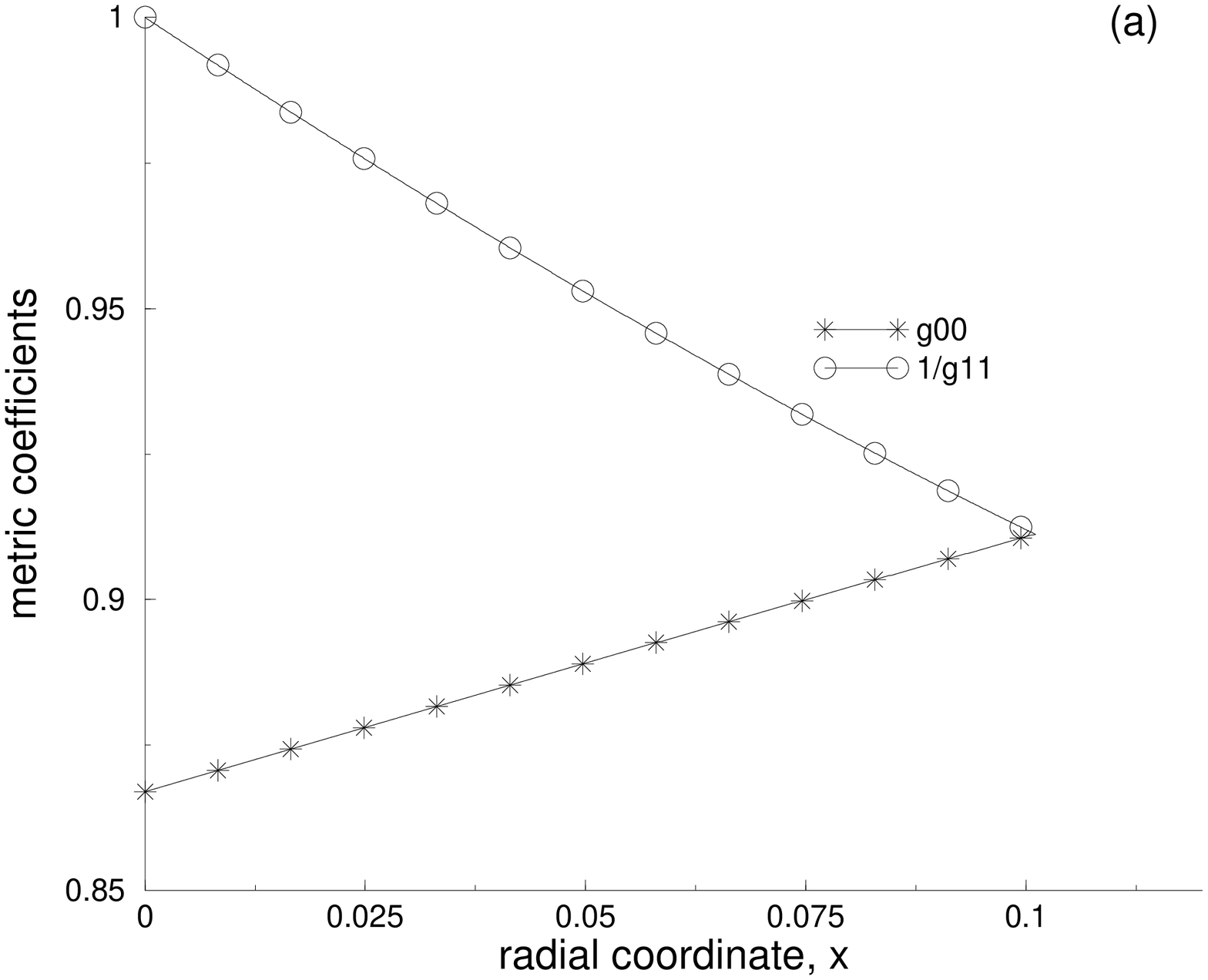}}
\rotatebox{-90}{\includegraphics{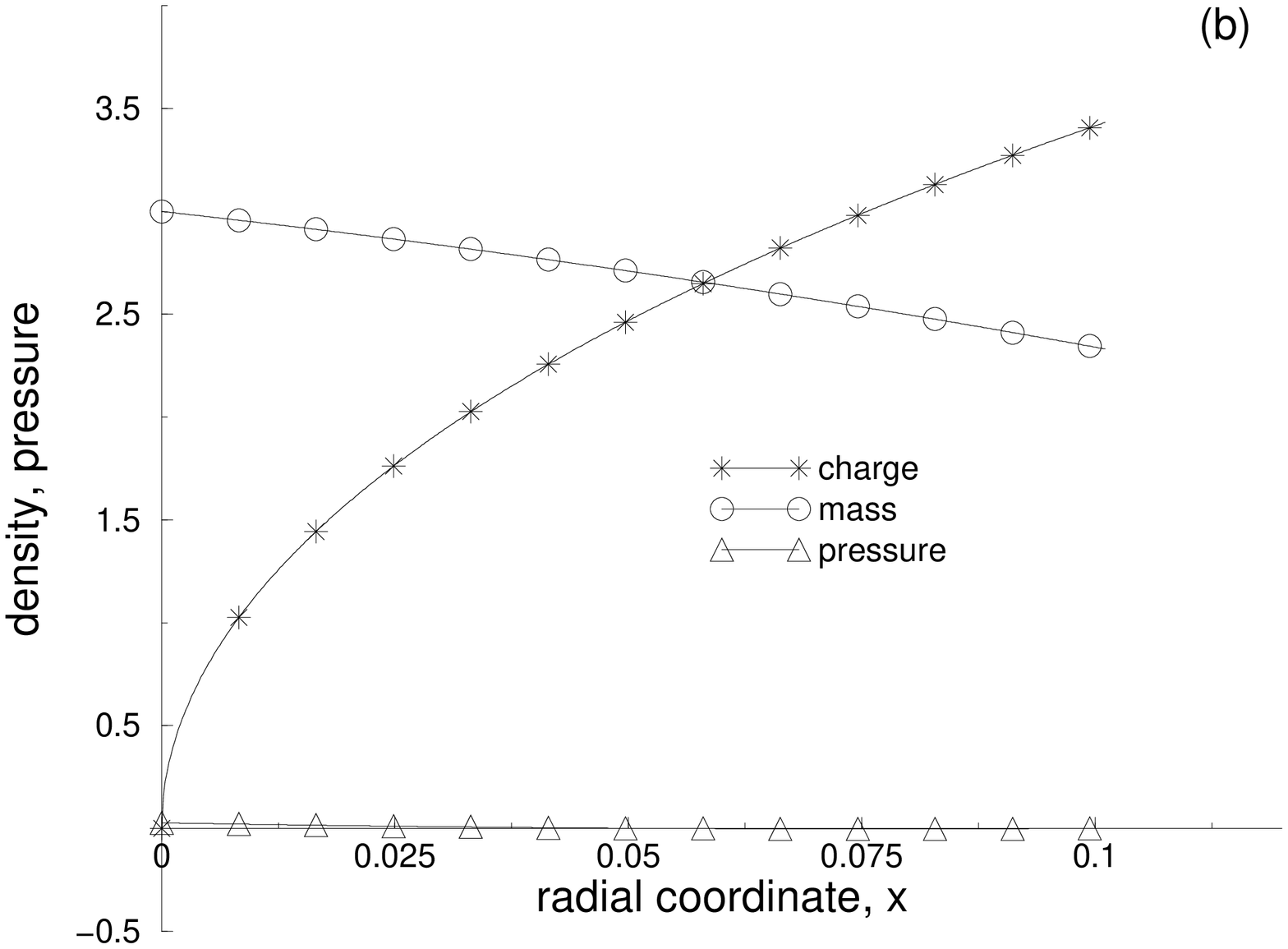}}}
\scalebox{.30}{\rotatebox{-90}{\includegraphics{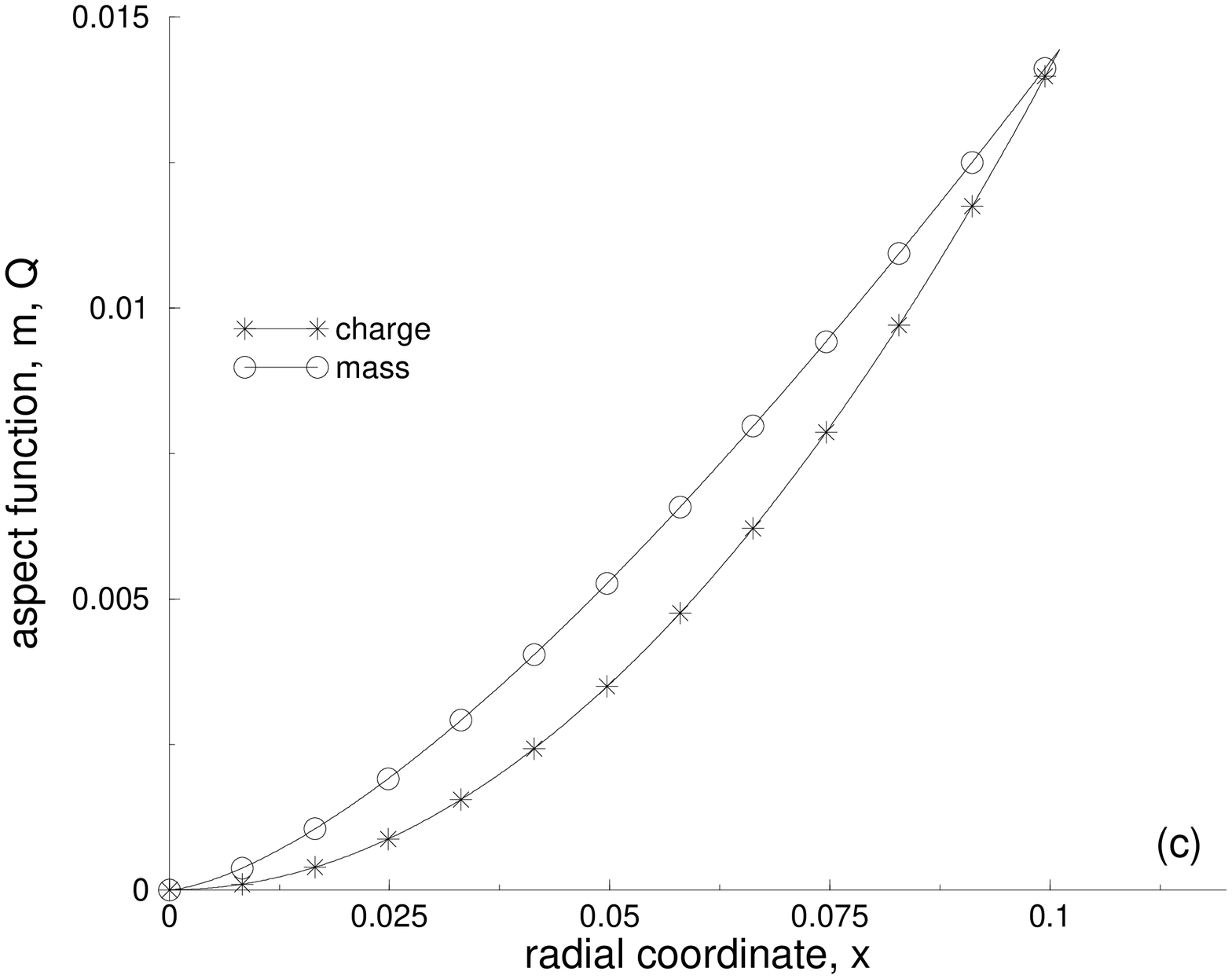}}
\rotatebox{-90}{\includegraphics{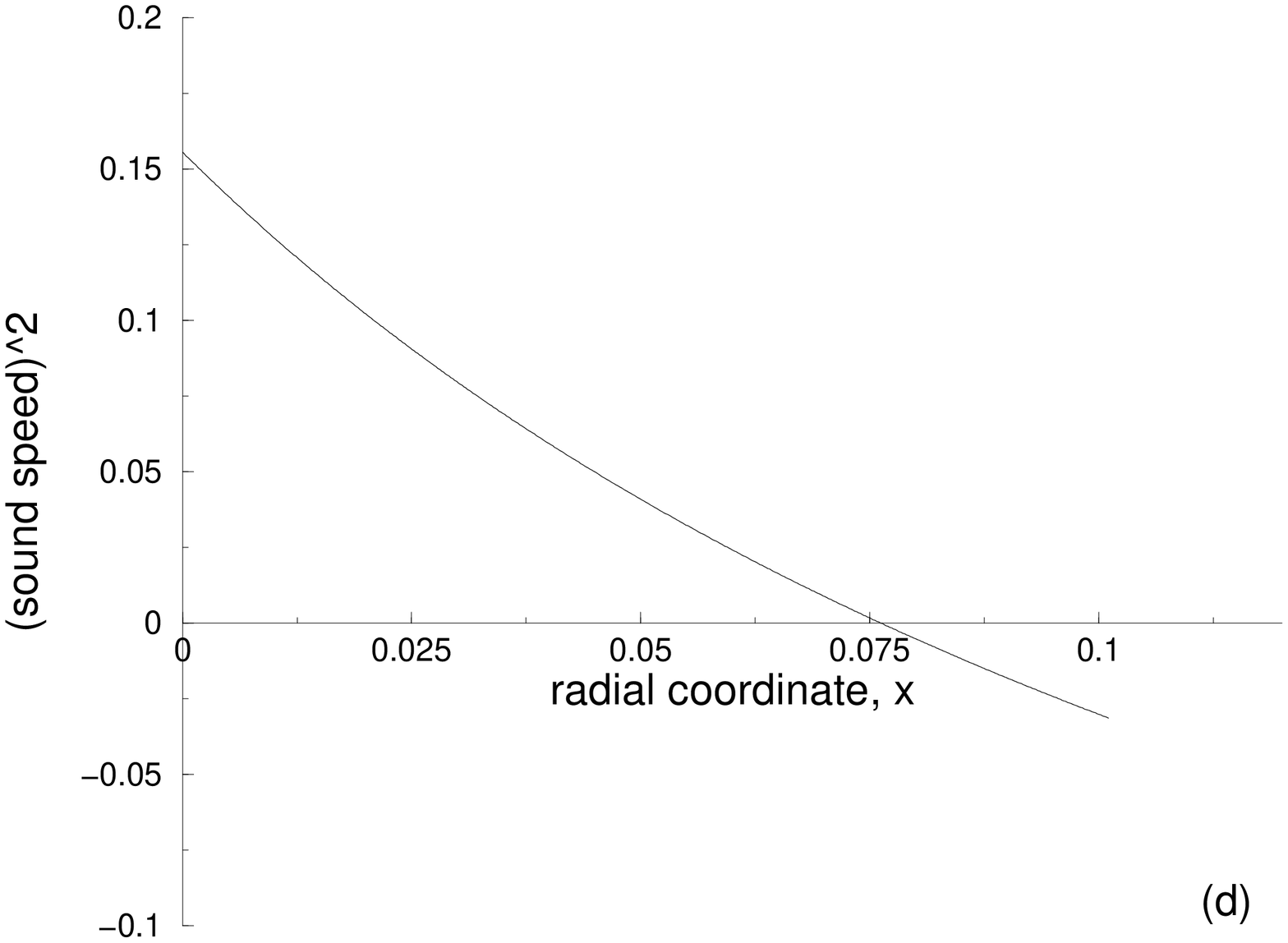}}}}
\end{center}
\caption{ Plots of: (a) the metric functions $y^2$ and $z$; (b) the pressure
$P$, the mass and charge densities $\kappa\rho$ and 
$\kappa\sigma$; (c) the mass and charge functions $m(r)$
and $q(r)$  and (d) $dP/d\rho$ for 
extreme Reissner-Nordstr\"{o}m solution and $b<0$.} 
\label{c-eRN-trig}
\end{figure}

Matching to the extreme Reissner-Nordstr\"{o}m solution results in
another situation where the pressure does not monotonically vanish
at the boundary, given in Figure (\ref{c-eRN-trig}) for the same
values as in the last example. The boundary is given from equation
(\ref{extremeRN-cond}):
\[
x_{0}=\left( \frac{\sqrt{-c}}{b}-\frac{1}{b}\sqrt{-c-ab}\right) ^{2}.
\]
\begin{figure}[h]
\begin{center}
{\scalebox{.30}{\rotatebox{-90}{\includegraphics{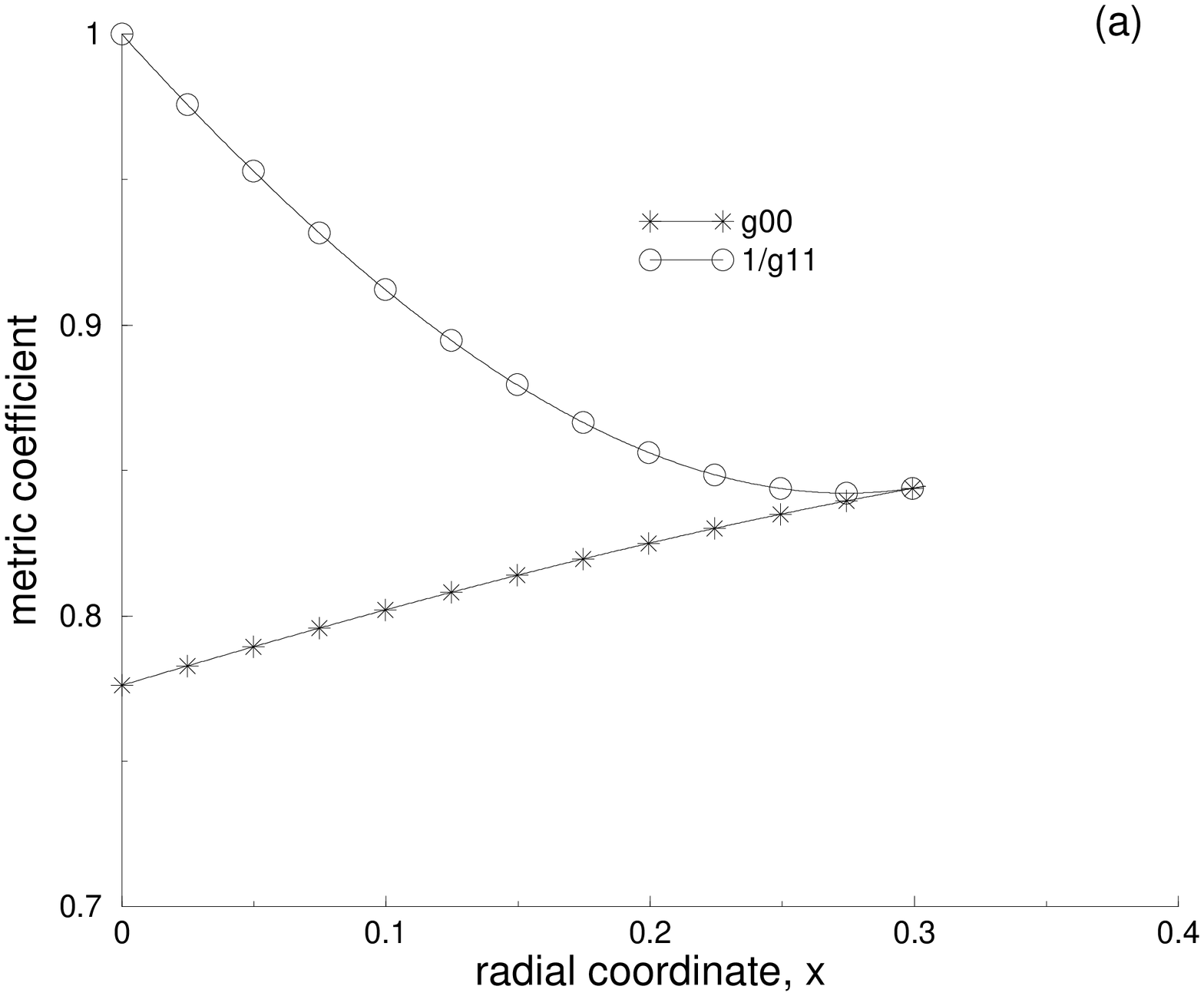}}
\rotatebox{-90}{\includegraphics{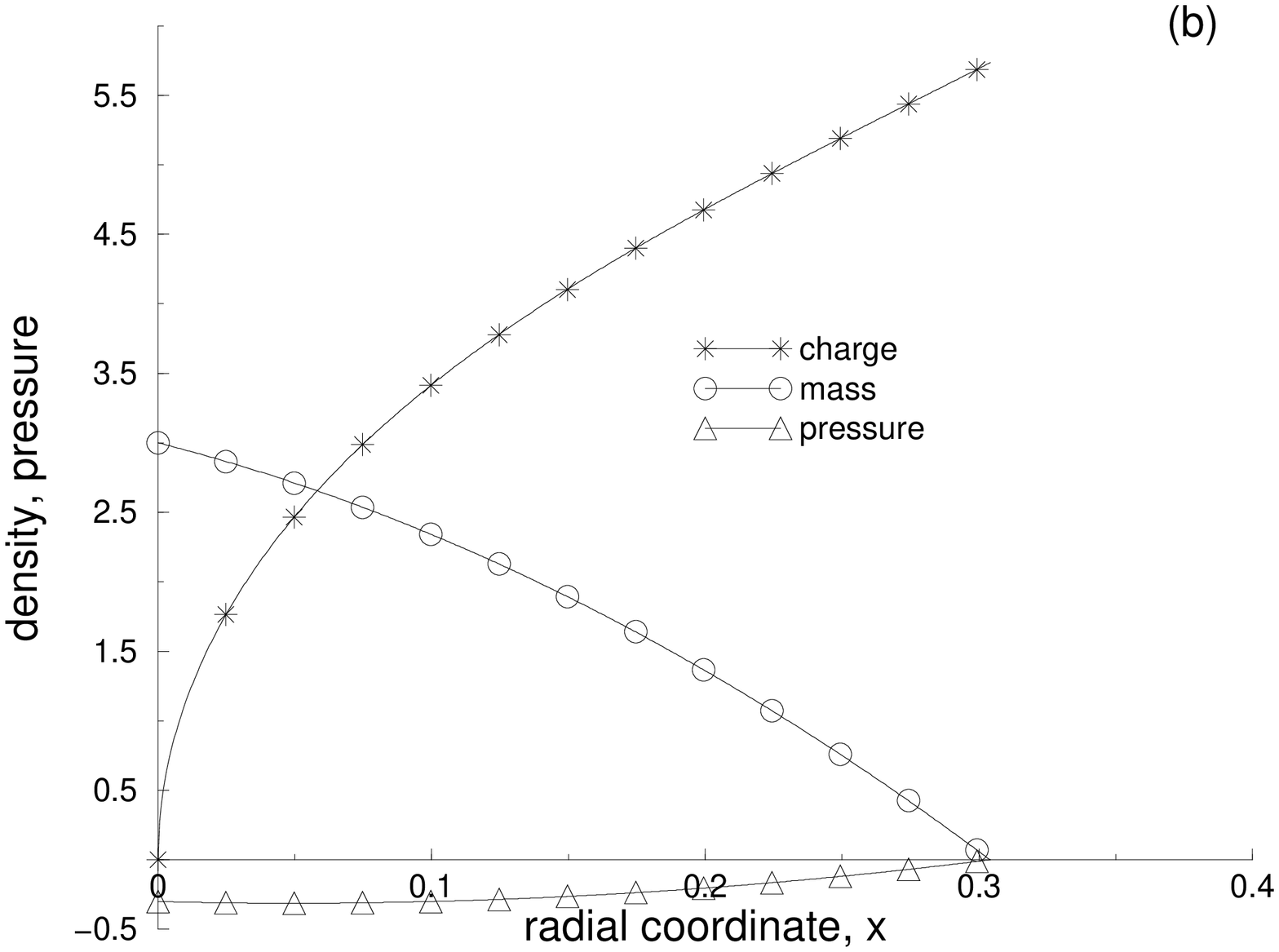}}}
\scalebox{.30}{\rotatebox{-90}{\includegraphics{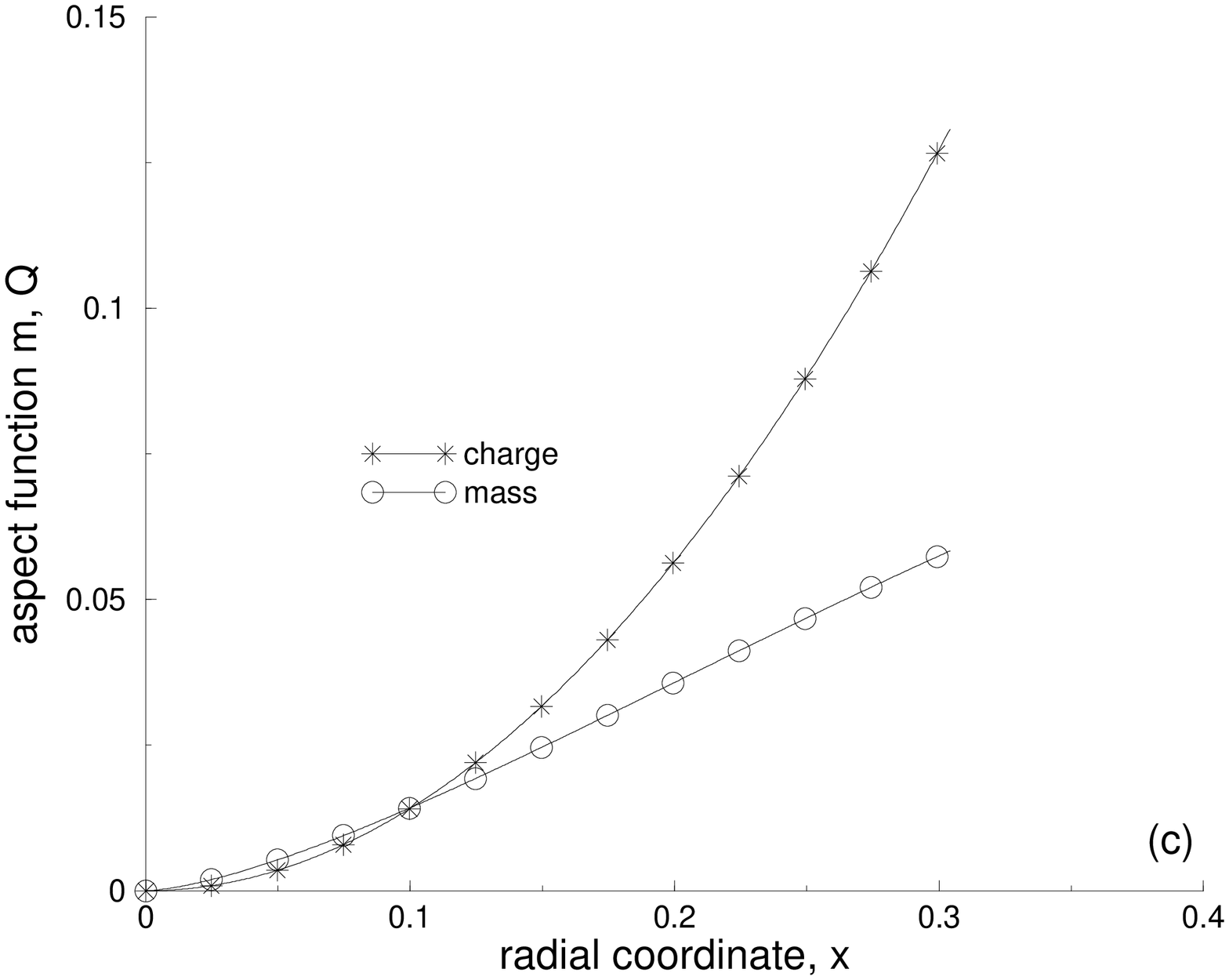}}
\rotatebox{-90}{\includegraphics{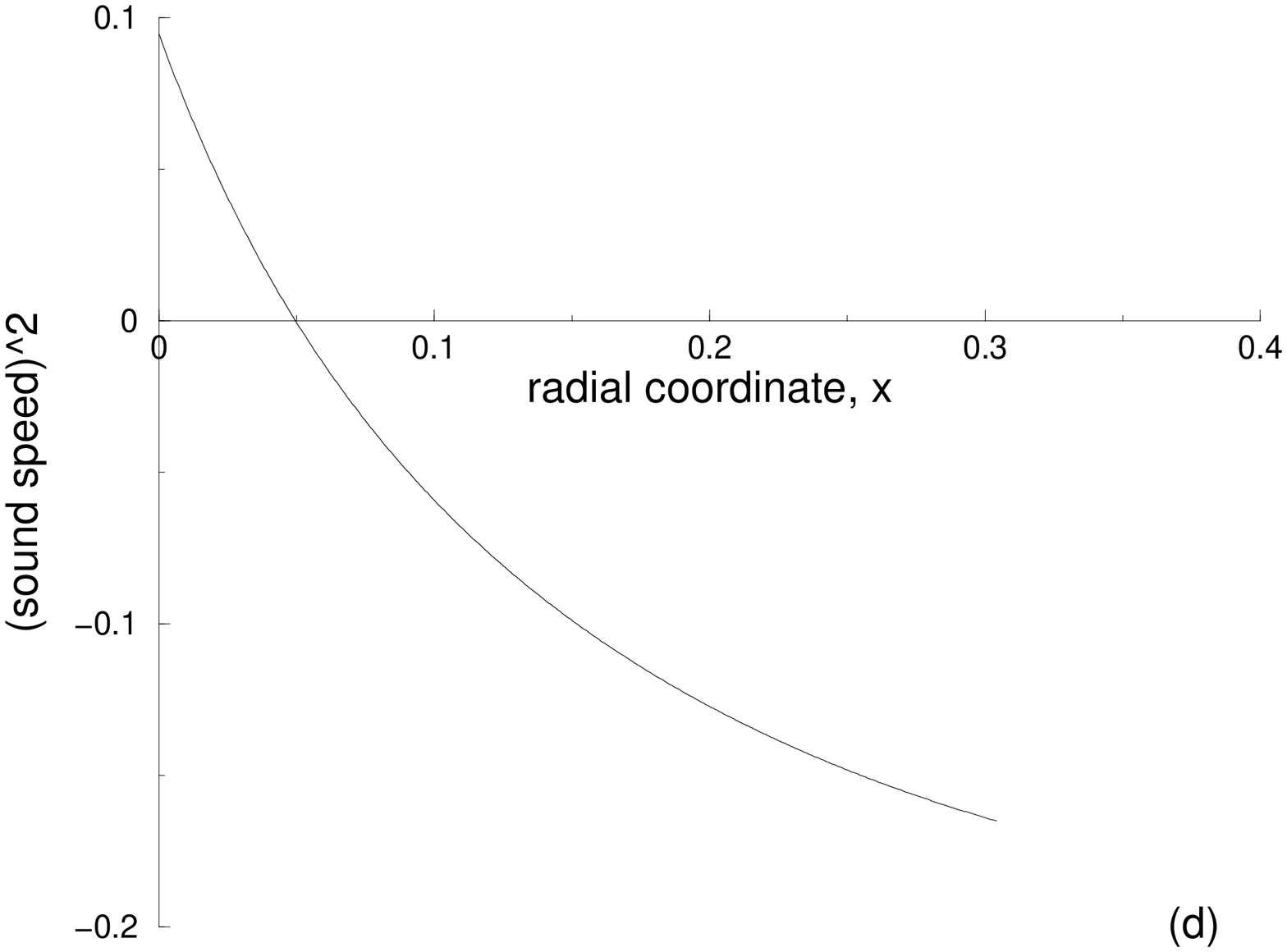}}}}
\end{center}
\caption{ Plots of: (a) the metric functions $y^2$ and $z$; (b) the pressure
$P$, the mass and charge densities $\kappa\rho$ and 
$\kappa\sigma$; (c) the mass and charge functions $m(r)$
and $q(r)$  and (d) $dP/d\rho$ for 
$b<0$ and $\protect\rho(x_{0})=0$.}
\label{c-rho0-trig}
\end{figure}

The speed of sound is fractional but does not remain positive as
illustrated in Figure (\ref{c-eRN-trig}).

The condition that the mass density goes to zero yields the
boundary
\begin{eqnarray*}
x_{0} &=&-\frac{\rho _{1}}{2\rho _{2}}+\frac{1}{2\rho _{2}}\sqrt{\rho
_{1}^{2}+4\rho _{0}\rho _{2}} \\
&=&\frac{5b}{16(-c)}+\frac{1}{16(-c)}\sqrt{25b^{2}+96a(-c)}.
\end{eqnarray*}
The junction condition (\ref{Junction-RN}) and equation (\ref{metric primes
equal}) are used to solve for the constants $C_{1}$ and $C_{2}$:
\begin{eqnarray*}
C_{1} &=&\sqrt{z(x_{0})}\sin \sqrt{d}\xi _{0}+\frac{z_{x}(x_{0})}{2\sqrt{d}}%
\cos \sqrt{d}\xi _{0} \\
C_{2} &=&\sqrt{z(x_{0})}\cos \sqrt{d}\xi _{0}-\frac{z_{x}(x_{0})}{2\sqrt{d}}%
\sin \sqrt{d}\xi _{0}.
\end{eqnarray*}
The pressure, from equation (\ref{P-eqn-xi}), is given by
\begin{equation}
\kappa P=4\sqrt{dz}\left( \frac{C_{1}\cos \sqrt{d}\xi -C_{2}\sin \sqrt{d}\xi
}{C_{1}\sin \sqrt{d}\xi +C_{2}\cos \sqrt{d}\xi }\right) -a-bx-2cx^{2}.
\label{P-eqn-trig-xi}
\end{equation}
The condition that the pressure be positive at the center leads to the
relation
\[
\frac{C_{1}}{C_{2}}>\frac{a}{4\sqrt{d}},
\]
which is not satisfied, in general. We show an example in Figure (\ref
{c-rho0-trig})with negative pressures where the following values were used

The speed of sound in this case, begins at a
positive fractional value but decreases through zero to negative values as
the pressure first decreases and then increases to zero at the boundary.

Finally One might ask which solutions (of the 127 

\section{Conclusions}
 
It has been shown that by giving the mass density and charge aspect functions
explicitly was fourth-order polynomials in the curvature radial coordinate
leads to a simple linear second-order differential equation with constant
coefficients that can be easily solved.  With an appropriate choice of
the constants appearing in the polynomial, some of the solutions can be made to
satisfy the requirements for begin physically acceptable.  Given the small
number of known solutions for both charged and neutral fluid spheres that obey
such criteria, the configurations of charge and mass found in these solutions
may be expected to form from a realistic collapse to a condensed spherical
object.
 
While matching to the external Reissner-Nordstrom solution causes no problems
since this only determines the values for the constants of integration, it has
been shown that with this family of solutions it is impossible to find
realistic configurations that lead to extreme Reissner-Nordstrom solutions.
In all cases where the solution was matched to the extreme Reissner-Nordstrom
vacuum solution, problems arose in maintaining a speed of sound that is
bounded by the speed of light.
 
It remains to be determined how stable these configurations are, and this will
be studied in the future.  Given that the equation of state for the fluid
has not been written in the form $p=p(\rho)$ does make the stability analysis
a bit more difficult since it is not known if one can use adiabaticity
arguments without directly checking them independently.  However a general
stability analysis of this and other known exact solutions should be interesting
from the point of view of trying to understand what charged fluid
configurations could possibly be formed from the gravitational collapse of
material that has an excess charge density.
 
Finally, one should be able to extend this method to other explicit functions
for both the mass density and charge aspect.  Higher order polynomials are
expected to produce similar results and one can expect that these would force
more of the charge to be distributed closer to the surface of the sphere.
Further studies on what the most general polynomials might be and the
relationships between the charge and mass distributions will be studied in the
future.

\subsection{Acknowledgments}   

DH would like to thank D. Giang for discussions concerning methods for
solving the Einstein-Maxwell system of equations. This research was funded
through an NSERC (Canada) Discovery Grant.
 
\section{References}

\begin{itemize}
 
\item [1] S.D.~Majumdar, Phys.~Rev., {\bf 72}, 390 (1947).
 
\item [2] A.~Papapetrou, Proc.~Roy.~Irish Acad., A {\bf 51}, 191 (1947).
 
\item [3] B.V. Ivanov, Phys.~Rev.~D {\bf 65}, 104001, (2002).
 
\item [4] M.S.R.~Delgaty and K.~Lake, Comput.~Phys.~Commun. {\bf 115}, 395, (1998)
[arXiv:gr-qc/9809013].
 
\item [5] M.R.~Finch and J.E.F.~Skea, unpublished preprint,\\
www.dft.if.euerj.br/users/Jim\_Skea/papers/pfrev.ps.
 
\item [5] R.C. Tolman, Phys.~Rev.~ {\bf 55}, 363, (1939).
 
\item [7] S. Bayin, Phys.~Rev.~D {\bf 18}, 2745, (1975).
\end{itemize}
 
\end{document}